\documentclass[]{aa}
\usepackage{hyperref}
\usepackage{comment}
\usepackage[dvipsnames]{xcolor}
\usepackage{mathtools}
\usepackage{caption}

\usepackage{xcolor}
\hypersetup{    
    colorlinks=true,
    citecolor = blue,
    linkcolor=blue,
    filecolor=magenta,      
    urlcolor=cyan,
}
\usepackage{multirow}

\usepackage{graphicx}
\usepackage{txfonts}

 \def\mso{\,\mathrm{M_\odot}}
 
 \def\lso{\,\mathrm{L_\odot}}
 \def\Msun{\,\mathrm{M_\odot}}
 
 \def\Lsun{\,\mathrm{L_\odot}}

 \def\kms{\, \mathrm{km\,s^{-1}}}
 \def\gunit{\mathrm{cm/s^2}}

 \def\Myr{\,\text{Myr}}
 \def\pc{\,\text{pc}}
 \def\kpc{\,\text{kpc}}
 \def\dex{\,\mathrm{dex}}
 \def\K{\,\mathrm{K}}
 
 \def\MK{\,\mathrm{MK}}

 \def\Mini{M_\mathrm{i}}
 \def\Teff{T_\mathrm{eff}}
 \def\logg{\log g}
 \def\logL{\log L}  
 \def\vsini{v\,\mathrm{sin}\,i}
 \def\veq{v_\mathrm{eq}}
 \def\vini{v_\mathrm{rot, i}}

 \def\fc{f_c}
 \def\fmu{f_{\mu}}
 \def\fnu{f_{\nu}}
 \def\pev{P_\mathrm{ev}}
 \def\ptl{P_\mathrm{T,L}}
 \def\ptg{P_\mathrm{T,g}}
 \def\pfc{P(\fc)}
 
 \def\pfcproduct{\pfc^\mathrm{product}}
 \def\pmax{\pfc^\mathrm{product}_\mathrm{max}}
 \def\pmax{\Pi}

 \def\logB{\log \epsilon \mathrm{(B)}}
 \def\logC{\log \epsilon \mathrm{(C)}}
 \def\logN{\log \epsilon \mathrm{(N)}}
 \def\logO{\log \epsilon \mathrm{(O)}}
 \def\logBi{\log \epsilon \mathrm{(B)}_\mathrm{i}}
 \def\logCi{\log \epsilon \mathrm{(C)}_\mathrm{i}}
 \def\logNi{\log \epsilon \mathrm{(N)}_\mathrm{i}}
 
 \def\logBBi{\log \mathrm{B/B_{i}}}

 \def\logNC{\log \mathrm{(N/C)}}
 \def\logNO{\log \mathrm{(N/O)}}
 \def\logNCNCi{\log \mathrm{(N/C)/(N/C)_{i}}}
 \def\logNONOi{\log \mathrm{(N/O)/(N/O)_{i}}}

\def\simle{\mathrel{\hbox{\rlap{\hbox{\lower4pt\hbox{$\sim$}}}\hbox{$<$}}}}
 \def\simgr{\mathrel{\hbox{\rlap{\hbox{\lower4pt\hbox{$\sim$}}}\hbox{$>$}}}}

\newcommand{\Fig}[1]{Fig.\,\ref{#1}}
\newcommand{\Figure}[1]{Figure\,\ref{#1}}

\newcommand{\Table}[1]{Table\,\ref{#1}}
\newcommand{\App}[1]{App.\,\ref{#1}}

\begin{document}

   \title {Boron depletion in Galactic early B-type stars reveals two different main sequence star populations}
    

   \author{Harim Jin \inst{1}
          \and Norbert Langer \inst{1,2} 
          \and Daniel J. Lennon\inst{3,4} 
          \and Charles R. Proffitt \inst{5} 
          }

   \institute{Argelander Institut für Astronomie,
              Auf dem Hügel 71, DE-53121 Bonn, Germany\\
              \email{hjin@astro.uni-bonn.de}
         \and
          Max-Planck-Institut für Radioastronomie, Auf dem Hügel 69, DE-53121 Bonn, Germany
         \and
          Instituto de Astrofísica de Canarias, 38 200 La Laguna, Tenerife, Spain
          \and
          Dpto. Astrofísica, Universidad de La Laguna, 38 205 La Laguna, Tenerife, Spain
          \and
        Space Telescope Science Institute, Baltimore, MD 21218, USA
    }

   \date{Received Month Day, 2024; accepted Month Day, 2024}

 
  \abstract
   {
   The evolution and fate of massive stars are thought to be affected by rotationally induced internal mixing. The surface boron abundance is a sensitive tracer of this in early B-type main sequence stars.
   }
   {
   We test current stellar evolution models of massive main sequence stars which include rotational mixing through a systematic study of their predicted surface boron depletion.
    }
   {We construct a dense grid of rotating single star models using MESA, for which we employ a new nuclear network which follows all the stable isotopes up to silicon, including lithium, beryllium, boron, as well as the radioactive isotope aluminium-26. We also compile the measured physical parameters of the 90 Galactic early B-type stars with boron abundance information. We then compare each observed stars with our models through a Bayesian analysis, which yields the mixing efficiency parameter with which the star is reproduced the best, and the probability that it is represented by the stellar models. 
    }
   {
   We find that about two-thirds of the sample stars are well represented by the stellar models, with the best agreement achieved for a rotational mixing efficiency of $\sim$50\% compared to the widely adopted value. The remaining one third of the stars, of which many are strongly boron depleted slow rotators, are largely incompatible with our models, for any rotational mixing efficiency. We investigate the observational incidence of binary companions and surface magnetic fields, and discuss their evolutionary implications.}
   {Our results confirm the concept of rotational mixing in radiative stellar envelopes. On the other hand, we find that a different boron depletion mechanism, and likely a different formation path, is required to explain about one-third of the sample stars. The large spread in the surface boron abundances of these stars may hold a clue to understanding their origin.
   }

   \maketitle
%

\section{Introduction}\label{sec:intro}
Massive stars are regarded as key agents for the chemical and dynamical evolution of star forming galaxies \citep{Timmes1995,MacLow2004}. 
It is therefore important to understand their evolution and explosions. Rotationally induced internal mixing has been recognized to significantly affect the evolution of massive stars \citep[e.g.,][]{Meynet1997, Heger2000a, Meynet2000, Limongi2018, Higgins2019}. Fast rotating stars can develop larger helium cores, enrich their surface with nuclear burning products, and in extreme cases, experience quasi-chemically homogeneous evolution. This may give rise to interesting evolutionary paths, such as the formation of rapidly rotating Wolf-Rayet stars and long Gamma-ray burst progenitors \citep{Yoon2005,Woosley2006,Aguilera2018}. 

However, the efficiency of rotational mixing remains uncertain. One way to constrain the efficiency is to consider the evolution of the surface abundances in stars. In massive stars, the surface nitrogen abundance has been used to calibrate the efficiency \citep{Heger2000b,Yoon2006,Brott2011a}. Nitrogen enhancement occurs when CNO-processed material is transported to the stellar surface. This enhancement is expected to be stronger for faster rotation and higher rotational mixing efficiency.

Furthermore, it has become evident in recent years that the majority of massive stars are born in close binary systems \citep{Sana2012}. In such binaries, the surface abundances may be altered by mass transfer \citep{Eldridge2008, Langer2012} or tides \citep{Song2018,Hastings2020a,Koenigsberger2021} while their surface rotational velocities are also affected \citep{Langer2020}. At the same time, spectral analyses of large samples of massive main sequence stars uncovered sub-populations of intrinsically slowly rotating stars which show a significant nitrogen enrichment \citep{Morel2006,Hunter2008b,Hunter2009,Rivero2012,Grin2017}.

In order to discriminate effects from rotational mixing and from binary processes, and to further constrain the rotational mixing efficiency, it is useful to consider the surface abundances of the light elements lithium, beryllium, and boron in massive stars. These elements are fragile in that they undergo proton capture at lower temperatures ($3\dots 6 \MK$) than those required for the CNO-cycle ($\sim17\MK$). In massive main sequence stars, the light elements are destroyed throughout the interior except for the outermost about one solar mass of the envelope \citep{Fliegner1996}. While lithium and beryllium cannot be observed in hot massive stars, boron measurements are possible in early B-type stars using boron absorption lines n the near UV \citep{Venn2002,Cunha2009,Kaufer2010}.


To investigate surface boron depletion in upper main sequence stars as a test of mixing was first proposed by \citet{Venn1996}. They found an anti-correlation between nitrogen enhancement and boron depletion in their stars, and \citet{Fliegner1996} showed that stellar models including rotational mixing could account for this trend. Since then, several sets of stellar evolution models have been used to probe rotational boron depletion \citep{Heger2000a,Brott2009,Mink2009,Frischknecht2010,Brott2011a}. Among these, two sets of stellar models have been confronted with boron observations. One consists of models computed using the stellar evolution code STERN \citep{Heger2000a}, and the other computed using the Geneva code (GENEC) \citep{Frischknecht2010}.

With STERN models from \citet{Heger2000a}, \citet{Mendel2006} showed that 12$\mso$ models with a range of rotational velocities could account for the observed dispersion of boron depletion. \citet{Venn2002} and \citet{Morel2008} used the same models and showed that the majority of their sample of stars followed the nitrogen enhancement and boron depletion trend predicted by rotating single star models. However, these works did not take into account the sample bias towards sharp-lined stars \citep{Kaufer2010}. On the other hand, \citet{Frischknecht2010} did consider the bias towards low $\vsini$ in their comparison, based on GENEC models. They compared the fraction of fast rotators that can deplete boron with the fraction of stars with boron depletion. They found that the numbers agreed well, but they did not take into account individual stellar ages in their analysis. 

Here, we present a Bayesian framework for testing rotational mixing comprehensively, considering all observationally derived stellar parameters simultaneously. Over the past decade, dedicated studies of boron abundances in stars within NGC\,3293 have substantially expanded the dataset of stars with boron abundance information \citep{Proffitt2016,Proffitt2024}. This dataset also includes a substantial number of relatively fast rotators, which is significant since rotational mixing has a more pronounced effect on fast rotators. We compute a new model grid of rotating single stars and conduct a quantitative study of the rotational mixing efficiency via our framework. In Section~\ref{sec_mod}, we explain the details of our new model grid. In Section~\ref{sec_obs}, we provide the dataset of stars with boron abundance information and discuss their properties and biases. In Section~\ref{sec_meth}, we present our Bayesian method for comparing stellar models and observed stars and for testing the rotational mixing efficiency. In Section~\ref{sec_res}, we present our main results.  We discuss the key uncertainties and the implications of our results in Section~\ref{sec_dis} and conclude the paper in Section~\ref{sec_con}.

\section{Stellar model description}\label{sec_mod}
We use the one-dimensional stellar evolution code MESA version 10398 \citep{Paxton2011,Paxton2013,Paxton2015,Paxton2018,Paxton2019} to calculate single star models. Unless noted otherwise, the same physical assumptions are adopted as in \citet{Brott2011a}. The input files including the initial abundances, opacity tables, nuclear network, zero-age main sequence models, etc. to reproduce the MESA calculations are available online along with the complete single star model grid.\footnote{https://doi.org/10.5281/zenodo.11203797}

\subsection{Initial abundances and opacity tables}
We adopt the proto-solar abundances provided by \citet{Asplund2021} as the initial abundances. These are corrected for the effects of atomic diffusion, radioactive decay, and nuclear burning that happened throughout the evolution of the Sun from the present-day solar photospheric values. Table~\ref{tab_abn} presents the list of the initial abundances for our MESA calculation. Even though the adopted initial abundances can differ from the initial abundances for the early B-type stars of our concern, this does not affect our analysis significantly since we do not use the absolute values of surface abundances (e.g., $\logB$) for comparing the stars and the stellar models, but the changes in them (e.g., $\logB - \logBi$, which hereafter referred to as $\logBBi$; similarly for other elements).

Table~\ref{tab_cno} compares the initial abundances adopted for four Galactic single star model grids, along with reference abundances for stars in NGC\,3293, the solar neighborhood, and the present-day Sun. BROTT GAL models \citep{Brott2011a} are calculated with STERN \citep{Heger2000b} based on the compilation of abundances obtained from \citet{Venn1995,Asplund2005,Hunter2007,Hunter2008a,Hunter2009}. GENEC models \citep{Frischknecht2010} are calculated with the Geneva stellar evolution code based on the photospheric abundances from \citet{Asplund2005} with modifications for neon. MIST models \citep{Choi2016} are calculated with MESA based on proto-solar abundances \citep{Asplund2009} for their $\mathrm{[Fe/H]=0.0}$ grid.

\begin{table}[]
\begin{center}
\caption{Initial abundances adopted.}
\label{tab_abn}
\begin{tabular}{cc|cc}
\hline \hline
Isotope & Mass fraction & Isotope & Mass fraction \\ \hline
$^{}\rm n$ & 0.000E+00 & $^{19}\rm F$ & 3.882E-07 \\
$^{1}\rm H$ & 7.121E-01 & $^{20}\rm Ne$ & 1.743E-03 \\
$^{2}\rm H$ & 2.364E-05 & $^{21}\rm Ne$ & 4.814E-06 \\
$^{3}\rm He$ & 3.391E-05 & $^{22}\rm Ne$ & 1.454E-04 \\
$^{4}\rm He$ & 2.725E-01 & $^{21}\rm Na$ & 0.000E+00 \\
$^{6}\rm Li$ & 4.059E-10 & $^{22}\rm Na$ & 0.000E+00 \\
$^{7}\rm Li$ & 9.449E-09 & $^{23}\rm Na$ & 3.105E-05 \\
$^{7}\rm Be$ & 0.000E+00 & $^{24}\rm Mg$ & 5.502E-04 \\
$^{9}\rm Be$ & 1.756E-10 & $^{25}\rm Mg$ & 7.215E-05 \\
$^{8}\rm B$ & 0.000E+00 & $^{26}\rm Mg$ & 8.419E-05 \\
$^{10}\rm B$ & 8.134E-10 & $^{25}\rm Al$ & 0.000E+00 \\
$^{11}\rm B$ & 3.645E-09 & $^{26}\rm Al^{g}$ & 0.000E+00 \\
$^{11}\rm C$ & 0.000E+00 & $^{26}\rm Al^{*}$ & 0.000E+00 \\
$^{12}\rm C$ & 2.815E-03 & $^{27}\rm Al$ & 5.911E-05 \\
$^{13}\rm C$ & 3.422E-05 & $^{27}\rm Si$ & 0.000E+00 \\
$^{12}\rm N$ & 0.000E+00 & $^{28}\rm Si$ & 6.878E-04 \\
$^{14}\rm N$ & 7.699E-04 & $^{29}\rm Si$ & 3.570E-05 \\
$^{15}\rm N$ & 1.890E-06 & $^{30}\rm Si$ & 2.440E-05 \\
$^{16}\rm O$ & 6.374E-03 & $^{40}\rm Ca$ & 7.248E-04 \\
$^{17}\rm O$ & 2.459E-06 & $^{56}\rm Fe$ & 1.226E-03 \\
$^{18}\rm O$ & 1.366E-05 &  &  \\ \hline
\end{tabular}
\end{center}
\tablefoot{
$^{26}\rm Al$ is considered in its ground state $^{26}\rm Al^{g}$ and in its first excited state  $^{26}\rm Al^{*}$. See text for more details.}
\end{table}

\begin{table*}[]
\caption{Initial abundances adopted for four Galactic single star model grids and reference abundances of concern. 
We use $\logC \equiv \log (\mathrm{C/H}) + 12$, $\logN \equiv \log (\mathrm{N/H}) + 12$, $\logO \equiv \log (\mathrm{O/H}) + 12$, $\logB \equiv \log (\mathrm{B/H}) + 12$ throughout the paper.}
\label{tab_cno}
\resizebox{\linewidth}{!}{%
\begin{tabular}{ccccccccc}
\hline \hline
 & $X$ & $Y$ & $Z$ & $\logC$ & $\logN$ & $\logO$ & $\logB$ & Ref. \\ \hline
BROTT GAL$^a$ & 0.7274 & 0.2638 & 0.0088 & 8.13 & 7.64 & 8.55 & 2.7 & \citet{Brott2011a} \\
GENEC$^b$ & 0.72 & 0.266 & 0.014 & 8.43 & 7.82 & 8.70 & 2.74 & \citet{Frischknecht2010} \\
MIST$^c$ & 0.7154 & 0.2703 & 0.0142 & 8.47 & 7.87 & 8.73 & ... & \citet{Choi2016} \\
This work$^d$ & 0.7121 & 0.2725 & 0.0154 & 8.52 & 7.89 & 8.75 & 2.76 & \citet{Asplund2021} \\ \hline
NGC\,3293 & ... & ... & ... & 7.97 $\pm$ 0.19 & 7.60 $\pm$ 0.15 & 8.65 $\pm$ 0.17 & ... & \citet{Hunter2009} \\
B-type stars$^e$ & 0.71 & 0.276 & 0.014 $\pm$ 0.002 & 8.33 $\pm$ 0.04 & 7.79 $\pm$ 0.04 & 8.76 $\pm$ 0.05 & ... & \citet{Nieva2012} \\
Present-day Sun$^f$ & 0.7438 & 0.2423 & 0.0139 & 8.46 & 7.83 & 8.69 & 2.7 & \citet{Asplund2021} \\ \hline
\end{tabular}}
\tablefoot{ 
\tablefoottext{a}{Own mixture from a Galactic cluster \citep{Hunter2007}, HII-regions \citep{Hunter2008a,Hunter2009}, A-supergiants \citep{Venn1995}, and present-day solar abundances \citep{Asplund2005};}
\tablefoottext{b}{\citet{Asplund2005};}
\tablefoottext{c}{Proto-solar abundances from \citet{Asplund2009} for their $\mathrm{[Fe/H]=0.0}$ grid;}
\tablefoottext{d}{Proto-solar abundances from \citet{Asplund2021};}
\tablefoottext{e}{Solar neighborhood ($< 500 \pc$) B-type stars;}
\tablefoottext{f}{Present-day solar abundances from \citet{Asplund2021};}
}
\end{table*}

Opacity tables are newly generated via the OPAL project website \citep{Iglesias1996} for the high temperature regime ($\log T \, \mathrm{[K]} > 4$). For the low temperature regime ($\log T \, \mathrm{[K]} < 4$), we adopt opacity tables by \citet{Ferguson2005} in which the effects of molecules are taken into account.

\subsection{Nuclear network}
We adopt the nuclear network used in STERN \citep{Heger2000b}, the same as BROTT GAL models. It consists of 77 reactions to follow the p-p chains, the CNO-, Ne-Na and Mg-Al cycles, and a simple treatment of helium, carbon, neon, and oxygen burning. In particular, this network includes reactions including the light elements (lithium, beryllium, and boron), which are often neglected due to their low abundances and negligible energy generation. The list of the employed reactions 
is provided in \App{app_com} along with a test run of STERN nuclear network against the MESA default nuclear network. The nuclear network adopted by GENEC models follows the evolution of all the stable isotopes of the light elements as our models, while that of MIST models does not. The evolution of the surface abundances of all the stable isotopes ranging from hydrogen to silicon is followed, and also that of $^{26}\rm Al$. $^{26}\rm Al$ is a radioactive isotope produced by the Mg-Al cycle, and it is considered with its ground state $^{26}\rm Al^{g}$ in its excited state $^{26}\rm Al^{*}$. Due to its long lifetime of $\sim\,1\Myr$, $^{26}\rm Al^{g}$ is a gamma-ray source in the Galactic plane \citep{Diehl1995}, thus its surface abundance is followed for future research.

\subsection{Mixing}\label{sec_mix}

Mixing of chemicals and angular momentum is treated as a diffusion process \citep{Paxton2013}. We adopt the input physics for convection, semiconvection, rotationally-induced instabilities, Spruit-Tayler dynamo as in BROTT GAL models, except for the overshooting; we adopt mass-dependent convective core overshooting as described in \citet{Hastings2021}, which adopted a linear increase of $\alpha_\mathrm{ov}$ from 0.1 at 1.66$\mso$ \citep{Claret2016} to 0.3 at 20$\mso$ \citep{Brott2011a}. Rotationally-induced instabilities encompass Eddington-Sweet circulation, dynamical shear instability, secular shear instability, and Goldreich-Schubert-Fricke instability \citep{Heger2000b}.

\begin{figure}
	\centering
	\includegraphics[width=\linewidth]{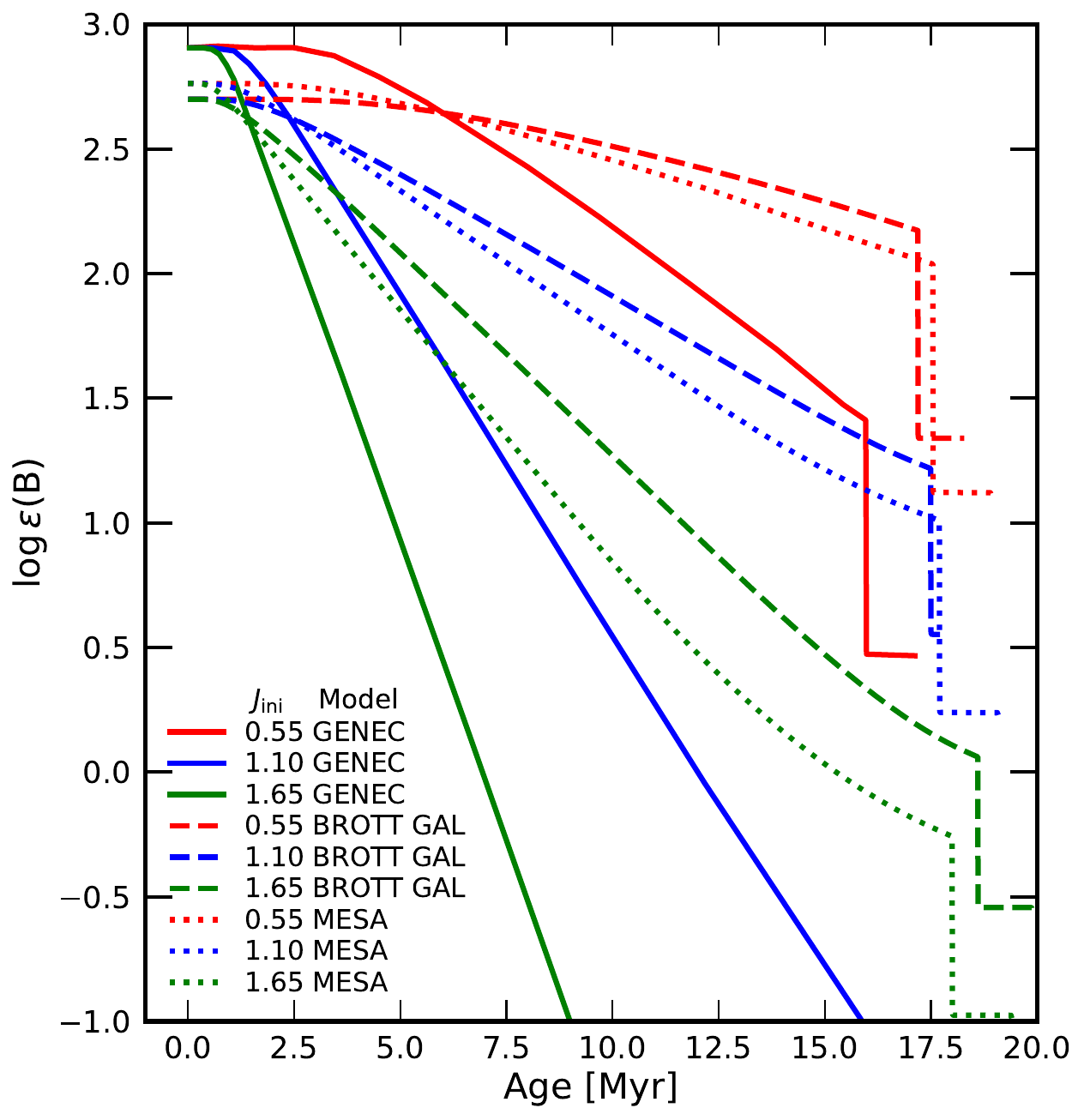}
	\caption{Surface boron abundance as a function of time for 12$\mso$ models with different initial angular momenta (in $10^{52}\,\mathrm{erg\,s}$) from different sets of stellar models. GENEC models from \citet{Frischknecht2010}, BROTT GAL models from \citet{Brott2011a}, and MESA models from this paper.}
	\label{fig_com}
\end{figure}

The Spruit-Tayler dynamo gives rise to a magnetic field, which effectively transports angular momentum via magnetic torques \citep{Heger2005,Suijs2008}. Consequently, our stellar models exhibit nearly rigid-body rotation during the main sequence evolution. Thus, the primary rotationally-induced instability governing chemical mixing in our models is Eddington-Sweet circulation, not the shear instabilities as in the GENEC models of \citet{Frischknecht2010}. \Figure{fig_com} compares its evolution in 12$\mso$ models from different calculations. Surface boron depletion is stronger for a higher initial rotation rate in all the models. While GENEC models predict a more rapid boron depletion compared to the BROTT GAL models and MESA models \citep[see also][]{Nandal2024}, those two show a quantitatively very similar evolution. We note that the magnetic angular momentum transport is supported by observations of many different types
of stars \citep[e.g.,][]{Suijs2008,Mosser2012,Takahashi2021,Schurmann2022}.

\begin{figure*}
	\centering
	\includegraphics[width=0.45\linewidth]{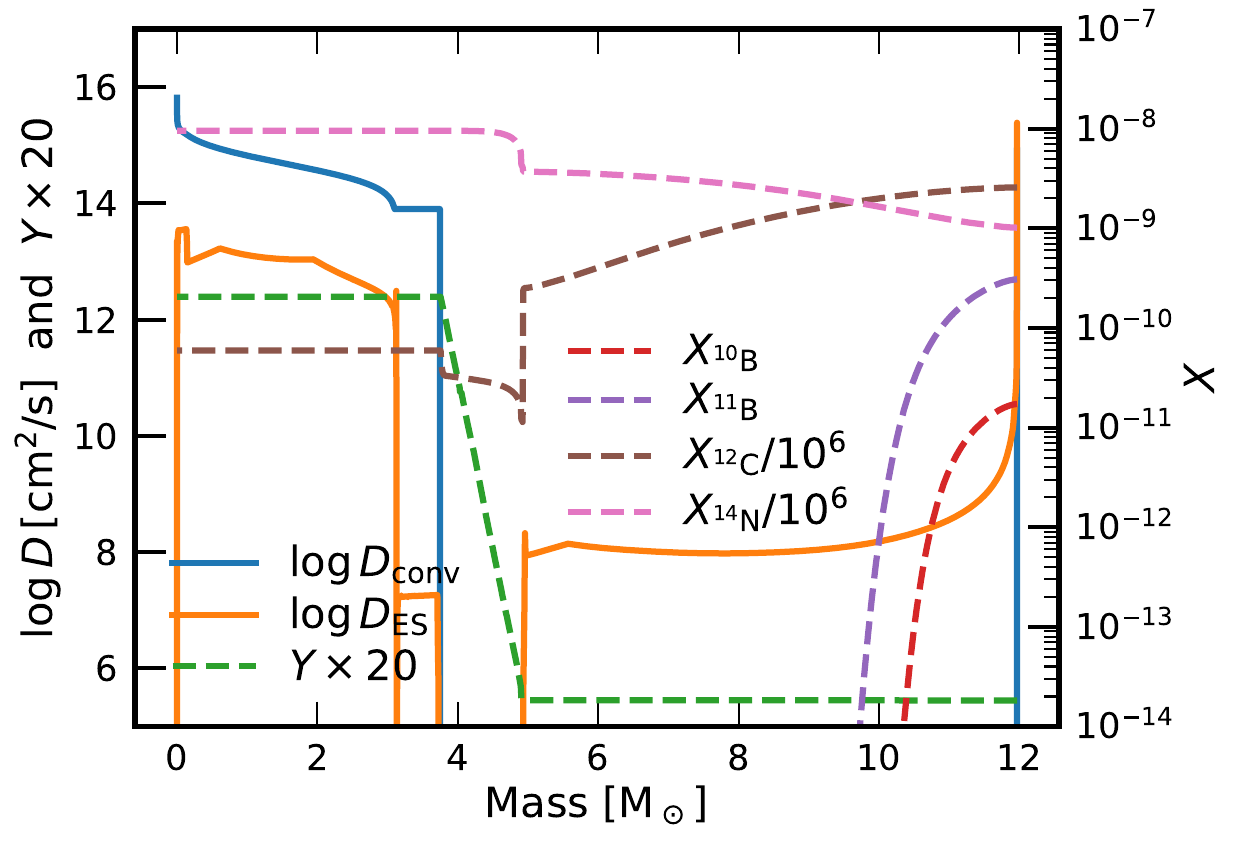}
	\includegraphics[width=0.45\linewidth]{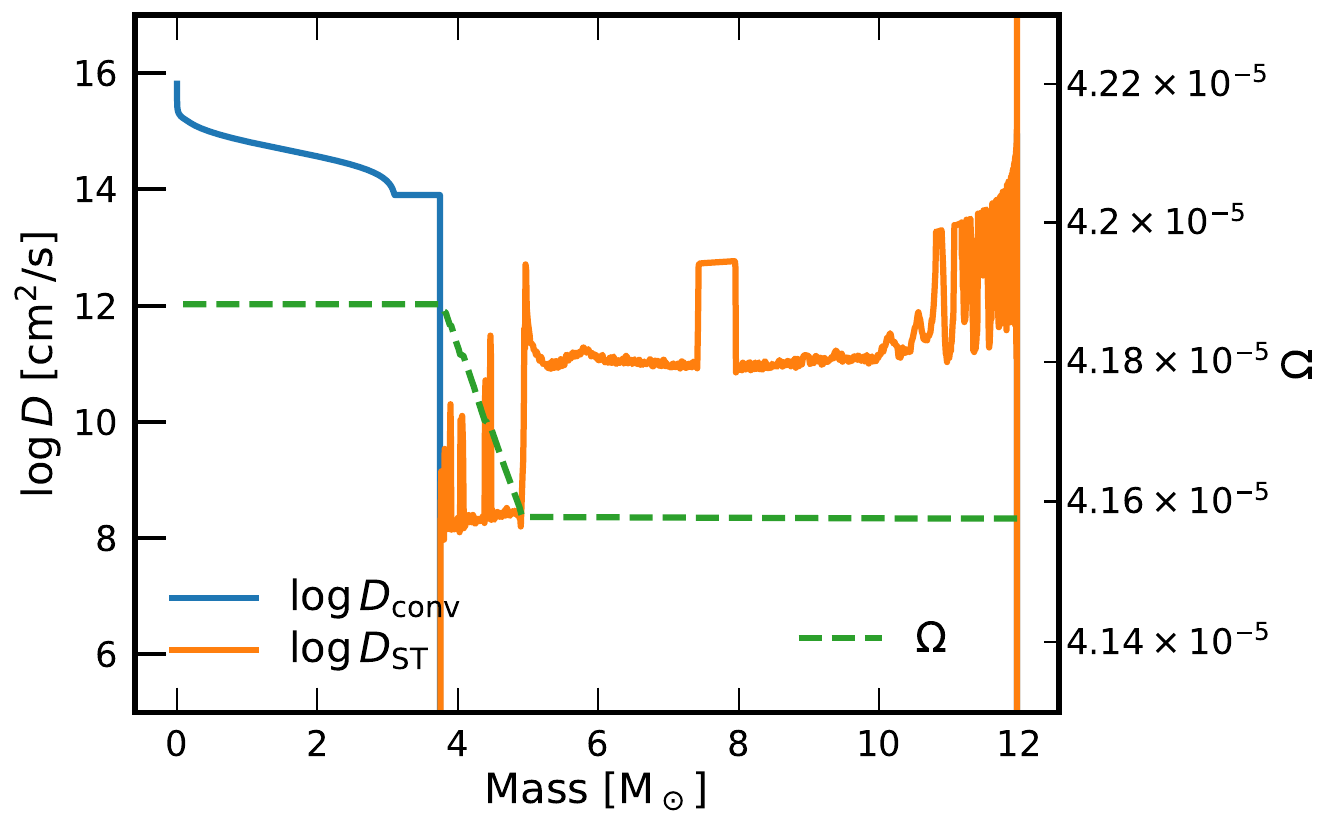}
	\caption{\textit{Left}: diffusion coefficients for chemical mixing and the helium abundance have values indicated by the left $y$-axis. Abundances of both boron isotopes, carbon, and nitrogen have values indicated by the right $y$-axis. The solid lines indicate diffusion coefficients, while the dashed lines indicate chemical abundances. Helium mass fraction $Y$ is multiplied by 20 for better visibility. \textit{Right}: diffusion coefficients for angular momentum transport (solid lines) and angular velocity (dashed line). All the profiles are from a 12$\mso$ star with an initial velocity of $200 \kms$ at 50\% central hydrogen depletion.}
	\label{fig_logD}
\end{figure*}

\Figure{fig_logD} elucidates the dominant mixing processes in our $12 \mso$ model during the main sequence evolution. The left panel shows diffusion coefficients for chemical mixing and abundances of interest. Above the convective core, the strengths of the other rotational instabilities than the Eddington-Sweet circulation are negligible, and their diffusion coefficients fall below the lower limit of the y-axis range.

\begin{figure*}
	\centering
	\includegraphics[width=0.48\linewidth]{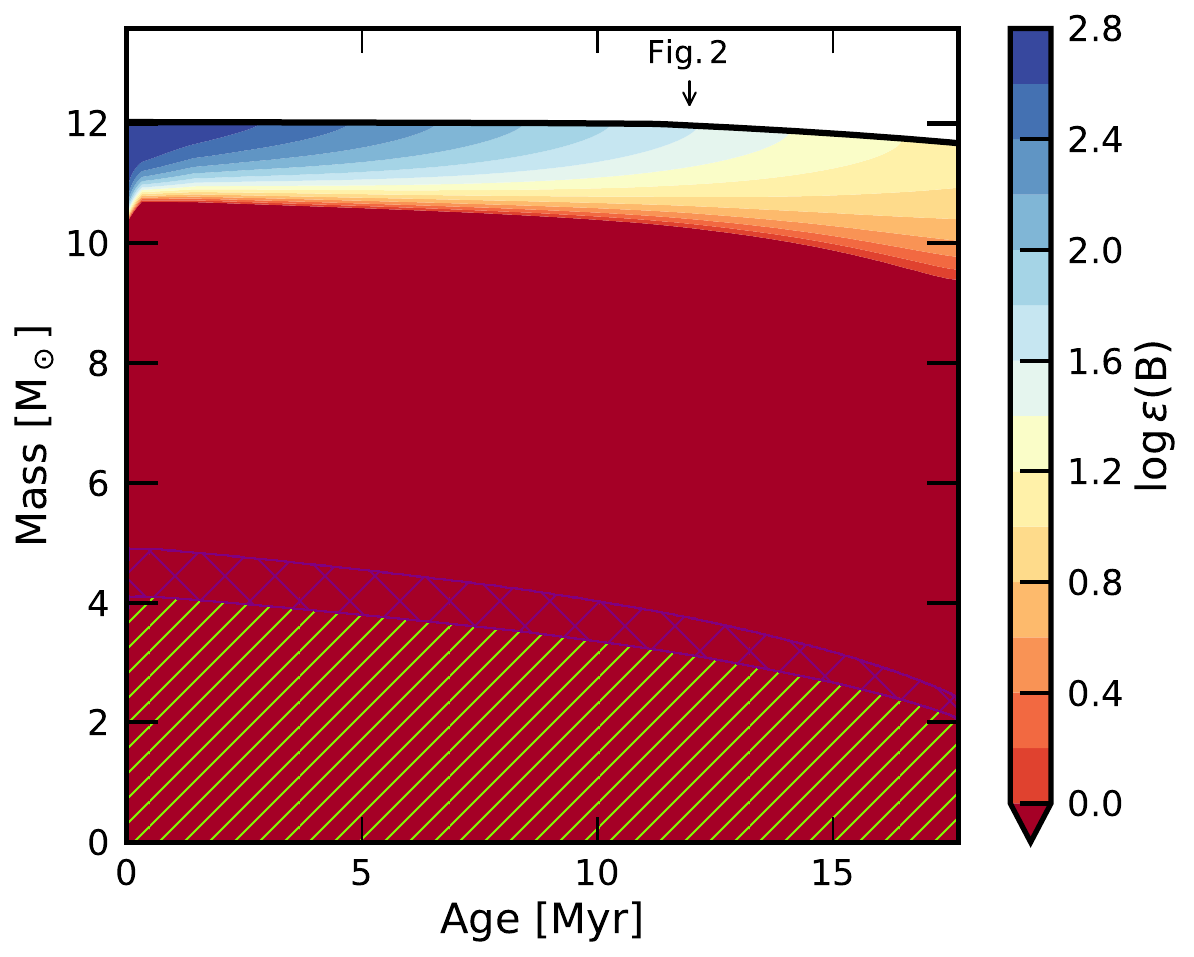}
	\includegraphics[width=0.48\linewidth]{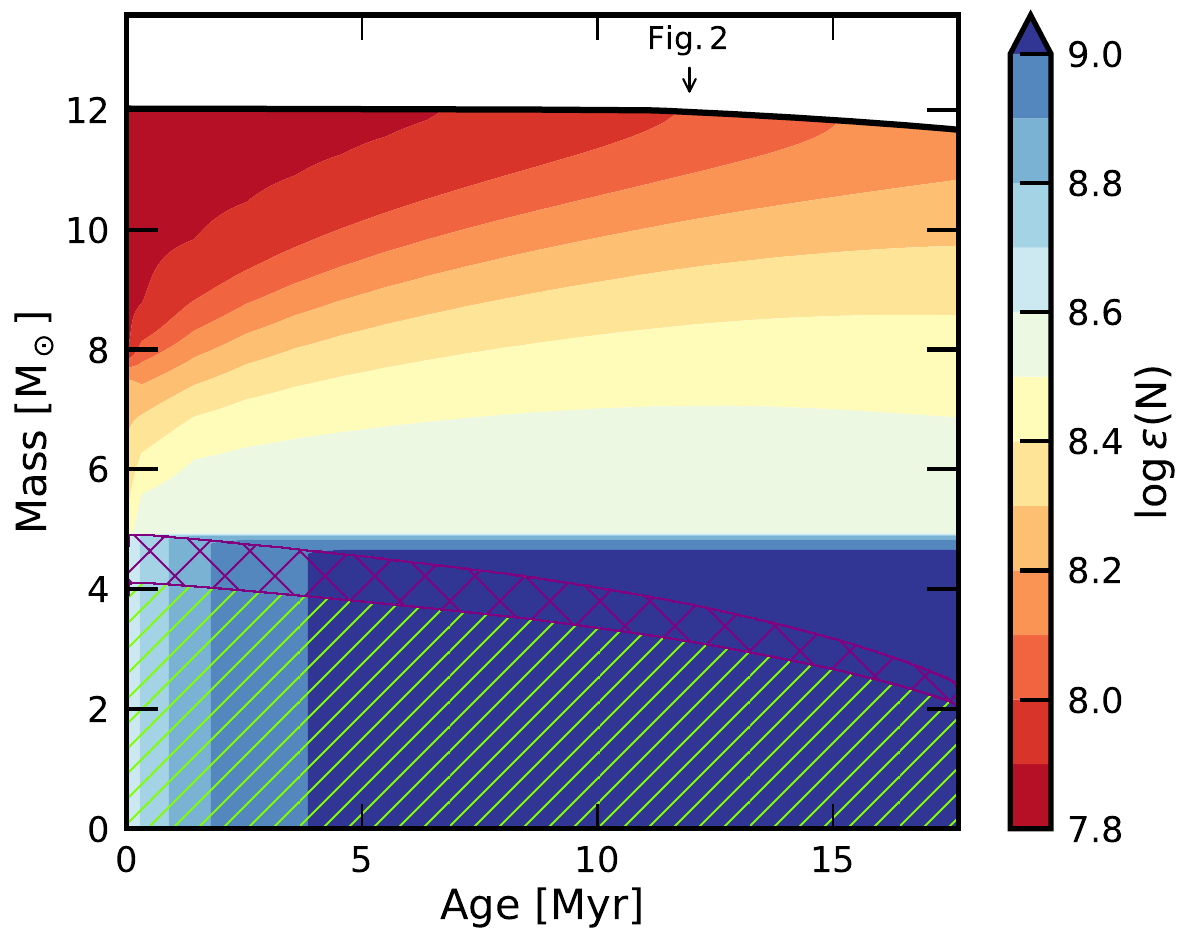}
	\caption{Kippenhahn diagrams showing the time evolution of the internal boron and nitrogen abundance for a 12$\mso$ star with an initial velocity of $200 \kms$ during core hydrogen burning. The green hatched area designates the convective core, while the purple cross-hatching designates the overshooting region. The arrows mark the time corresponding to \Fig{fig_logD}.}
	\label{fig_Kipp}
\end{figure*}

The boron isotopes ($^{10}\rm B$, $^{11}\rm B$) are depleted in the interior deeper than $\sim\,1.5 \mso$ below the surface, and their abundances at the surface have decreased from their initial values ($X_\textrm{B10}=8.1\times10^{-10}$; $X_\textrm{B11}=3.6\times10^{-9}$) due to mixing. Carbon and nitrogen abundances at the surface have only slightly changed from their initial values ($X_\textrm{C12}=2.8\times10^{-3}$; $X_\textrm{N14}=7.7\times10^{-4}$) due to the hydrogen-helium gradient at mass coordinate of $4-5 \mso$, which hinders the outward transport of CNO-processed material from the core. \Figure{fig_Kipp} demonstrates the effect of rotational mixing on the internal chemical structure. 

The right panel of \Fig{fig_logD} presents the diffusion coefficients for angular momentum transport, along with the angular velocity profile. The highly efficient angular momentum transport throughout the star leads to nearly rigid-body rotation, making shear instabilities insignificant in our models.

\subsection{Stellar wind mass-loss}

Our stellar wind mass-loss prescription is mostly the same as in \citet{Brott2011a}. For the stars with hydrogen-rich envelope, in which the surface hydrogen mass fraction $X_\mathrm{s}$ is above 0.7, the mass-loss rate of \citet{Vink2001} is used for hot stars, and the maximum of this rate and that of \citet{Nieuwenhuijzen1990} is used for cooler stars. The boundary between the two regimes is defined by the bi-stability jump at $\sim 25\, \mathrm{kK}$, at which a jump in the iron opacity is expected.

We only address the stars with hydrogen-rich envelopes in this study, but we will describe the full wind prescription adopted in our stellar model calculations for completeness. For the stars with $X_\mathrm{s}<0.7$, a compilation of mass-loss rates for Wolf-Rayet stars is used. For the stars with a relatively high surface hydrogen abundance ($0.2<X_\mathrm{s}<0.4$), the wind mass-loss rate from \citet{Nugis2000} enhanced by a factor of about 2 based on the clumping factor adopted in \citet{Hamann2006} is used. For the stars in which hydrogen is depleted at the surface ($X_\mathrm{s}<10^{-5}$), we adopt the prescription of \citet{Yoon2017}, whereby the mass-loss rate obtained from \citet{Hamann2006} and \citet{Hainich2014} is used for the hydrogen-free WN phase (during which the surface helium mass fraction $Y_\mathrm{s}$ is above 0.9) and that of \citet{Tramper2016} for the WC phase ($Y_\mathrm{s}<0.9$). These mass-loss rates can impact our stellar models after the core hydrogen burning phase if they strip their envelope.

Mass-loss rates are interpolated linearly in between different recipes for smooth transitions. The enhancement of mass-loss by rotation is modeled as in \citet{Yoon2005}. Since most of the observed stars considered in this study are main sequence stars with evolutionary masses of $\lesssim 20 \mso$ (Sect.\,\ref{sec_obs}), stellar wind mass-loss is expected to have had a negligible impact on their evolution (Sect.\,\ref{sec_grid}).

\subsection{Stellar model grid}\label{sec_grid}
\begin{figure*}
	\centering
	\includegraphics[width=0.45\linewidth]{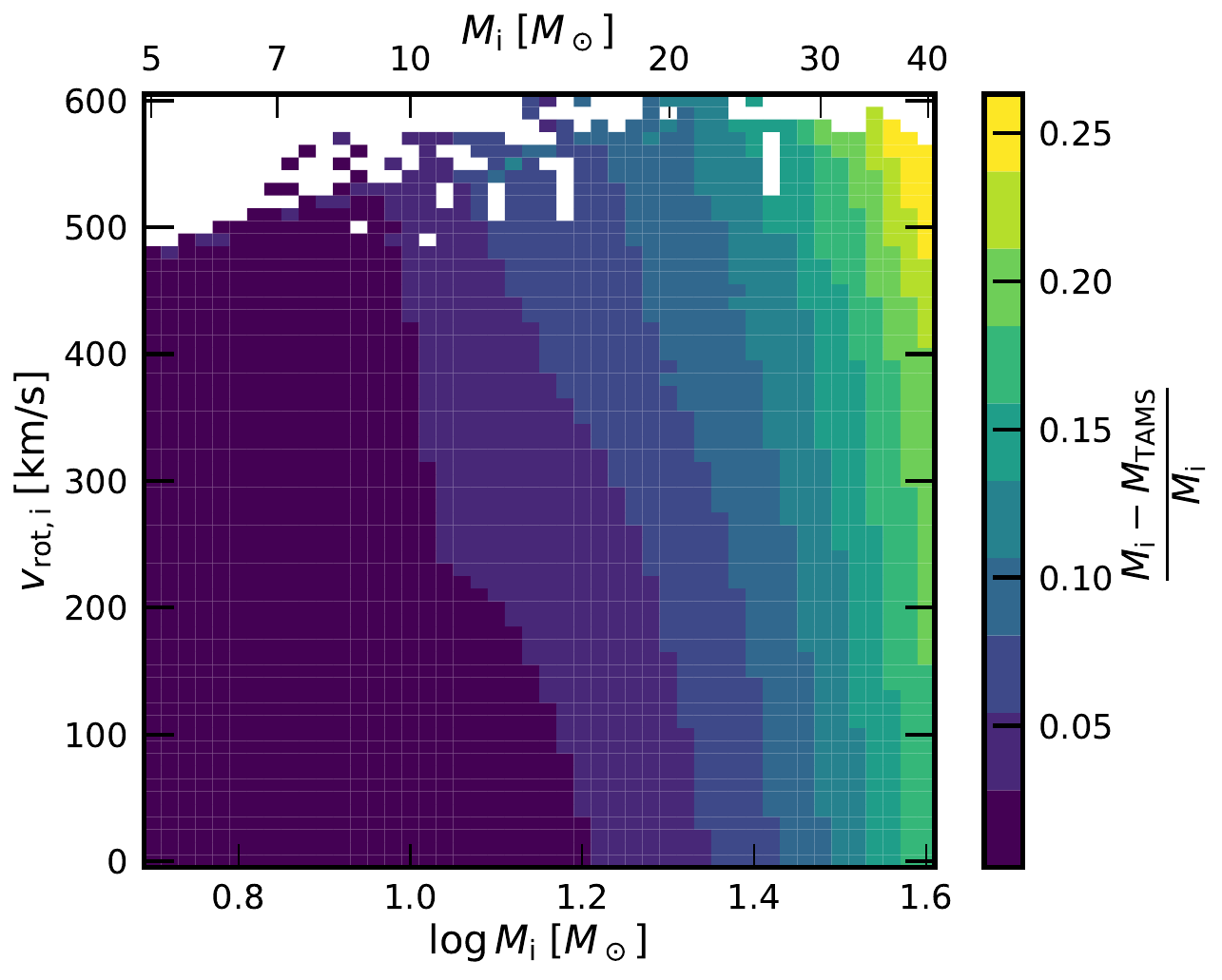}
	\includegraphics[width=0.45\linewidth]{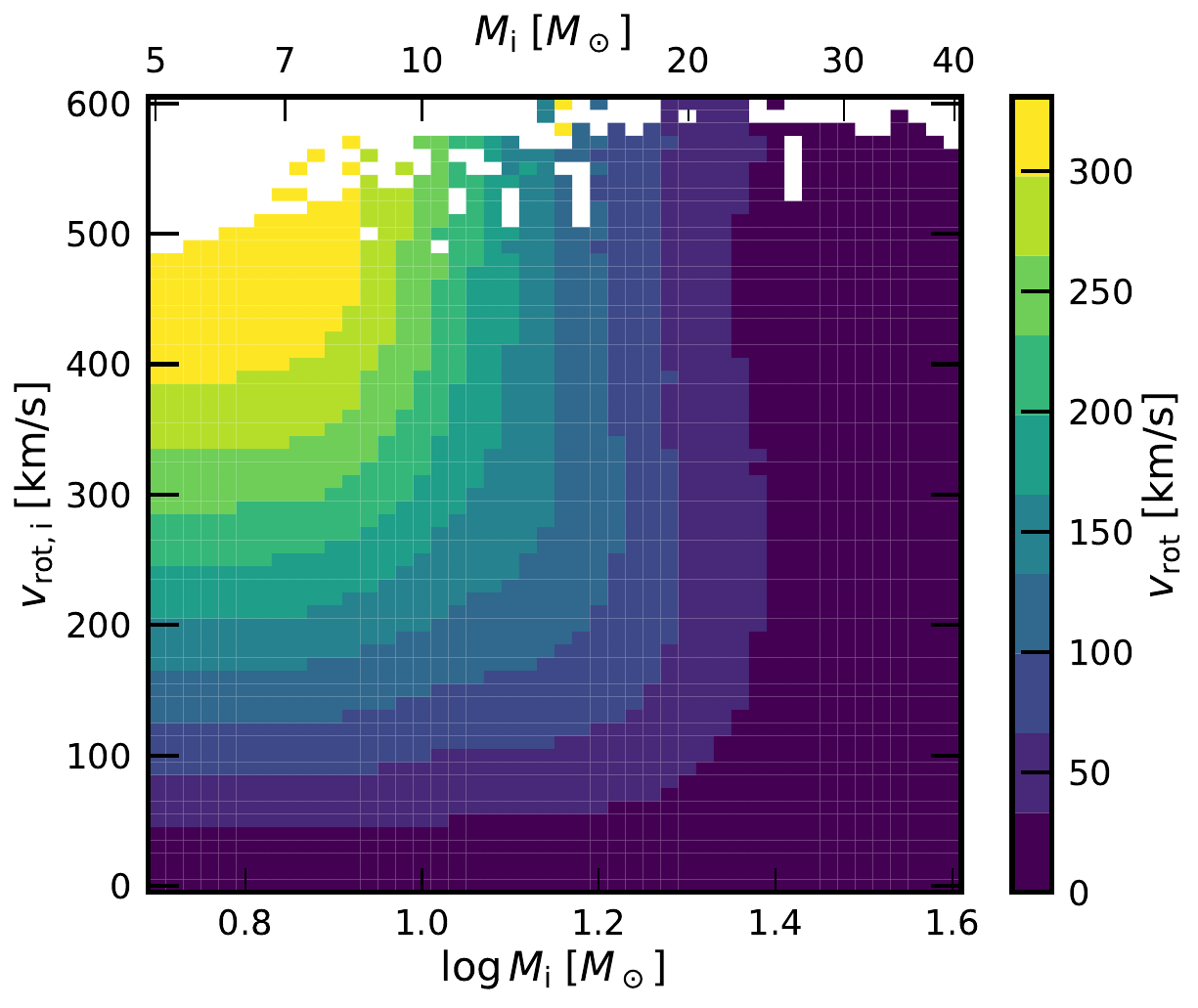}
	\includegraphics[width=0.45\linewidth]{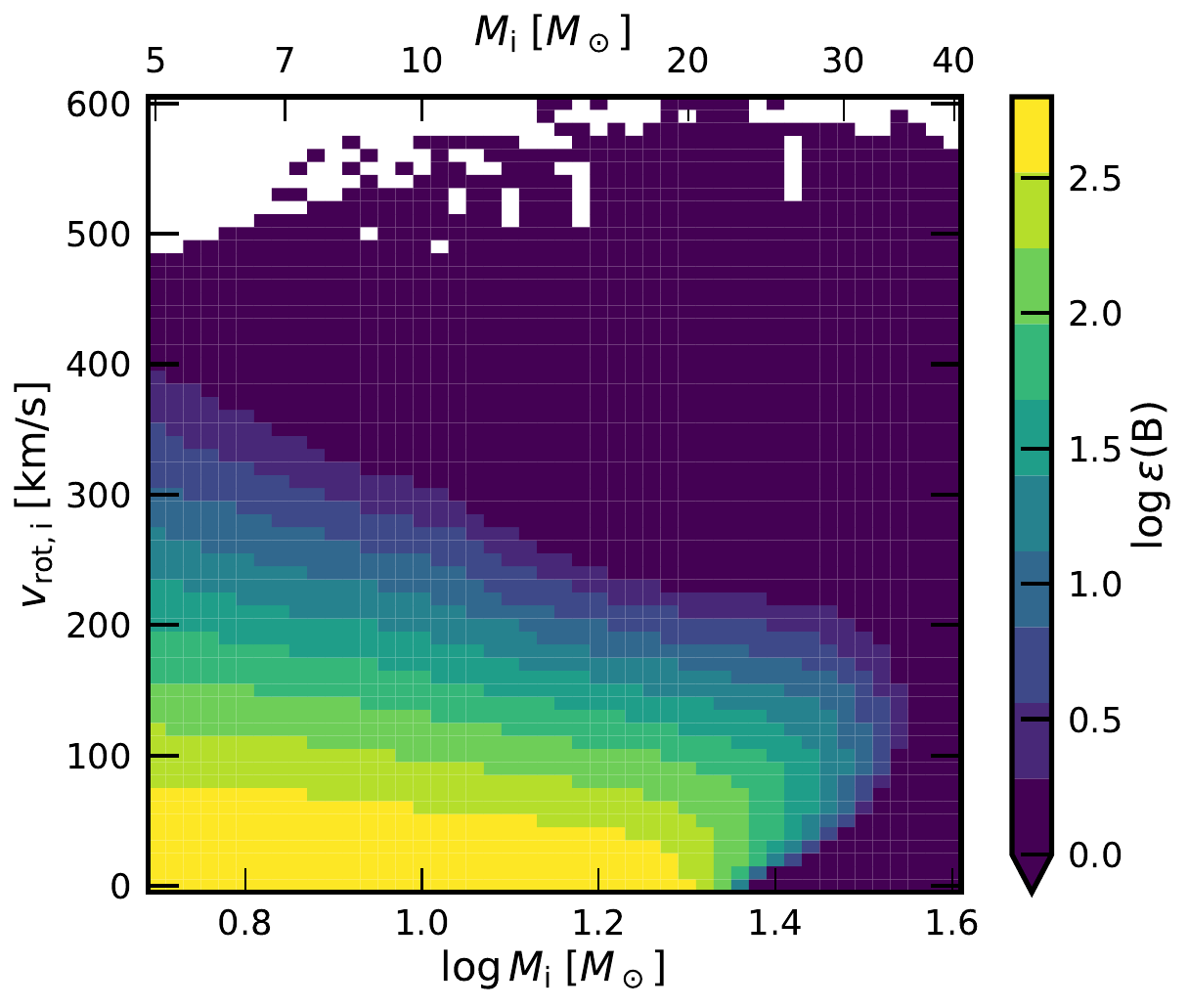}
	\includegraphics[width=0.45\linewidth]{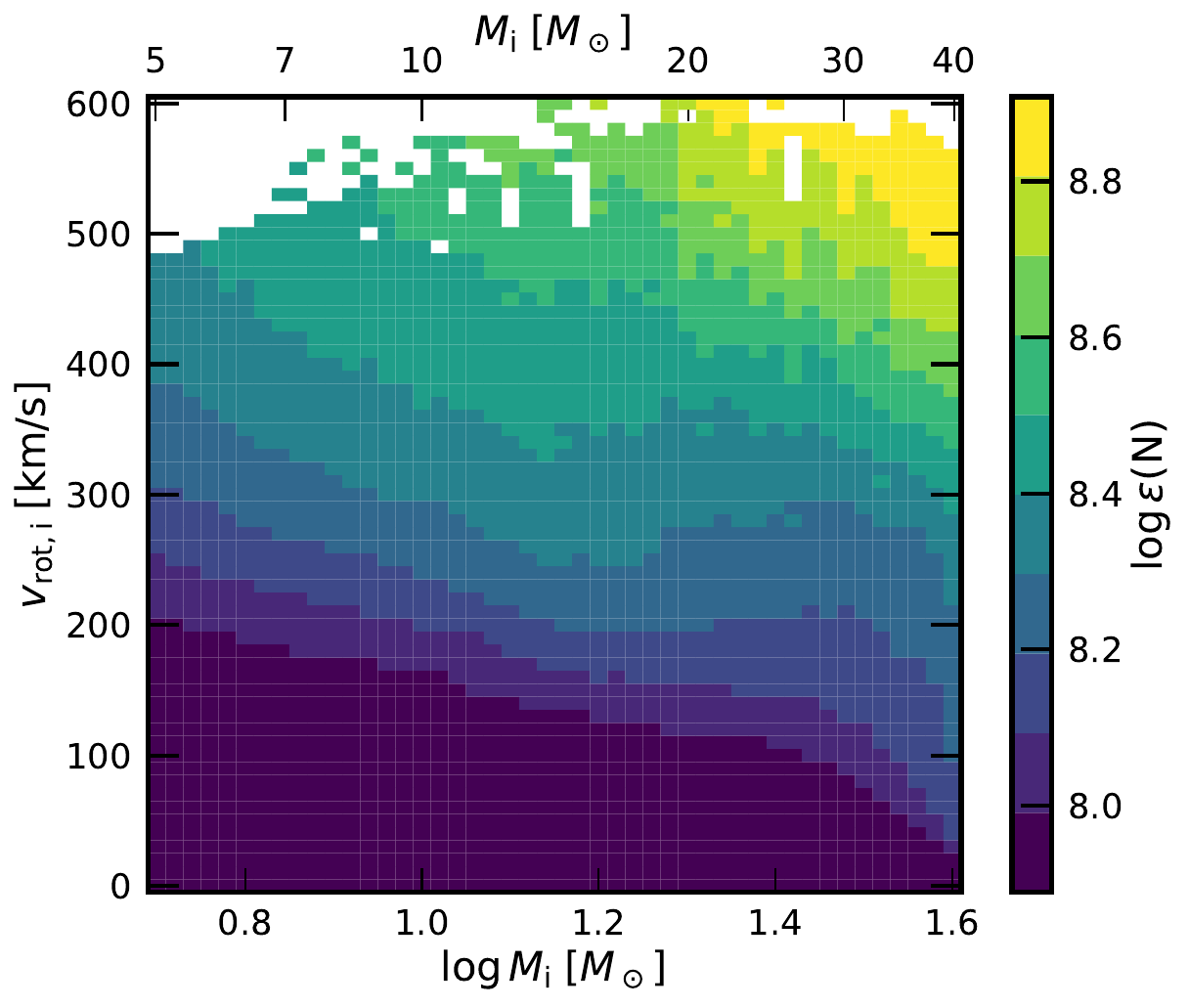}
	 \caption{Mass (top left), rotational velocity (top right), surface boron abundance (bottom left), and surface nitrogen abundance (bottom right) at terminal-age main sequence for all models in our single star grid. See fig.\,5.17 of \citet{Marchant2017} for the Large Magellanic Cloud metallicity (LMC) counterpart for the surface nitrogen abundance.}
	\label{fig_grid}
\end{figure*}

Our single star model grid contains $\sim 2600$ evolutionary sequences which span initial masses of $5-40\mso$ ($\log M_\textrm{i}[\mso]=0.70-1.60$ with $\Delta \log M_\textrm{i}[\mso]=0.02$) and initial rotational velocities of $0-600\kms$ with $\Delta \vini = 10\kms$ (if the break-up velocity is lower than $600 \kms$, we adopt the break-up velocity as the maximum initial rotational velocity). Here, the initial rotational velocities refer to the zero-age main sequence stage. The mass range covers the evolutionary masses of our set of observed stars (see below). The models are computed until core carbon depletion, except for the models that undergo the asymptotic giant branch phase. \Figure{fig_grid} shows the mass, rotational velocity, and surface boron and nitrogen abundances at terminal-age main sequence for the whole grid of stellar models. The lack of a few models near the break-up velocity is due to numerical problems.

Mass and rotational velocity at the terminal-age main sequence are significantly reduced for the initially more massive stars ($M_\mathrm{i} \gtrsim 25 \mso$ or $\log M_\mathrm{i}[\mso] \gtrsim 1.4$). This is because of the wind mass loss and accompanying angular momentum loss which is more pronounced for more massive stars. Surface boron depletion and nitrogen enhancement show a trend that initially more rapidly rotating stars and/or initially more massive stars show larger changes. Notably, we see that when the nitrogen abundance increases by $0.1 \dex$, the boron abundance decreases by $1 \dex$. This shows the usefulness of boron for testing rotational mixing, as the surface boron abundance is more sensitive to rotational mixing compared to nitrogen. 

\begin{figure}
	\centering
	\includegraphics[width=0.9\linewidth]{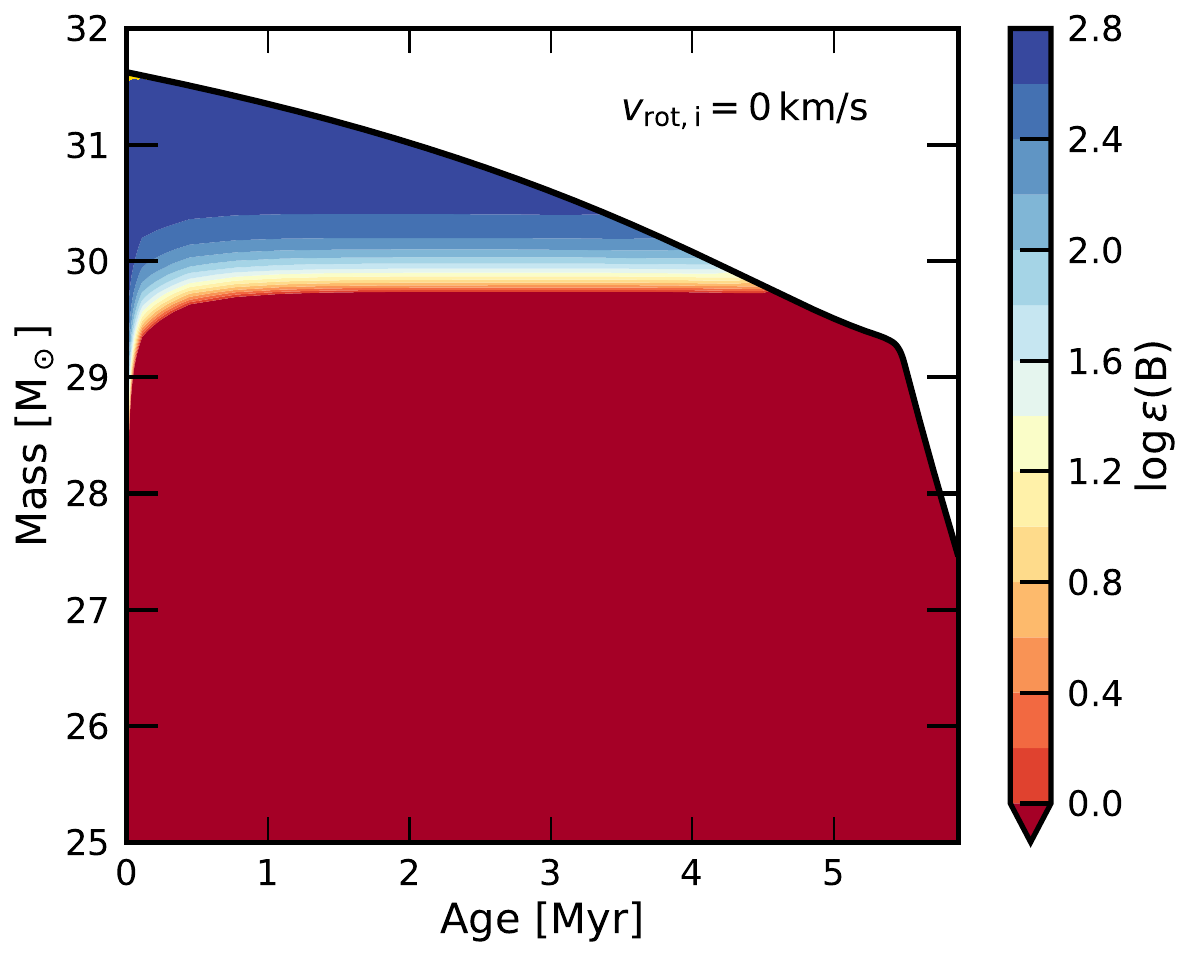}
	\includegraphics[width=0.9\linewidth]{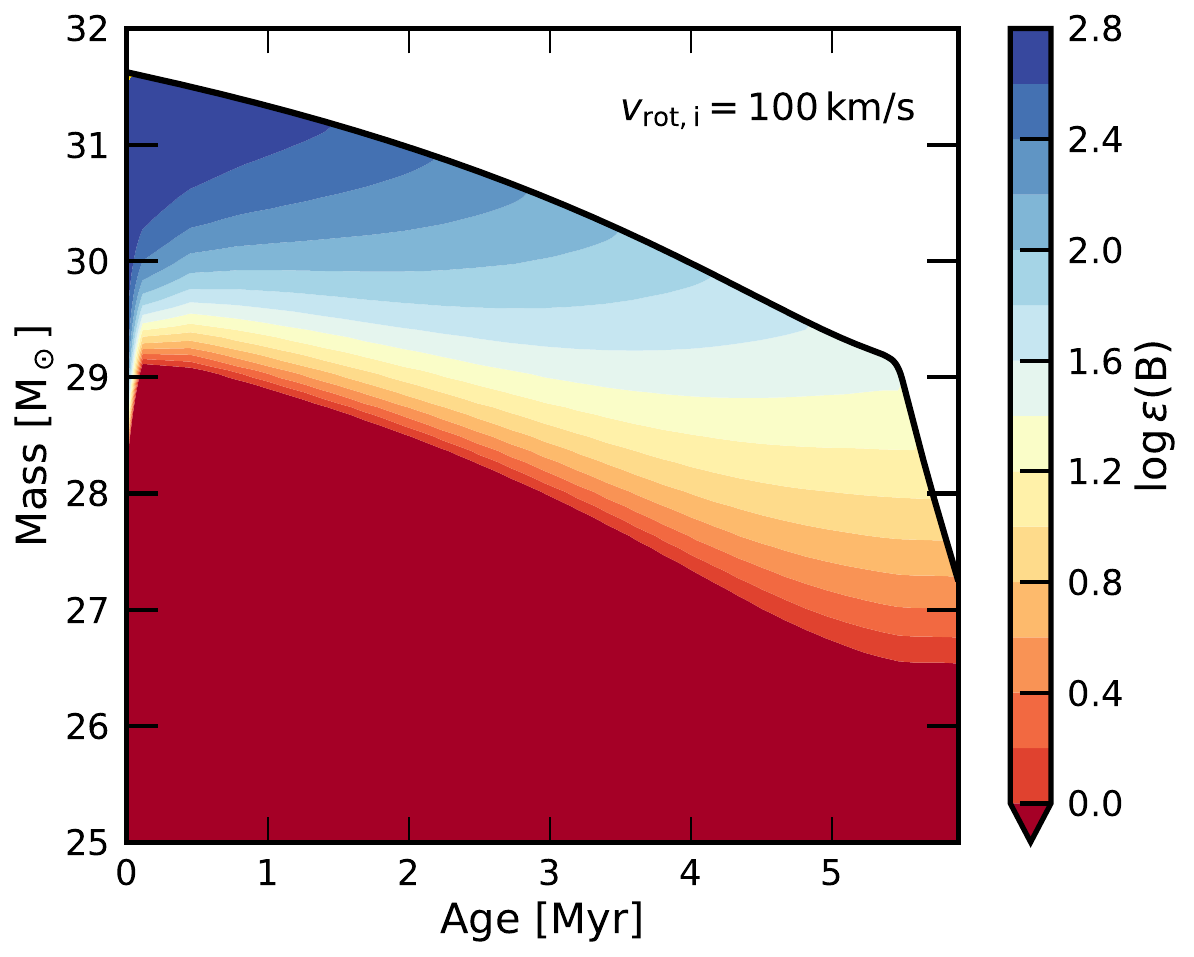}
	\includegraphics[width=0.9\linewidth]{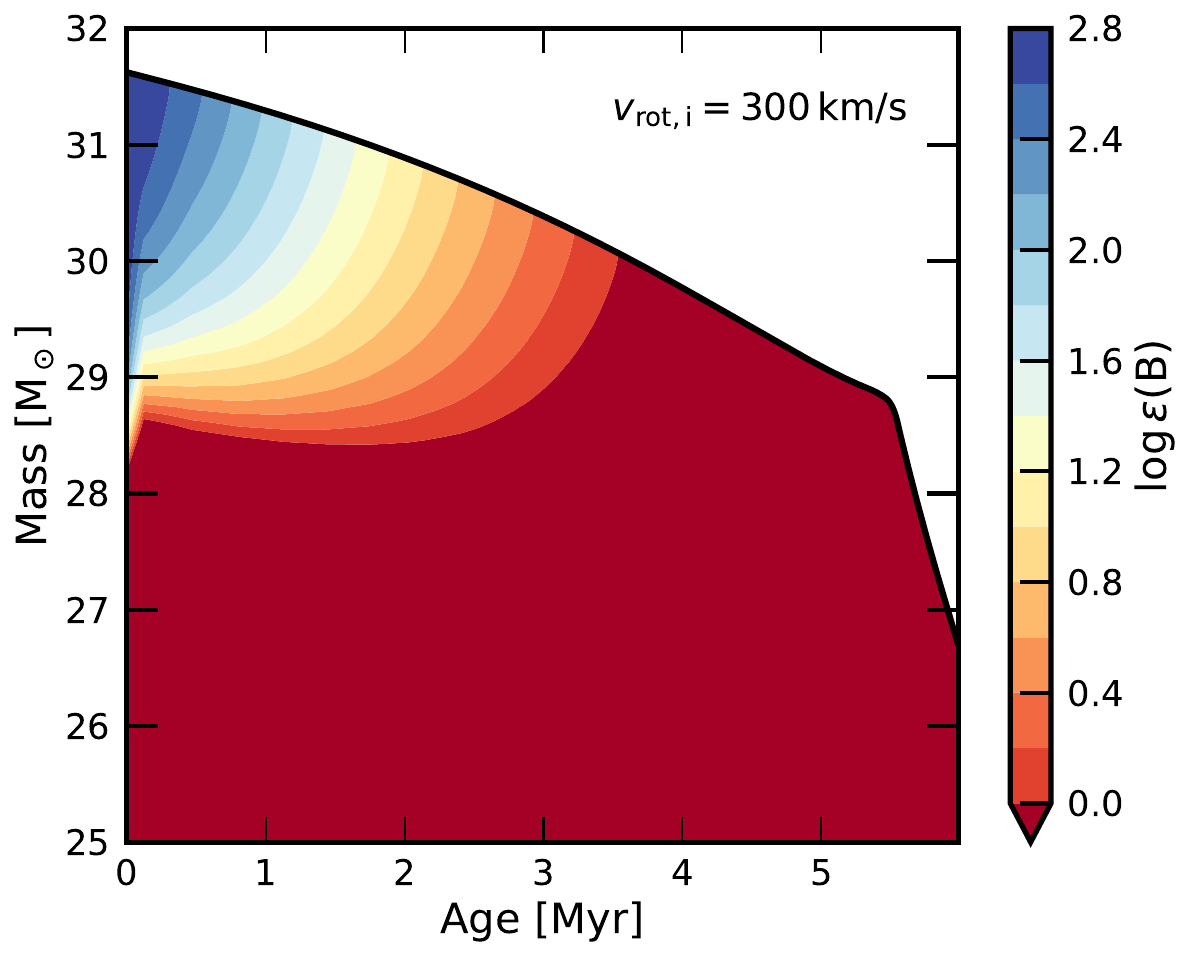}
	 \caption{Kippenhahn diagrams of a $\sim\,32 \mso$ star with initial rotational velocities of $0, 100, 300 \kms$, showing the interior boron abundance. This manifests the combined effect of wind mass loss and rotational mixing.}
	\label{fig_wind}
\end{figure}

An interesting and exceptional case can be seen in the surface boron abundance of stars with $M_\mathrm{i} \gtrsim 25 \mso$, showing a non-monotonic trend with $\vini$. This is a result of the combined effect of stellar wind mass loss and rotational mixing. \Figure{fig_wind} shows the evolution of the interior boron abundance in the envelope of a $\sim\,32 \mso$ star for different initial rotational velocities. When a star is not rotating, the outermost region of a mass coordinate of $M \gtrsim \, 30 \mso$ keeps the pristine boron abundance until wind mass loss digs into the boron-depleted region. Since the star expands during the main sequence evolution, the temperature corresponding to the same mass coordinate decreases with time. Thus the region of a mass coordinate of $M \gtrsim \, 30 \mso$ remains always cooler than the boron destruction temperature. In the case of mild rotation ($\vini = 100 \kms$), rotational mixing connects the boron-rich outer region and the boron-depleted inner region. The boron-rich matter that is transported slightly inwards (a mass coordinate of $M\sim 27\mso$ at terminal-age main sequence) may avoid boron destruction temperature and mass stripping. In the case of fast rotation ($\vini = 300 \kms$), stronger rotational mixing depletes boron in the outermost region quickly such that no boron survives in the envelope.

\section{Observational dataset}\label{sec_obs}

A primary observational constraint regarding stellar boron is that its spectral lines are located in the ultraviolet region. The key lines used for boron abundance determinations in early B-type stars are \ion{B}{II} $\lambda1362$ resonance line and \ion{B}{III} $\lambda2066$ resonance doublet line. Since the ultraviolet spectrum is susceptible to atmospheric absorption, spectra must be taken from space telescopes like the International Ultraviolet Explorer and the Hubble Space Telescope. More than that, these resonance lines often suffer from blending with other metallic lines in the spectrum, making it challenging to isolate boron lines and measure their strengths accurately. Additionally, non-local thermodynamic equilibrium effects can significantly impact the observed line profiles and intensities. These challenges make the determination of boron abundances in stars a complex and intricate task.

\begin{figure*}
	\centering
	\includegraphics[width=0.48\linewidth]{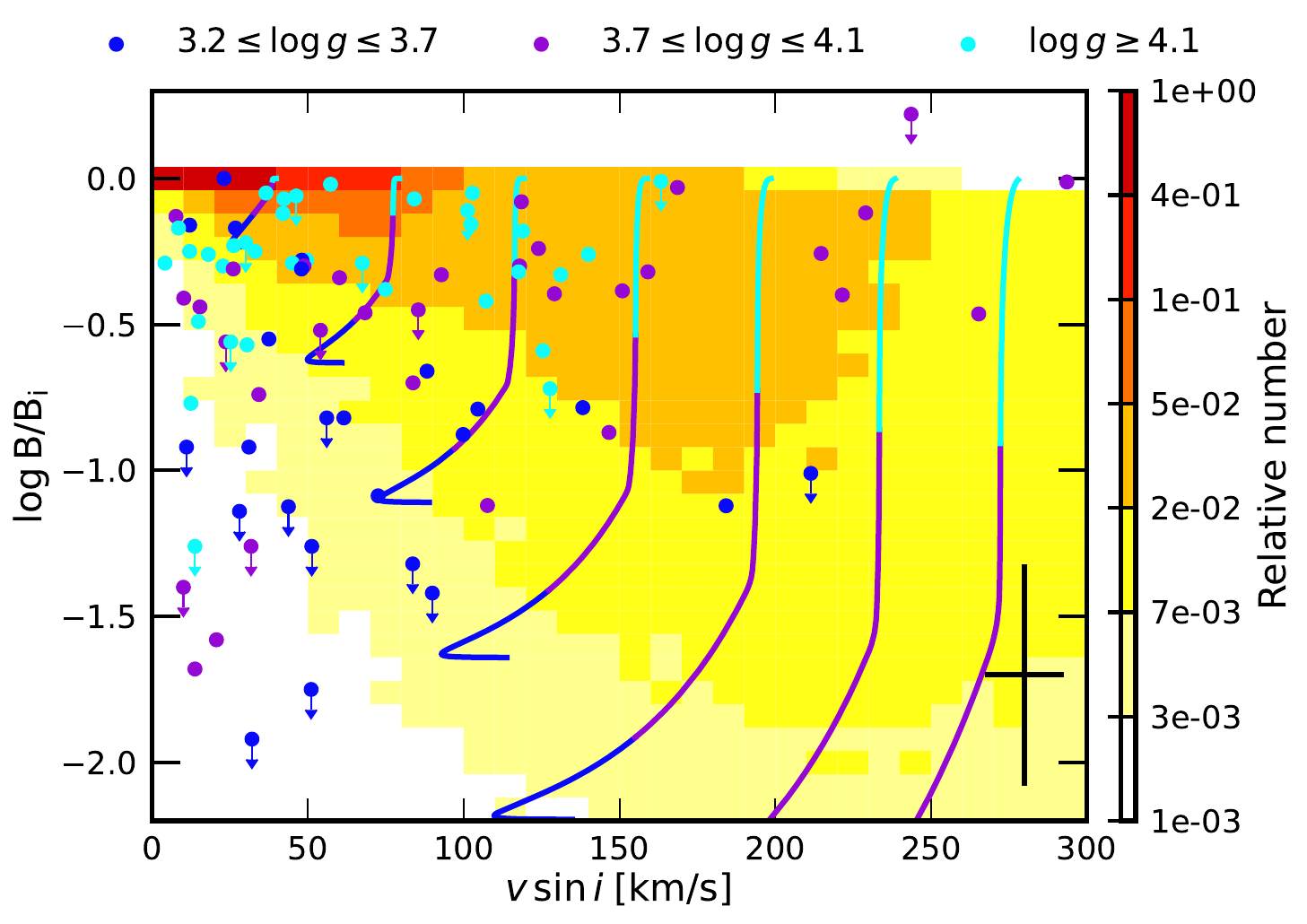}
	\includegraphics[width=0.48\linewidth]{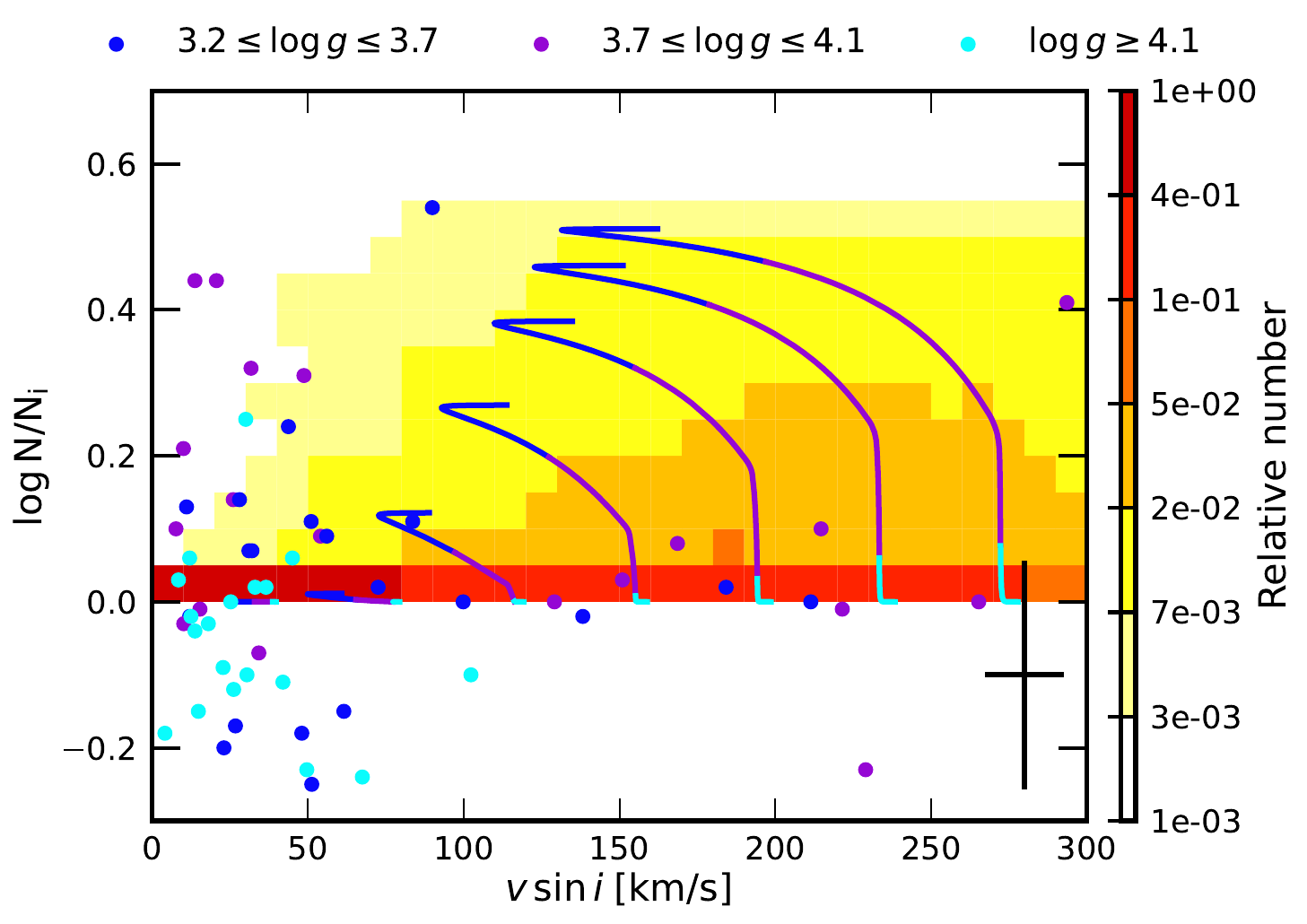}
	 \caption{Surface abundances as a function of projected rotational velocity \citep[dubbed Hunter diagram;][]{Hunter2008b} for the boron depletion factor (left) and for the nitrogen enhancement factor (right) are shown. The background contour is the predicted number distribution from a single star model population. Evolutionary tracks for 12 $\mso$ stars with different initial rotational velocities are shown, whose surface rotational velocities are multiplied by $\pi/4$. The crosses in the lower right corners show the typical errors in the observed quantities. See \citet{Brott2011b} for the LMC metallicity counterpart for the Hunter diagram for nitrogen.}
	\label{fig_Hunter}
\end{figure*}

The majority of boron abundance analyses in B-type stars to date have been conducted on sharp-lined stars with projected rotational velocities below $50 \kms$ \citep[see \Fig{fig_Hunter} for their $\vsini$ distribution]{Venn1996,Proffitt2001,Venn2002}. Stars with low $\vsini$ ought to be either intrinsic slow rotators or are observed nearly pole-on. Furthermore, most of the stars targeted for boron abundance analyses have been limited to the solar vicinity, within a distance of $\lesssim 500 \pc$, to ensure high signal-to-noise observations. See \App{app_field} to refer to their newly derived distances and luminosities.

The limited dataset of stars with boron abundance analyses has been extended, owing to dedicated studies on relatively fast rotators with an average projected rotational velocity of approximately $130 \kms$ \citep{Proffitt2015,Proffitt2016,Proffitt2024}. The expanded dataset encompasses both field stars \citep{Proffitt2015} and stars within the Galactic open cluster NGC\,3293 \citep{Proffitt2016,Proffitt2024}, which is $2.3 \kpc$ away from us. Since fast rotators are expected to have experienced stronger rotational mixing compared to their slowly rotating counterparts, they allow for a more critical test of rotational mixing. Also, some stars in NGC\,3293 are near the cluster's turn-off, which indicates that they are highly evolved stars and thus expected to have undergone significant rotational mixing. Additionally, stars within a cluster are thought to share a common origin, having formed from the same molecular cloud simultaneously. As a result, these stars share similar conditions, including their initial chemical composition, age, and distance from us. This uniform nature of cluster stars makes them an excellent testbed for studying rotational mixing \citep{Proffitt2024}.

\begin{figure}
	\centering
	\includegraphics[width=\linewidth]{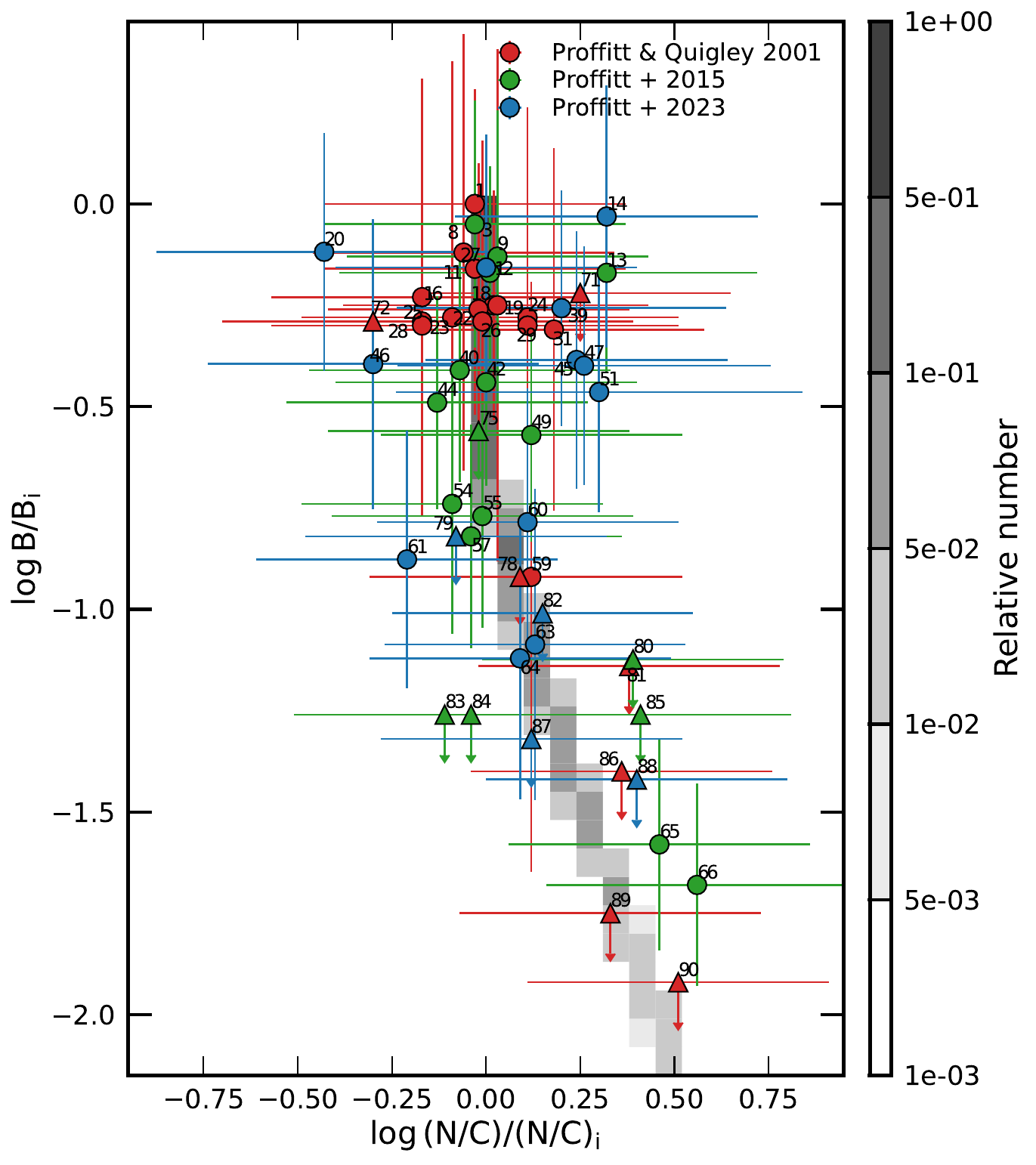}
	 \caption{Boron depletion factor against nitrogen enhancement factor for the stars in our dataset. Each data point is color-coded based on the reference from which the surface boron abundance is obtained. The background contour is the predicted number distribution from a single star model population.} 
	\label{fig_BN}
\end{figure}

The dataset of B-type stars with boron information is presented in Table~\ref{tab_obs}. To our knowledge, this is a complete list of stars with boron abundance analysis so far, except for cool F-, G-dwarfs \citep[e.g.,][]{Cunha2000, Smith2001,Boesgaard2004, Boesgaard2005}, two A-type stars \citep[HD\,46300, HD\,87737;][]{Venn1996}, a post-AGB-type star \citep[PG0832+676;][]{Venn2002}, and two stars in the Small Magellanic Cloud \citep[AV 304, NGC 346-637;][]{Brooks2002}. \Figure{fig_BN} presents the boron depletion factor $\logBBi$ as a function of the nitrogen enhancement factor $\logNCNCi$. The initial carbon and nitrogen abundances are assumed to be $\logCi=8.33$, $\logNi=7.79$ for the stars in the solar neighborhood and $\logCi=7.97$, $\logNi=7.57$ for the stars in NGC\,3293 (see Table~\ref{tab_cno}). See below for our choice of the initial boron abundance. The population of single star models, of which predictions are presented in the figure, consists of early B-type main sequence stars, and is based on the \citet{Salpeter1955} initial mass function, the rotational velocity distribution from \citet{Dufton2013}, and a constant star formation rate. 

\begin{figure*}
	\centering
	\includegraphics[width=0.48\linewidth]{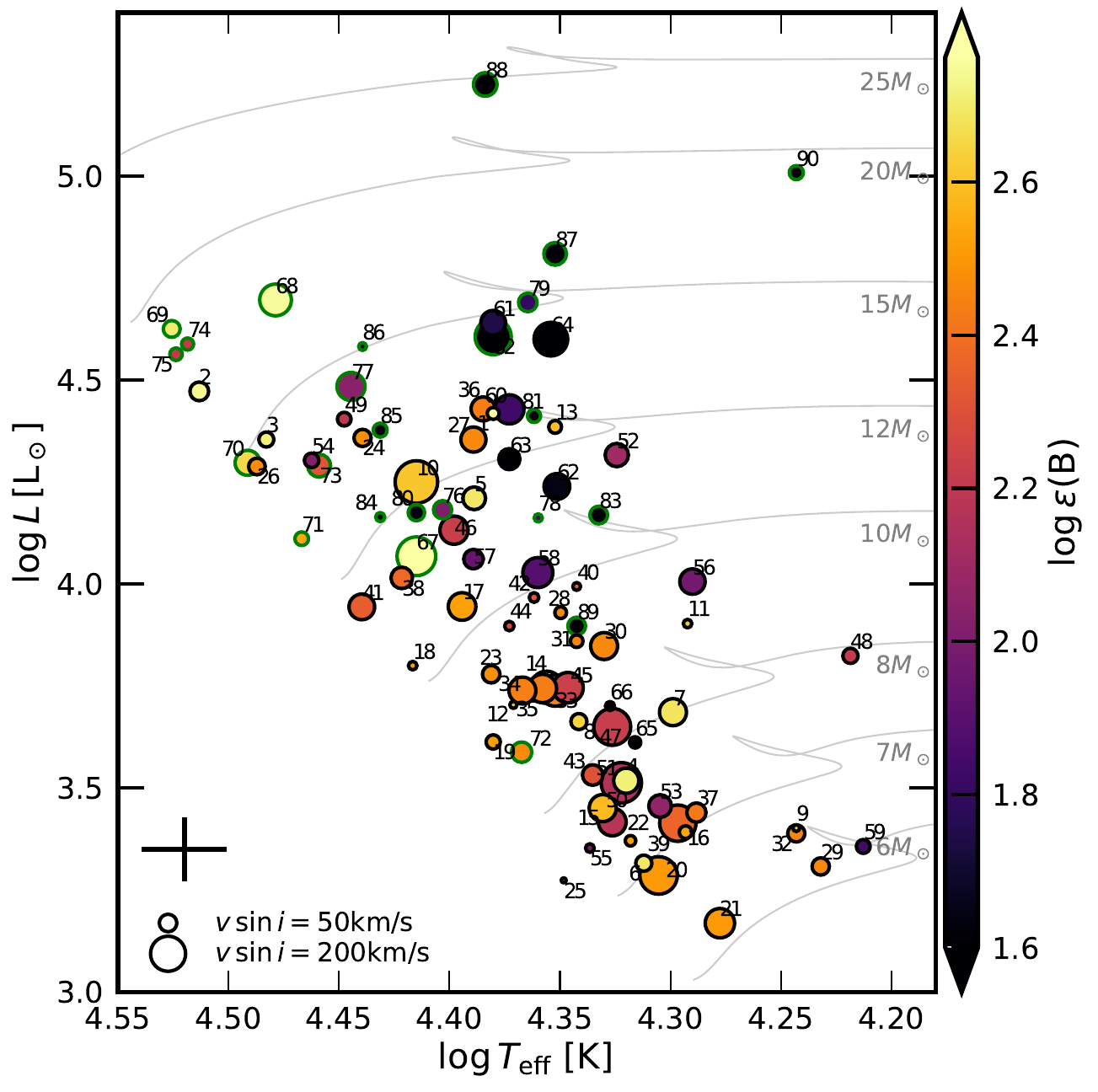}
	\includegraphics[width=0.48\linewidth]{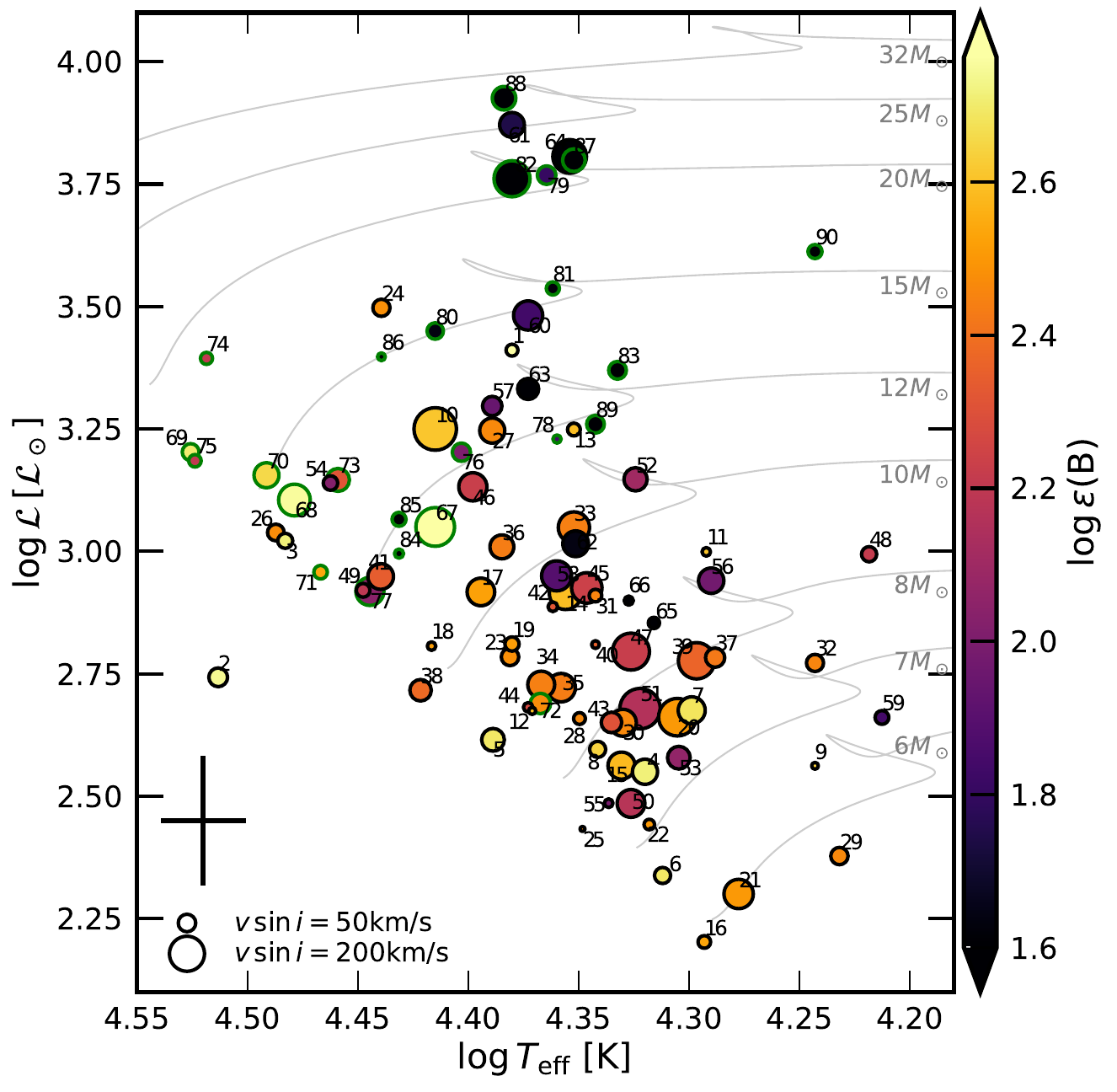}
	\caption{Stars in our dataset on the HRD (left) and sHRD (right). The size of the circles is proportional to the projected rotational velocity of each star. The inner color of the circles corresponds to the measured surface boron abundance (stars\,1...66), and stars upper limits on the boron abundance (stars\,67...90) are marked by circles with a green edge. Evolutionary tracks of non-rotating single star models are also shown. The crosses in the lower left corners show the typical errors in the observed quantities.}
	\label{fig_HRD}
\end{figure*}

\Figure{fig_HRD} shows the location of our sample stars in the Hertzsprung-Russel diagram (HRD) and the spectroscopic HRD (sHRD) \citep{Langer2014}. In the sHRD, the y-axis uses the quantity $ \mathcal{L} =  (T_\mathrm{eff}^4 / g)$, which is proportional to the stellar luminosity-to-mass ratio, and which can be determined from stellar spectra independent of distance and extinction. The expected  trend that apparently more massive and more evolved stars have lower surface boron abundances can be observed (e.g., Stars\,64, 87, 88). However, exceptions like Stars\,65, 84, where low surface boron abundance is observed despite their relatively low mass and early age, do exist.

\begin{table*}[]
\captionsetup{font=scriptsize}
\caption{List of our sample of stars. Columns $\logB$...$\vsini$ present empirical parameters based on the mentioned references. If $\veq$ is available, it is shown in parenthesis in the $\vsini$ column. In some occasions, $\vsini$ is larger than $\veq$ since the former reflects the total line broadening which may include turbulence, pulsations, and rotation \citep{Morel2008}. Columns Mass...$\pev$ / $\ptl$ / $\ptg$ / $\pmax$ are the results from our Bayesian analysis (see text). In particular, columns Mass, Age, $\vini$ are the results for the mixing efficiency $\fc$ which leads to the maximum probability $\Pi$. Values of $\pev$, $\ptl$, $\ptg$, and $\pmax$ that are smaller than 0.05, 0.1, 0,1, and 0.6, respectively, are marked in red. The last column presents information about the longitudinal magnetic field strengths (preceded by $^\mathrm{m}$) and binary properties (preceded by $^\mathrm{b}$), if known, with their references. Magnetic field values are given in Gauss, where boldface values highlight stars in which all the corresponding references report field detection, while the plane values highlight stars in which not all the corresponding references report field detection. When multiple measurements exist, we show the value with the highest significance. ``Non-det.'' means that no magnetic field was detected, while ``N/A'' implies that spectropolarimetric data was taken but a magnetic field measurement was not attempted \citep{Wade2016}. Binary properties show binary type / $P_\mathrm{orb}$ / $e$ / $(R/R_\mathrm{RL})_\mathrm{max}$, with binary type as SB1, SB2, or EB (eclipsing binary), and $(R/R_\mathrm{RL})_\mathrm{max}$ denoting the maximum possible ratio of the stellar radius over the Roche lobe radius (see text). Most of the orbital periods ($P_\mathrm{orb}$) and eccentricities ($e$) are obtained via SB9: The ninth catalog of spectroscopic binary orbits \citep{Pourbaix2004}. Stars in NGC\,3293 are marked in bold (except for the last column).}
\label{tab_obs}
\resizebox{\linewidth}{!}{%
\begin{tabular}{ccccccccccccccccccccc}
\hline \hline
No. & Star & $\logB$ & $\logC$ & $\logN$ & $\logO$ & \begin{tabular}[c]{@{}c@{}}$\Teff$\\ {[}kK]\end{tabular} & \begin{tabular}[c]{@{}c@{}}$\logg$\\ {[}cm/s$^2$]\end{tabular} & \begin{tabular}[c]{@{}c@{}}$\logL$\\ {[}$\Lsun$]\end{tabular} & \begin{tabular}[c]{@{}c@{}}$\vsini$\\ {[}km/s]\end{tabular} & Ref. & \begin{tabular}[c]{@{}c@{}}Mass\\ {[}$\Msun$]\end{tabular} & \begin{tabular}[c]{@{}c@{}}Age\\ {[}Myr]\end{tabular} & \begin{tabular}[c]{@{}c@{}}$\vini$\\ {[}km/s]\end{tabular} & $\pev$ / $\ptl$ / $\ptg$ / $\pmax$ & Magnetic field ($^\mathrm{m}$) Binary ($^\mathrm{b}$) \\ \hline
1 & HD 44743 & 2.76 ± 0.20 & 8.16 ± 0.11 & 7.59 ± 0.14 & 8.62 ± 0.18 & 24.0 ± 1.0 & 3.50 ± 0.15 & 4.42 ± 0.06 & 23 (31) & 1,2,3 & 12.6$^{+0.8}_{-0.8}$ & 13.8$^{+2.0}_{-1.2}$ & 45$^{+25}_{-25}$ & 0.54 / 0.80 / 0.60 / 0.99 & $\prescript{\mathrm{m}}{(18,19,20,21,22)}{}$-26 ± 3 \\
2 & HD~37023 & 2.74 ± 0.40 &  &  &  & 32.6 & 4.70 & 4.47 ± 0.07 & 57 & 1 & 16.7$^{+1.4}_{-1.0}$ & 0.0$^{+0.3}_{-0.0}$ & 65$^{+25}_{-15}$ & \textcolor{red}{0.00} / 0.57 / \textcolor{red}{0.00} / 0.98 & $\prescript{\mathrm{m}}{(23)}{}$N/A $\prescript{\mathrm{b}}{(29)}{}$SB2 / 20 / 0.68 / 0.5 \\
3 & HD~34816 & 2.71 ± 0.23 & 8.38 ± 0.05 & 7.81 ± 0.15 & 8.71 ± 0.09 & 30.4 & 4.30 & 4.35 ± 0.07 & 37 & 1,4,5 & 14.6$^{+1.1}_{-0.7}$ & 1.1$^{+2.4}_{-1.1}$ & 45$^{+25}_{-25}$ & 0.20 / 0.81 / 0.29 / 0.83 &  \\
4 & HD~142669 & 2.71 ± 0.34 &  &  &  & 20.9 & 4.12 & 3.52 ± 0.06 & 103 & 5 & 7.7$^{+0.4}_{-0.4}$ & 17.2$^{+6.6}_{-11.9}$ & 115$^{+35}_{-25}$ & 0.54 / 0.88 / 0.77 / 0.95 & $\prescript{\mathrm{m}}{(23)}{}$N/A $\prescript{\mathrm{b}}{(30)}{}$SB1 / 4.0 / 0.27 / 0.6 \\
5 & HD~63578 & 2.69 ± 0.23 &  &  &  & 24.5 & 4.33 & 4.21 ± 0.08 & 84 & 5 & 11.4$^{+1.3}_{-1.0}$ & 7.0$^{+4.1}_{-3.3}$ & 95$^{+25}_{-25}$ & \textcolor{red}{0.00} / 0.76 / 0.27 / 0.83 &  \\
6 & HD~112092 & 2.69 ± 0.40 &  &  &  & 20.5 & 4.30 & 3.32 ± 0.07 & 42 & 1 & 6.9$^{+0.3}_{-0.3}$ & 5.9$^{+10.6}_{-5.9}$ & 55$^{+25}_{-25}$ & 0.31 / 0.81 / 0.41 / 0.87 & $\prescript{\mathrm{m}}{(23)}{}$N/A \\
7 & HD~212978 & 2.68 ± 0.30 &  &  &  & 19.9 & 3.91 & 3.69 ± 0.08 & 119 & 5 & 8.2$^{+0.4}_{-1.0}$ & 24.5$^{+5.8}_{-5.8}$ & 185$^{+55}_{-55}$ & 0.43 / 0.76 / 0.63 / 0.87 &  \\
8 & HD~36351 & 2.64 ± 0.50 & 8.28 ± 0.04 & 7.68 ± 0.03 & 8.52 ± 0.06 & 22.0 & 4.16 & 3.66 ± 0.07 & 42 & 1,6 & 8.7$^{+0.4}_{-0.7}$ & 13.6$^{+7.3}_{-7.3}$ & 55$^{+25}_{-25}$ & 0.32 / 0.86 / 0.82 / 0.80 & $\prescript{\mathrm{b}}{(31)}{}$SB2 / --- / --- / --- \\
9 & HD~160762 & 2.63 ± 0.30 & 8.40 ± 0.07 & 7.89 ± 0.12 & 8.80 ± 0.09 & 17.5 & 3.80 & 3.40 ± 0.07 & 7.6 & 1,4,5 & 6.5$^{+0.4}_{-0.4}$ & 42.5$^{+11.0}_{-7.9}$ & 15$^{+25}_{-15}$ & 0.77 / 0.76 / 0.62 / 0.68 & $\prescript{\mathrm{m}}{(24)}{}$Non-det. $\prescript{\mathrm{b}}{(32)}{}$SB1 / 114 / 0.43 / 0.1 \\
\textbf{10} & \textbf{3293-015} & \textbf{2.61 ± 0.21} &  &  &  & \textbf{26.0 ± 1.5} & \textbf{3.80 ± 0.15} & \textbf{4.25 ± 0.07} & \textbf{294} & \textbf{7} & \textbf{12.4$^{+1.7}_{-1.2}$} & \textbf{10.3$^{+2.7}_{-3.8}$} & \textbf{325$^{+55}_{-45}$} & \textbf{0.57 / 0.79 / 0.54 / 0.75} &  \\
11 & HD~35039 & 2.60 ± 0.30 & 8.34 ± 0.10 & 7.77 ± 0.09 & 8.79 ± 0.07 & 19.6 ± 0.2 & 3.56 ± 0.07 & 3.90 ± 0.06 & 12 & 1,8 & 8.9$^{+0.3}_{-0.2}$ & 26.4$^{+1.9}_{-1.9}$ & 25$^{+25}_{-25}$ & 0.55 / 0.90 / 0.86 / 0.88 &  \\
12 & HD~35299 & 2.59 ± 0.17 & 8.35 ± 0.09 & 7.82 ± 0.08 & 8.84 ± 0.09 & 23.5 & 4.20 & 3.70 ± 0.06 & 8.4 & 1,4,5 & 9.3$^{+0.4}_{-0.9}$ & 8.3$^{+5.7}_{-5.7}$ & 15$^{+25}_{-15}$ & 0.58 / 0.87 / 0.70 / \textcolor{red}{0.42} &  \\
13 & BD+56 576 & 2.59 ± 0.14 & 7.84 ± 0.18 & 7.62 ± 0.22 & 8.34 ± 0.14 & 22.5 & 3.55 & 4.39 ± 0.07 & 27 & 5,6 & 12.1$^{+0.7}_{-0.7}$ & 16.0$^{+2.3}_{-2.3}$ & 45$^{+35}_{-25}$ & 0.66 / 0.85 / 0.66 / 0.89 & $\prescript{\mathrm{b}}{(33)}{}$SB1,EB / 26 / 0.30 / 0.4 \\
\textbf{14} & \textbf{3293-023} & \textbf{2.59 ± 0.26} & \textbf{7.72 ± 0.25} & \textbf{7.57 ± 0.23} & \textbf{8.69 ± 0.36} & \textbf{22.7 ± 1.0} & \textbf{3.90 ± 0.10} & \textbf{3.75 ± 0.06} & \textbf{169} & \textbf{7,9} & \textbf{9.0$^{+0.4}_{-0.4}$} & \textbf{18.0$^{+4.3}_{-6.0}$} & \textbf{205$^{+35}_{-25}$} & \textbf{0.35 / 0.87 / 0.64 / 0.92} &  \\
15 & HD~121790 & 2.58 ± 0.32 &  &  &  & 21.4 & 4.15 & 3.45 ± 0.06 & 119 & 5 & 7.9$^{+0.4}_{-0.6}$ & 10.1$^{+10.1}_{-7.9}$ & 135$^{+105}_{-25}$ & 0.85 / 0.85 / 0.83 / 1.02 & $\prescript{\mathrm{m}}{(23)}{}$N/A \\
16 & HD~36430 & 2.53 ± 0.50 & 8.38 ± 0.03 & 7.67 ± 0.09 & 8.57 ± 0.04 & 19.6 & 4.36 & 3.39 ± 0.07 & 26 & 1,6 & 7.1$^{+0.4}_{-0.4}$ & 9.6$^{+9.6}_{-9.6}$ & 35$^{+25}_{-25}$ & \textcolor{red}{0.02} / 0.85 / 0.24 / 0.65 &  \\
17 & HD~64722 & 2.52 ± 0.32 &  &  &  & 24.8 & 4.05 & 3.95 ± 0.06 & 124 & 5 & 10.3$^{+1.0}_{-0.4}$ & 10.8$^{+3.6}_{-5.6}$ & 185$^{+55}_{-55}$ & 0.85 / 0.90 / 0.72 / 0.95 &  \\
18 & HD~36959 & 2.51 ± 0.20 & 8.37 ± 0.11 & 7.85 ± 0.09 & 8.70 ± 0.06 & 26.1 ± 0.2 & 4.25 ± 0.07 & 3.80 ± 0.05 & 12 & 1,8 & 10.4$^{+0.3}_{-0.5}$ & 1.0$^{+2.3}_{-1.0}$ & 25$^{+25}_{-25}$ & 0.48 / 0.49 / 0.60 / \textcolor{red}{0.46} & $\prescript{\mathrm{m}}{(23)}{}$N/A \\
19 & HD~37744 & 2.51 ± 0.60 & 8.32 ± 0.07 & 7.81 ± 0.10 & 8.70 ± 0.07 & 24.0 ± 0.4 & 4.10 ± 0.10 & 3.61 ± 0.05 & 33 & 1,8 & 9.0$^{+0.3}_{-0.5}$ & 5.2$^{+3.8}_{-3.8}$ & 45$^{+25}_{-25}$ & 0.13 / 0.59 / 0.86 / 0.83 &  \\
\textbf{20} & \textbf{3293-038} & \textbf{2.50 ± 0.22} & \textbf{8.16 ± 0.26} & \textbf{7.57 ± 0.29} & \textbf{8.61 ± 0.41} & \textbf{20.2 ± 1.0} & \textbf{3.95 ± 0.15} & \textbf{3.29 ± 0.07} & \textbf{229} & \textbf{7} & \textbf{6.7$^{+0.7}_{-0.5}$} & \textbf{18.9$^{+10.2}_{-13.1}$} & \textbf{265$^{+55}_{-45}$} & \textbf{0.23 / 0.82 / 0.58 / 0.83} &  \\
21 & HD~37017 & 2.50 ± 0.38 &  &  &  & 19.0 & 4.20 & 3.17 ± 0.08 & 140 & 5 & 6.4$^{+0.5}_{-0.5}$ & 14.1$^{+14.1}_{-14.1}$ & 185$^{+45}_{-35}$ & 0.60 / 0.85 / 0.73 / 0.93 & $\prescript{\mathrm{m}}{(22)}{}$\textbf{-1800 ± 300} $\prescript{\mathrm{b}}{(34)}{}$SB1 / 19 / 0.31 / 0.2 \\
22 & HD~122980 & 2.50 ± 0.30 & 8.32 ± 0.09 & 7.76 ± 0.08 & 8.72 ± 0.06 & 20.8 ± 0.3 & 4.22 ± 0.05 & 3.37 ± 0.04 & 18 & 1,4 & 7.2$^{+0.2}_{-0.2}$ & 13.3$^{+3.7}_{-4.8}$ & 25$^{+25}_{-25}$ & 0.34 / 0.95 / 0.85 / \textcolor{red}{0.39} & $\prescript{\mathrm{m}}{(23)}{}$N/A \\
23 & HD~37209 & 2.48 ± 0.60 & 8.19 ± 0.04 & 7.56 ± 0.05 & 8.51 ± 0.08 & 24.1 & 4.13 & 3.78 ± 0.06 & 50 & 1,6 & 9.5$^{+0.4}_{-0.7}$ & 9.6$^{+5.8}_{-5.8}$ & 65$^{+25}_{-25}$ & 0.87 / 0.88 / 0.83 / 0.73 &  \\
24 & HD~111123 & 2.48 ± 0.20 & 8.04 ± 0.10 & 7.61 ± 0.17 & 8.59 ± 0.16 & 27.5 ± 1.0 & 3.65 ± 0.20 & 4.36 ± 0.09 & 48 (2) & 1,10,11 & 13.3$^{+0.8}_{-0.5}$ & 10.7$^{+1.6}_{-2.3}$ & 35$^{+15}_{-25}$ & 0.20 / 0.78 / 0.50 / \textcolor{red}{0.02} & $\prescript{\mathrm{m}}{(20,21)}{}$Non-det. $\prescript{\mathrm{b}}{(35)}{}$SB1 / 1828 / 0.38 / 0.0 \\
25 & HD~36629 & 2.47 ± 0.30 & 8.32 ± 0.03 & 7.61 ± 0.03 & 8.32 ± 0.06 & 22.3 & 4.35 & 3.27 ± 0.07 & 4.1 & 1,6 & 7.2$^{+0.5}_{-0.5}$ & 0.8$^{+4.0}_{-0.8}$ & 15$^{+15}_{-15}$ & 0.10 / 0.19 / 0.24 / \textcolor{red}{0.45} & $\prescript{\mathrm{m}}{(21)}{}$Non-det. \\
26 & HD~37020 & 2.47 ± 0.40 & 8.40 ± 0.07 & 7.85 ± 0.09 & 8.70 ± 0.10 & 30.7 ± 0.3 & 4.30 ± 0.08 & 4.29 ± 0.08 & 45 & 1,8 & 14.7$^{+0.9}_{-0.6}$ & 0.4$^{+1.5}_{-0.4}$ & 55$^{+25}_{-25}$ & 0.28 / 0.58 / 0.29 / 0.65 & $\prescript{\mathrm{m}}{(23)}{}$N/A $\prescript{\mathrm{b}}{(36)}{}$SB1 / 65 / 0.63 / 0.2 \\
\textbf{27} & \textbf{3293-012} & \textbf{2.46 ± 0.26} & \textbf{7.86 ± 0.17} & \textbf{7.45 ± 0.13} & \textbf{8.83 ± 0.25} & \textbf{24.5 ± 1.0} & \textbf{3.70 ± 0.10} & \textbf{4.35 ± 0.06} & \textbf{102} & \textbf{7,9} & \textbf{12.5$^{+0.9}_{-0.9}$} & \textbf{13.4$^{+2.2}_{-2.2}$} & \textbf{185$^{+65}_{-85}$} & \textbf{0.79 / 0.80 / 0.69 / 1.09} &  \\
28 & HD~37356 & 2.46 ± 0.40 & 8.41 ± 0.03 & 7.70 ± 0.04 & 8.44 ± 0.05 & 22.4 & 4.13 & 3.93 ± 0.08 & 23 & 1,6 & 9.5$^{+1.0}_{-0.6}$ & 15.3$^{+4.0}_{-5.7}$ & 35$^{+25}_{-25}$ & 0.06 / 0.76 / 0.80 / \textcolor{red}{0.51} &  \\
29 & HD~41753 & 2.46 ± 0.50 & 8.53 ± 0.15 & 8.10 ± 0.14 & 8.57 ± 0.05 & 17.1 & 3.94 & 3.31 ± 0.08 & 49 & 1,6 & 6.5$^{+0.4}_{-1.0}$ & 42.9$^{+11.1}_{-7.9}$ & 65$^{+25}_{-25}$ & 0.37 / 0.75 / 0.71 / 0.78 & $\prescript{\mathrm{b}}{(37)}{}$SB1 / 131 / 0.64 / 0.2 \\
30 & HD~109668 & 2.46 ± 0.33 &  &  &  & 21.4 & 4.06 & 3.85 ± 0.11 & 118 & 5 & 8.8$^{+0.5}_{-0.8}$ & 16.9$^{+5.6}_{-5.6}$ & 185$^{+65}_{-65}$ & 0.13 / 0.71 / 0.70 / 0.89 & $\prescript{\mathrm{m}}{(23)}{}$N/A $\prescript{\mathrm{b}}{(31)}{}$SB2 / --- / --- / --- \\
31 & HD~29248 & 2.45 ± 0.40 & 8.29 ± 0.13 & 7.93 ± 0.09 & 8.78 ± 0.09 & 22.0 ± 0.3 & 3.85 ± 0.05 & 3.86 ± 0.06 & 26 (6) & 1,4,12 & 9.0$^{+0.3}_{-0.1}$ & 20.9$^{+1.0}_{-1.0}$ & 25$^{+15}_{-15}$ & 0.75 / 0.86 / 0.84 / \textcolor{red}{0.18} & $\prescript{\mathrm{m}}{(18,20,21,25)}{}$Non-det. \\
32 & HD~38622 & 2.45 ± 0.37 &  &  &  & 17.5 & 3.59 & 3.39 ± 0.08 & 48 & 5 & 6.4$^{+0.5}_{-0.5}$ & 49.3$^{+7.9}_{-7.9}$ & 65$^{+35}_{-15}$ & 0.20 / 0.75 / 0.62 / 0.94 &  \\
33 & HD~120324 & 2.44 ± 0.31 &  &  &  & 22.5 & 3.75 & 3.74 ± 0.09 & 159 & 5 & 9.1$^{+0.5}_{-0.5}$ & 21.4$^{+4.3}_{-6.0}$ & 205$^{+35}_{-35}$ & 0.09 / 0.83 / 0.64 / 0.88 & $\prescript{\mathrm{m}}{(21,26)}{}$Non-det. \\
34 & HD~202347 & 2.44 ± 0.32 &  &  &  & 23.3 & 4.13 & 3.74 ± 0.07 & 118 & 5 & 9.2$^{+0.7}_{-1.0}$ & 9.9$^{+5.3}_{-6.8}$ & 185$^{+55}_{-75}$ & 0.70 / 0.90 / 0.82 / 0.92 &  \\
35 & HD~64740 & 2.43 ± 0.30 &  &  &  & 22.8 & 4.10 & 3.74 ± 0.06 & 131 & 5 & 9.2$^{+0.4}_{-0.4}$ & 12.1$^{+5.7}_{-5.7}$ & 185$^{+55}_{-45}$ & 0.59 / 0.88 / 0.76 / 0.89 & $\prescript{\mathrm{m}}{(22)}{}$\textbf{-820 ± 90} \\
36 & HD~68243 & 2.43 ± 0.19 &  &  &  & 24.3 & 3.92 & 4.43 ± 0.07 & 93 & 5 & 12.8$^{+0.9}_{-0.9}$ & 12.7$^{+1.3}_{-3.1}$ & 115$^{+25}_{-25}$ & \textcolor{red}{0.03} / 0.78 / 0.66 / 0.92 & $\prescript{\mathrm{m}}{(23)}{}$N/A $\prescript{\mathrm{b}}{(38)}{}$SB1 / 1.5 / 0.06 / 1.8 \\
37 & HD~45546 & 2.42 ± 0.29 &  &  &  & 19.4 & 3.76 & 3.44 ± 0.09 & 60 & 5 & 6.9$^{+0.4}_{-0.4}$ & 36.3$^{+7.9}_{-7.9}$ & 75$^{+25}_{-15}$ & 0.20 / 0.78 / 0.62 / 0.95 & $\prescript{\mathrm{m}}{(23)}{}$N/A \\
38 & HD~24131 & 2.38 ± 0.22 &  &  &  & 26.4 & 4.36 & 4.02 ± 0.06 & 75 & 5 & 11.4$^{+0.5}_{-0.9}$ & 1.9$^{+3.5}_{-1.9}$ & 85$^{+25}_{-25}$ & 0.06 / 0.88 / 0.17 / \textcolor{red}{0.41} &  \\
\textbf{39} & \textbf{3293-030} & \textbf{2.36 ± 0.21} & \textbf{7.86 ± 0.27} & \textbf{7.57 ± 0.27} & \textbf{8.65 ± 0.37} & \textbf{19.8 ± 1.0} & \textbf{3.80 ± 0.15} & \textbf{3.41 ± 0.08} & \textbf{215} & \textbf{7} & \textbf{7.1$^{+0.4}_{-0.4}$} & \textbf{26.8$^{+11.0}_{-11.0}$} & \textbf{255$^{+65}_{-35}$} & \textbf{0.24 / 0.85 / 0.53 / 0.89} &  \\
40 & HD~886 & 2.35 ± 0.19 & 8.37 ± 0.08 & 7.76 ± 0.07 & 8.73 ± 0.11 & 22.0 & 3.95 & 3.99 ± 0.08 & 10 & 1,4,5 & 9.4$^{+1.0}_{-0.6}$ & 18.6$^{+4.0}_{-4.0}$ & 25$^{+25}_{-25}$ & 0.18 / 0.76 / 0.65 / \textcolor{red}{0.23} & $\prescript{\mathrm{m}}{(18,24,27)}{}$46 ± 11 $\prescript{\mathrm{b}}{(25)}{}$SB1 / 6.8 / --- / \textgreater{} 0.5 \\
41 & HD~143018 & 2.34 ± 0.35 &  &  &  & 27.5 & 4.20 & 3.94 ± 0.09 & 107 & 5 & 11.2$^{+1.3}_{-0.9}$ & 1.6$^{+3.5}_{-1.6}$ & 115$^{+35}_{-15}$ & 0.43 / 0.41 / 0.67 / 0.69 & $\prescript{\mathrm{m}}{(23)}{}$N/A $\prescript{\mathrm{b}}{(39)}{}$SB2,EB / 1.6 / 0.0 / 0.7 \\
42 & HD~216916 & 2.32 ± 0.16 & 8.32 ± 0.07 & 7.78 ± 0.10 & 8.78 ± 0.08 & 23.0 & 3.95 & 3.97 ± 0.07 & 15 & 1,4,5 & 9.6$^{+1.0}_{-0.6}$ & 17.5$^{+3.8}_{-5.3}$ & 25$^{+35}_{-25}$ & 0.51 / 0.78 / 0.65 / \textcolor{red}{0.25} & $\prescript{\mathrm{m}}{(18)}{}$Non-det. $\prescript{\mathrm{b}}{(40)}{}$EB / 12 / 0.05 / 0.3 \\
43 & HD~121743 & 2.30 ± 0.29 &  &  &  & 21.6 & 4.08 & 3.53 ± 0.06 & 68 & 5 & 7.7$^{+0.4}_{-0.4}$ & 16.9$^{+7.9}_{-7.9}$ & 85$^{+25}_{-25}$ & 0.78 / 0.90 / 0.72 / 0.63 & $\prescript{\mathrm{m}}{(22)}{}$\textbf{330 ± 90} \\
44 & HD~35337 & 2.27 ± 0.17 & 8.31 ± 0.09 & 7.64 ± 0.10 & 8.55 ± 0.17 & 23.6 & 4.20 & 3.90 ± 0.06 & 15 & 1,5,6 & 9.8$^{+0.7}_{-0.4}$ & 11.3$^{+4.2}_{-5.3}$ & 25$^{+25}_{-25}$ & 0.12 / 0.83 / 0.70 / \textcolor{red}{0.14} & $\prescript{\mathrm{b}}{(32)}{}$SB1 / 293 / 0.15 / 0.0 \\
\textbf{45} & \textbf{3293-024} & \textbf{2.24 ± 0.25} & \textbf{7.75 ± 0.23} & \textbf{7.62 ± 0.25} & \textbf{8.78 ± 0.35} & \textbf{22.2 ± 1.0} & \textbf{3.85 ± 0.10} & \textbf{3.75 ± 0.06} & \textbf{151} & \textbf{7,9} & \textbf{8.9$^{+0.4}_{-0.4}$} & \textbf{19.3$^{+6.4}_{-4.6}$} & \textbf{195$^{+45}_{-25}$} & \textbf{0.34 / 0.82 / 0.64 / 0.87} &  \\
\textbf{46} & \textbf{3293-019} & \textbf{2.23 ± 0.30} & \textbf{8.26 ± 0.29} & \textbf{7.59 ± 0.25} & \textbf{8.58 ± 0.24} & \textbf{25.0 ± 1.0} & \textbf{3.85 ± 0.10} & \textbf{4.13 ± 0.06} & \textbf{129} & \textbf{7,9} & \textbf{11.4$^{+0.5}_{-0.9}$} & \textbf{13.3$^{+2.5}_{-3.4}$} & \textbf{185$^{+55}_{-25}$} & \textbf{0.63 / 0.81 / 0.68 / 0.90} & $\prescript{\mathrm{b}}{(9)}{}$SB2 / --- / --- / --- \\
\textbf{47} & \textbf{3293-028} & \textbf{2.22 ± 0.22} & \textbf{7.69 ± 0.33} & \textbf{7.52 ± 0.30} & \textbf{8.69 ± 0.43} & \textbf{21.2 ± 1.0} & \textbf{3.90 ± 0.15} & \textbf{3.65 ± 0.07} & \textbf{222} & \textbf{7} & \textbf{8.3$^{+0.4}_{-0.4}$} & \textbf{19.3$^{+8.0}_{-8.0}$} & \textbf{275$^{+55}_{-45}$} & \textbf{0.64 / 0.80 / 0.57 / 0.94} &  \\
48 & HD~180163 & 2.21 ± 0.60 &  &  &  & 16.5 & 3.27 & 3.82 ± 0.08 & 37 & 1 & 7.9$^{+0.4}_{-0.4}$ & 33.3$^{+5.2}_{-3.7}$ & 335$^{+95}_{-45}$ & 0.09 / 0.28 / 0.11 / \textcolor{red}{0.48} & $\prescript{\mathrm{b}}{(32)}{}$SB1 / 56 / 0.53 / 0.4 \\
49 & HD~22951 & 2.19 ± 0.17 & 8.11 ± 0.21 & 7.69 ± 0.32 & 8.42 ± 0.17 & 28.0 & 4.26 & 4.40 ± 0.07 & 30 & 1,5,6 & 14.0$^{+1.2}_{-0.7}$ & 5.0$^{+2.7}_{-2.7}$ & 45$^{+25}_{-25}$ & \textcolor{red}{0.02} / 0.80 / 0.46 / \textcolor{red}{0.34} & $\prescript{\mathrm{b}}{(4)}{}$SB2? / --- / --- / --- \\
50 & HD~110879 & 2.17 ± 0.38 &  &  &  & 21.2 & 4.21 & 3.42 ± 0.08 & 125 & 5 & 7.7$^{+0.4}_{-0.7}$ & 9.2$^{+7.1}_{-7.1}$ & 185$^{+45}_{-65}$ & 0.66 / 0.84 / 0.69 / 0.78 & $\prescript{\mathrm{m}}{(23)}{}$N/A $\prescript{\mathrm{b}}{(31)}{}$SB2 / --- / --- / --- \\
\textbf{51} & \textbf{3293-031} & \textbf{2.16 ± 0.22} & \textbf{7.66 ± 0.33} & \textbf{7.70 ± 0.37} &  & \textbf{21.0 ± 1.0} & \textbf{4.00 ± 0.15} & \textbf{3.51 ± 0.07} & \textbf{265} & \textbf{7} & \textbf{7.7$^{+0.4}_{-0.4}$} & \textbf{17.1$^{+8.0}_{-10.2}$} & \textbf{305$^{+45}_{-45}$} & \textbf{0.67 / 0.91 / 0.63 / 0.82} &  \\
52 & HD~188252 & 2.10 ± 0.25 &  &  &  & 21.1 & 3.54 & 4.32 ± 0.08 & 88 & 5 & 11.1$^{+0.9}_{-0.7}$ & 16.8$^{+3.8}_{-1.6}$ & 185$^{+65}_{-25}$ & 0.39 / 0.74 / 0.66 / 0.91 &  \\
53 & HD~214263 & 2.06 ± 0.36 &  &  &  & 20.2 & 4.03 & 3.46 ± 0.07 & 84 & 5 & 7.1$^{+0.4}_{-0.4}$ & 23.5$^{+11.0}_{-11.0}$ & 105$^{+25}_{-25}$ & 0.81 / 0.85 / 0.66 / 0.73 &  \\
54 & HD~36960 & 2.02 ± 0.25 & 8.35 ± 0.09 & 7.72 ± 0.11 & 8.67 ± 0.08 & 29.0 & 4.10 & 4.30 ± 0.06 & 34 & 1,4,5 & 13.8$^{+1.0}_{-1.0}$ & 5.4$^{+2.3}_{-2.8}$ & 45$^{+25}_{-25}$ & 0.91 / 0.91 / 0.83 / \textcolor{red}{0.34} & $\prescript{\mathrm{m}}{(21)}{}$Non-det. \\
55 & HD~36285 & 1.99 ± 0.19 & 8.32 ± 0.07 & 7.77 ± 0.09 & 8.80 ± 0.10 & 21.7 & 4.25 & 3.35 ± 0.06 & 12 & 1,5,8 & 7.2$^{+0.7}_{-0.5}$ & 8.3$^{+8.3}_{-8.3}$ & 25$^{+25}_{-25}$ & 0.48 / 0.53 / 0.56 / \textcolor{red}{0.07} &  \\
56 & HD~56139 & 1.97 ± 0.37 &  &  &  & 19.5 & 3.61 & 4.01 ± 0.08 & 105 & 5 & 9.3$^{+0.7}_{-0.7}$ & 23.3$^{+5.6}_{-4.0}$ & 185$^{+45}_{-35}$ & 0.46 / 0.72 / 0.65 / 0.89 & $\prescript{\mathrm{m}}{(21)}{}$Non-det. \\
57 & HD~214993 & 1.94 ± 0.19 & 8.22 ± 0.12 & 7.64 ± 0.18 & 8.42 ± 0.23 & 24.5 & 3.65 & 4.06 ± 0.07 & 62 (45) & 1,2,5,13 & 10.6$^{+0.6}_{-0.6}$ & 16.9$^{+2.9}_{-2.9}$ & 85$^{+25}_{-15}$ & 0.06 / 0.78 / 0.70 / \textcolor{red}{0.44} & $\prescript{\mathrm{m}}{(18)}{}$Non-det. \\
58 & HD~106490 & 1.89 ± 0.33 &  &  &  & 22.9 & 3.88 & 4.03 ± 0.10 & 147 & 5 & 9.9$^{+1.1}_{-1.1}$ & 16.9$^{+4.0}_{-4.0}$ & 205$^{+35}_{-35}$ & 0.60 / 0.71 / 0.64 / 0.93 & $\prescript{\mathrm{m}}{(23)}{}$N/A \\
59 & HD~184171 & 1.84 ± 0.70 & 8.28 ± 0.13 & 7.86 ± 0.25 & 8.69 ± 0.05 & 16.3 & 3.58 & 3.36 ± 0.08 & 31 & 1,6 & 6.1$^{+0.5}_{-0.5}$ & 53.3$^{+12.0}_{-8.6}$ & 55$^{+25}_{-25}$ & 0.48 / 0.74 / 0.59 / \textcolor{red}{0.48} &  \\
\textbf{60} & \textbf{3293-008} & \textbf{1.84 ± 0.26} & \textbf{7.83 ± 0.23} & \textbf{7.59 ± 0.15} & \textbf{8.77 ± 0.32} & \textbf{23.6 ± 1.0} & \textbf{3.40 ± 0.10} & \textbf{4.43 ± 0.06} & \textbf{138} & \textbf{7,9} & \textbf{12.8$^{+0.7}_{-0.7}$} & \textbf{14.4$^{+1.6}_{-1.6}$} & \textbf{235$^{+105}_{-45}$} & \textbf{0.23 / 0.83 / 0.70 / 0.82} & $\prescript{\mathrm{b}}{(9)}{}$SB1 / --- / --- / --- \\
\textbf{61} & \textbf{3293-004} & \textbf{1.74 ± 0.25} & \textbf{8.17 ± 0.28} & \textbf{7.55 ± 0.09} & \textbf{8.75 ± 0.19} & \textbf{24.0 ± 1.0} & \textbf{3.04 ± 0.10} & \textbf{4.64 ± 0.06} & \textbf{100} & \textbf{7,9} & \textbf{14.4$^{+1.0}_{-0.7}$} & \textbf{12.5$^{+1.4}_{-1.0}$} & \textbf{205$^{+55}_{-25}$} & \textbf{\textcolor{red}{0.00} / 0.84 / 0.60 / 0.77} &  \\
62 & HD~160578 & 1.64 ± 0.39 &  &  &  & 22.5 & 3.78 & 4.24 ± 0.07 & 108 & 5 & 11.1$^{+0.9}_{-0.9}$ & 16.1$^{+2.8}_{-2.8}$ & 185$^{+55}_{-25}$ & 0.20 / 0.82 / 0.64 / 0.90 & $\prescript{\mathrm{m}}{(19,21)}{}$Non-det. \\
\textbf{63} & \textbf{3293-010} & \textbf{1.53 ± 0.33} & \textbf{7.85 ± 0.21} & \textbf{7.45 ± 0.08} & \textbf{8.82 ± 0.23} & \textbf{23.6 ± 1.0} & \textbf{3.55 ± 0.10} & \textbf{4.31 ± 0.06} & \textbf{73} & \textbf{7,9} & \textbf{11.8$^{+0.8}_{-0.8}$} & \textbf{15.1$^{+2.3}_{-1.4}$} & \textbf{105$^{+25}_{-25}$} & \textbf{0.41 / 0.80 / 0.69 / 0.71} &  \\
\textbf{64} & \textbf{3293-005} & \textbf{1.50 ± 0.28} & \textbf{7.89 ± 0.24} & \textbf{7.56 ± 0.08} & \textbf{8.75 ± 0.30} & \textbf{22.6 ± 1.5} & \textbf{3.00 ± 0.10} & \textbf{4.60 ± 0.08} & \textbf{184} & \textbf{7} & \textbf{13.2$^{+1.2}_{-0.9}$} & \textbf{14.3$^{+1.8}_{-1.8}$} & \textbf{385$^{+55}_{-95}$} & \textbf{\textcolor{red}{0.01} / 0.70 / 0.72 / \textcolor{red}{0.04}} &  \\
65 & HD~3360 & 1.18 ± 0.17 & 8.31 ± 0.08 & 8.23 ± 0.07 & 8.80 ± 0.08 & 20.7 & 3.80 & 3.61 ± 0.07 & 21 (55) & 1,4,5,14 & 7.7$^{+0.6}_{-0.4}$ & 28.2$^{+8.2}_{-5.8}$ & 135$^{+35}_{-25}$ & 0.31 / 0.79 / 0.63 / \textcolor{red}{0.20} & $\prescript{\mathrm{m}}{(14,22,24)}{}$\textbf{-25 ± 2} \\
66 & HD~16582 & 1.08 ± 0.15 & 8.21 ± 0.09 & 8.23 ± 0.08 & 8.79 ± 0.07 & 21.3 & 3.80 & 3.70 ± 0.08 & 14 (28) & 1,4,5,15 & 8.4$^{+0.6}_{-0.4}$ & 26.7$^{+4.8}_{-4.8}$ & 95$^{+25}_{-25}$ & 0.36 / 0.78 / 0.63 / \textcolor{red}{0.00} & $\prescript{\mathrm{m}}{(18,20,21,24,26)}{}$-75 ± 14 \\
67 & HD~108249 & $<$ 2.98 &  &  &  & 26.0 & 4.00 & 4.07 ± 0.07 & 244 & 5 & 11.6$^{+0.7}_{-0.7}$ & 10.5$^{+2.7}_{-2.7}$ & 325$^{+115}_{-45}$ & 0.68 / 0.90 / 0.69 / $<$ 0.89 &  \\
68 & HD 143275 & $<$2.75 &  &  &  & 30.1 & 4.20 & 4.70 ± 0.13 & 163 & 5 & 16.4$^{+2.1}_{-2.1}$ & 3.8$^{+2.0}_{-2.4}$ & 195$^{+35}_{-35}$ & \textcolor{red}{0.04}~/ 0.70 / 0.65 / $<$ 0.91 & $\prescript{\mathrm{m}}{(18)}{}$Non-det. $\prescript{\mathrm{b}}{(41)}{}$SB1 / 3864 / 0.94 / 0.1 \\
69 & HD~214680 & $<$ 2.70 &  &  &  & 33.6 & 4.29 & 4.63 ± 0.07 & 46 & 1 & 19.1$^{+1.1}_{-1.9}$ & 0.7$^{+1.6}_{-0.7}$ & 55$^{+25}_{-15}$ & 0.23 / 0.78 / 0.29 / $<$ 0.84 & $\prescript{\mathrm{m}}{(24)}{}$204 ± 55 \\
70 & HD~93030 & $<$ 2.65 &  &  &  & 31.0 & 4.20 & 4.30 ± 0.10 & 101 & 5 & 15.1$^{+0.8}_{-1.4}$ & 1.0$^{+2.2}_{-1.0}$ & 115$^{+25}_{-25}$ & 0.46 / 0.47 / 0.64 / $<$ 1.26 & $\prescript{\mathrm{m}}{(21)}{}$Non-det. $\prescript{\mathrm{b}}{(42)}{}$SB1 / 2.2 / --- / \textgreater{} 0.6 \\
71 & HD~37042 & $<$ 2.54 & 8.33 ± 0.11 & 8.04 ± 0.08 & 8.75 ± 0.08 & 29.3 ± 0.3 & 4.30 ± 0.05 & 4.11 ± 0.07 & 30 & 1,8 & 12.9$^{+0.8}_{-0.5}$ & 0.1$^{+1.1}_{-0.1}$ & 35$^{+35}_{-15}$ & 0.22 / 0.41 / 0.27 / $<$ 0.72 & $\prescript{\mathrm{m}}{(24)}{}$Non-det. \\
72 & HD~37481 & $<$ 2.47 & 8.39 & 7.55 ± 0.02 & 8.75 ± 0.08 & 23.3 & 4.17 & 3.59 ± 0.06 & 67 & 1,6 & 8.6$^{+0.6}_{-0.8}$ & 7.3$^{+7.3}_{-7.3}$ & 85$^{+25}_{-25}$ & 0.68 / 0.66 / 0.78 / $<$ 0.62 &  \\
73 & HD~108248 & $<$ 2.31 &  &  &  & 28.8 & 4.08 & 4.29 ± 0.07 & 85 & 5 & 14.2$^{+0.6}_{-1.1}$ & 5.6$^{+2.4}_{-3.5}$ & 195$^{+35}_{-35}$ & 0.86 / 0.91 / 0.79 / $<$ 0.78 & $\prescript{\mathrm{m}}{(23)}{}$N/A \\
74 & HD~34078 & $<$ 2.2 &  & 7.25 ± 0.09 & 8.34 ± 0.25 & 33.0 & 4.07 & 4.59 ± 0.06 & 24 & 1,5,16 & 17.7$^{+1.8}_{-1.3}$ & 3.3$^{+1.4}_{-1.7}$ & 195$^{+35}_{-35}$ & 0.50 / 0.83 / 0.82 / $<$ \textcolor{red}{0.36} & $\prescript{\mathrm{m}}{(24,28)}{}$Non-det. \\
75 & HD~36512 & $<$ 2.2 & 8.35 ± 0.14 & 7.79 ± 0.11 & 8.75 ± 0.09 & 33.4 & 4.30 & 4.56 ± 0.07 & 25 & 1,4,5 & 17.7$^{+1.9}_{-1.3}$ & 0.5$^{+2.0}_{-0.5}$ & 35$^{+25}_{-25}$ & 0.26 / 0.64 / 0.26 / $<$ \textcolor{red}{0.19} & $\prescript{\mathrm{b}}{(43)}{}$SB1 / --- / --- / --- \\
\textbf{76} & \textbf{3293-016} & $<$ \textbf{2.0} &  & \textbf{7.66 ± 0.33} &  & \textbf{25.3 ± 1.0} & \textbf{3.80 ± 0.10} & \textbf{4.18 ± 0.06} & \textbf{54} & \textbf{7} & \textbf{11.4$^{+1.0}_{-0.6}$} & \textbf{13.7$^{+2.4}_{-3.3}$} & \textbf{185$^{+25}_{-45}$} & \textbf{0.45 / 0.82 / 0.68 /~}$<$~\textbf{0.96} &  \\
77 & HD~24760 & $<$ 2.04 &  &  &  & 27.8 & 4.25 & 4.48 ± 0.07 & 128 & 5 & 14.4$^{+2.1}_{-0.7}$ & 5.8$^{+1.9}_{-2.5}$ & 205$^{+115}_{-35}$ & \textcolor{red}{0.01} / 0.81 / 0.49 / $<$ 0.68 & $\prescript{\mathrm{m}}{(23)}{}$N/A $\prescript{\mathrm{b}}{(44)}{}$SB1 / 14 / 0.46 / 0.5 \\
78 & HD~74575 & $<$ 1.84 & 8.37 ± 0.10 & 7.92 ± 0.10 & 8.79 ± 0.08 & 22.9 ± 0.3 & 3.60 ± 0.05 & 4.16 ± 0.06 & 11 & 1,4 & 11.2$^{+0.2}_{-0.2}$ & 16.9$^{+0.8}_{-1.1}$ & 155$^{+35}_{-45}$ & 0.13 / 0.85 / 0.85 / $<$~\textcolor{red}{ 0.37} & $\prescript{\mathrm{m}}{(20,21,26)}{}$-219 ± 60 \\
\textbf{79} & \textbf{3293-007} & $<$ \textbf{1.8} & \textbf{8.13 ± 0.22} & \textbf{7.50 ± 0.08} & \textbf{8.71 ± 0.17} & \textbf{23.2 ± 1.0} & \textbf{3.08 ± 0.10} & \textbf{4.69 ± 0.06} & \textbf{56} & \textbf{7,9} & \textbf{14.9$^{+1.0}_{-1.0}$} & \textbf{12.1$^{+1.4}_{-1.4}$} & \textbf{395$^{+45}_{-55}$} & \textbf{\textcolor{red}{0.05} / 0.88 / 0.65 /~}$<$~\textbf{0.68} &  \\
80 & HD~50707 & $<$ 1.64 & 8.18 ± 0.10 & 8.03 ± 0.15 & 8.64 ± 0.18 & 26.0 & 3.60 & 4.18 ± 0.07 & 44 & 1,5,11 & 12.0$^{+1.0}_{-0.7}$ & 13.1$^{+2.2}_{-2.2}$ & 465$^{+45}_{-75}$ & \textcolor{red}{0.02} / 0.83 / 0.67 / $<$ 1.03 & $\prescript{\mathrm{m}}{(20,21,26)}{}$149 ± 19 \\
81 & HD~52089 & $<$ 1.62 & 8.09 ± 0.12 & 7.93 ± 0.14 & 8.44 ± 0.18 & 23.0 ± 1.0 & 3.30 ± 0.15 & 4.41 ± 0.06 & 28 & 1,11 & 12.7$^{+0.7}_{-0.9}$ & 14.8$^{+1.9}_{-1.9}$ & 395$^{+55}_{-45}$ & 0.19 / 0.86 / 0.59 / $<$ 0.87 & $\prescript{\mathrm{m}}{(20,21,22,26)}{}$-156 ± 18 \\
\textbf{82} & \textbf{3293-006} & $<$ \textbf{1.6} & \textbf{7.81 ± 0.23} & \textbf{7.37 ± 0.16} & \textbf{8.84 ± 0.31} & \textbf{24.0 ± 1.5} & \textbf{3.15 ± 0.10} & \textbf{4.61 ± 0.07} & \textbf{211} & \textbf{7} & \textbf{14.2$^{+0.5}_{-0.9}$} & \textbf{13.9$^{+0.8}_{-1.3}$} & \textbf{585$^{+15}_{-15}$} & \textbf{\textcolor{red}{0.05} / 0.78 / 0.61 /~}$<$\textbf{\textcolor{red}{~0.35}} &  \\
83 & HD~30836 & $<$ 1.50 & 8.19 ± 0.09 & 7.54 ± 0.15 & 8.40 ± 0.29 & 21.5 & 3.35 & 4.17 ± 0.09 & 51 & 1,5,11 & 10.2$^{+0.6}_{-0.4}$ & 19.2$^{+3.5}_{-2.5}$ & 175$^{+35}_{-25}$ & 0.12 / 0.79 / 0.66 / $<$ 0.68 & $\prescript{\mathrm{m}}{(23)}{}$N/A $\prescript{\mathrm{b}}{(45)}{}$SB1 / 9.5 / 0.03 / 0.5 \\
84 & HD~36591 & $<$ 1.50 & 8.33 ± 0.08 & 7.75 ± 0.09 & 8.75 ± 0.11 & 27.0 & 4.12 & 4.16 ± 0.09 & 14 & 1,4,5 & 12.6$^{+0.7}_{-1.7}$ & 7.3$^{+3.1}_{-3.1}$ & 185$^{+35}_{-35}$ & 0.57 / 0.88 / 0.84 / $<$\textcolor{red}{~0.16} &  \\
85 & HD~205021 & $<$ 1.50 & 8.24 ± 0.06 & 8.11 ± 0.11 & 8.64 ± 0.13 & 27.0 & 4.05 & 4.38 ± 0.08 & 32 (26) & 1,4,5,17 & 13.5$^{+0.5}_{-0.5}$ & 9.2$^{+2.4}_{-2.4}$ & 35$^{+25}_{-15}$ & 0.19 / 0.80 / 0.74 / $<$~\textcolor{red}{0.00} & $\prescript{\mathrm{m}}{(17,18,22,24)}{}$ \textbf{76 ± 6} $\prescript{\mathrm{b}}{(46)}{}$SB1 / 11 / 0.52 / 0.7 \\
86 & HD~46328 & $<$ 1.36 & 8.18 ± 0.12 & 8.00 ± 0.17 & 8.59 ± 0.17 & 27.5 ± 1.0 & 3.75 ± 0.15 & 4.58 ± 0.07 & 10 & 1,2 & 15.3$^{+0.7}_{-1.1}$ & 10.0$^{+1.4}_{-2.0}$ & 115$^{+35}_{-35}$ & 0.80 / 0.82 / 0.58 / $<$~\textcolor{red}{0.33} & $\prescript{\mathrm{m}}{(18,19,20,21,22,24)}{}$350 ± 10 \\
\textbf{87} & \textbf{3293-003} & $<$ \textbf{1.3} & \textbf{7.95 ± 0.22} & \textbf{7.52 ± 0.06} & \textbf{8.75 ± 0.20} & \textbf{22.5 ± 1.0} & \textbf{3.00 ± 0.10} & \textbf{4.81 ± 0.06} & \textbf{84} & \textbf{7,9} & \textbf{16.2$^{+0.8}_{-0.8}$} & \textbf{10.6$^{+1.0}_{-0.7}$} & \textbf{275$^{+25}_{-35}$} & \textbf{0.09 / 0.72 / 0.76 /~}$<$\textbf{~0.75} &  \\
\textbf{88} & \textbf{3293-002} & $<$ \textbf{1.2} & \textbf{8.10 ± 0.22} & \textbf{8.08 ± 0.12} & \textbf{8.67 ± 0.15} & \textbf{24.2 ± 1.0} & \textbf{3.00 ± 0.10} & \textbf{5.22 ± 0.05} & \textbf{90} & \textbf{7,9} & \textbf{23.6$^{+1.3}_{-1.8}$} & \textbf{6.7$^{+0.8}_{-0.5}$} & \textbf{285$^{+35}_{-25}$} & \textbf{0.57 / 0.82 / 0.56 /~}$<$~\textbf{\textcolor{red}{0.30}} &  \\
89 & HD~35468 & $<$ 1.01 & 8.11 ± 0.09 & 7.90 ± 0.16 & 8.16 ± 0.27 & 22.0 ± 1.0 & 3.50 ± 0.20 & 3.90 ± 0.06 & 51 & 1,11 & 9.5$^{+0.6}_{-0.6}$ & 19.2$^{+5.6}_{-4.0}$ & 325$^{+65}_{-55}$ & 0.13 / 0.79 / 0.48 / $<$ 0.88 & $\prescript{\mathrm{m}}{(23)}{}$N/A $\prescript{\mathrm{b}}{(4)}{}$SB2? / --- / --- / --- \\
90 & HD~51309 & $<$ 0.84 & 7.89 ± 0.19 & 7.86 ± 0.34 & 8.25 ± 0.44 & 17.5 ± 1.0 & 2.75 ± 0.15 & 5.01 ± 0.20 & 32 & 1,11 & 20.5$^{+3.8}_{-2.1}$ & 7.4$^{+2.0}_{-1.2}$ & 395$^{+35}_{-35}$ & \textcolor{red}{0.00} / \textcolor{red}{0.04} / \textcolor{red}{0.06} / $<$~\textcolor{red}{0.47} &  \\ \hline
\end{tabular}}
\captionsetup{labelformat=empty}
\caption{Key to references: 
[1] \citet{Proffitt2001}; 
[2] \citet{Morel2006}; 
[3] \citet{Mazmumdar2006}; 
[4] \citet{Nieva2012}; 
[5] \citet{Proffitt2015}; 
[6] \citet{Venn2002}; 
[7] \citet{Hunter2009,Proffitt2024}; 
[8] \citet{Simon2010,Nieva2011}; 
[9] \citet{Proffitt2016}; 
[10] \citet{Dupret2004}; 
[11] \citet{Morel2008}; 
[12] \citet{Pamyatnykh2004}; 
[13] \citet{Aerts1996}; 
[14] \citet{Neiner2003,Briquet2016}; 
[15] \citet{Aerts2006}; 
[16] \citet{Gies1992}; 
[17] \citet{Henrichs2013};
[18] \citet{Rudy1978};
[19] \citet{Hubrig2006};
[20] \citet{Hubrig2009};
[21] \citet{Bagnulo2015};
[22] \citet{Shultz2018};
[23] \citet{Wade2016};
[24] \citet{Schnerr2008};
[25] \citet{Schnerr2006};
[26] \citet{Bagnulo2012};
[27] \citet{Butkovskaya2007};
[28] \citet{Grunhut2017};
[29] \citet{Vitrichenko2002};
[30] \citet{Levato1987};
[31] \citet{Chini2012};
[32] \citet{Abt1978};
[33] \citet{Southworth2004};
[34] \citet{Leone1999};
[35] \citet{Aerts1998};
[36] \citet{Stickland2000};
[37] \citet{Ebbighausen1960};
[38] \citet{Hernandez1980};
[39] \citet{Stickland1996};
[40] \citet{Lehmann2001};
[41] \citet{Miroshnichenko2001};
[42] \citet{Lloyd1995};
[43] \citet{Burssens2020};
[44] \citet{Libich2006};
[45] \citet{Luyten1936};
[46] \citet{Fitch1969}
\\
The typical 1-$\sigma$ error for $\Teff$ ($\logg$) of 1000$\K$ (0.1$\gunit$) is adopted when not provided.}
\end{table*}

\section{Testing rotational mixing: our method}\label{sec_meth}

\subsection{Initial boron abundance}\label{sec_logBi}
It is not certain that our stars were born with the proto-solar boron abundances. The direct determination of the initial boron abundance in B-type stars has an inherent challenge due to boron depletion throughout stellar evolution. Assuming a reasonable value for the initial boron abundance is crucial for estimating the degree of boron depletion, which affects estimating the strength of rotational mixing. 

In F- and G-type dwarf stars where the surface boron abundance is thought to have remained constant because their surface beryllium content is not depleted, a correlation between boron and oxygen abundances has been identified \citep{Smith2001}. This correlation suggests that the boron abundance scales with the oxygen abundance according to $\logB=(1.39\pm0.08) \times \logO- (9.62\pm1.38)$. This correlation is thought to originate from the boron production mechanisms, i.e., cosmic-ray spallation of CNO nuclei, and the $\nu$-process in supernovae.

If we consider the boron-oxygen correlation, the challenge of estimating the initial boron abundance boils down to estimating the initial oxygen abundance in our stars. In the solar neighborhood ($< 500 \pc$), chemical homogeneity has been found in unevolved B-type stars \citep{Przybilla2008,Nieva2012} (see Table~\ref{tab_cno}). Given that most of the stars in our dataset are B-type stars in the solar neighborhood without nitrogen enhancement --- which precedes any oxygen depletion; see \citet{Brott2011a} ---, it is reasonable to assume that their surface oxygen abundances reflect their initial values. The observed chemical homogeneity in the solar neighborhood suggests that their initial surface oxygen abundance should be homogeneous, implying that their initial boron abundance should be homogeneous too, based on the boron-oxygen correlation. The oxygen abundance of $\logO=8.76 \pm 0.05$ determined by \citet{Nieva2012} is very similar to the proto-solar value $\logO=8.75$ from \citet{Asplund2021}. Therefore, we adopt the proto-solar boron abundance $\logB=2.76$ as the initial boron abundance for our stars in the solar neighborhood. Note that this value is consistent with the meteoritic boron abundance of $\logB=2.78 \pm 0.05$ \citep{Zhai1994}, further validating this choice.

The stars in NGC\,3293 show a somewhat lower surface oxygen abundance of $\logO=8.65 \pm 0.17$ \citep{Hunter2009} compared to the solar neighborhood. Consequently, we assume a slightly lower initial boron abundance of $\logB = 2.62$ for these stars, based on the boron-oxygen correlation. We do not use individual surface oxygen abundances to derive the initial boron abundances due to their large uncertainties. We adopt an uncertainty of 0.20 \citep{Cunha1999,Asplund2021} for both initial boron abundances. The impact of the adopted initial boron abundance will be addressed in Section~\ref{sec_dis}. 

\subsection{Rotational mixing parameters}

The rotational mixing scheme implemented in MESA incorporates several physical parameters. The value of $\fc$ (\texttt{am\_D\_mix\_factor} in MESA) determines how strong rotationally-induced instabilities contribute to mixing, $\fmu$ (\texttt{am\_gradmu\_factor} in MESA) determines the strength of the inhibiting effect of the mean molecular weight gradient on rotational mixing, $\fnu$ (\texttt{am\_nu\_factor} in MESA) determines the efficiency of the angular momentum transfer is via magnetic torques.

\begin{figure*}
	\centering
	\includegraphics[width=0.30\linewidth]{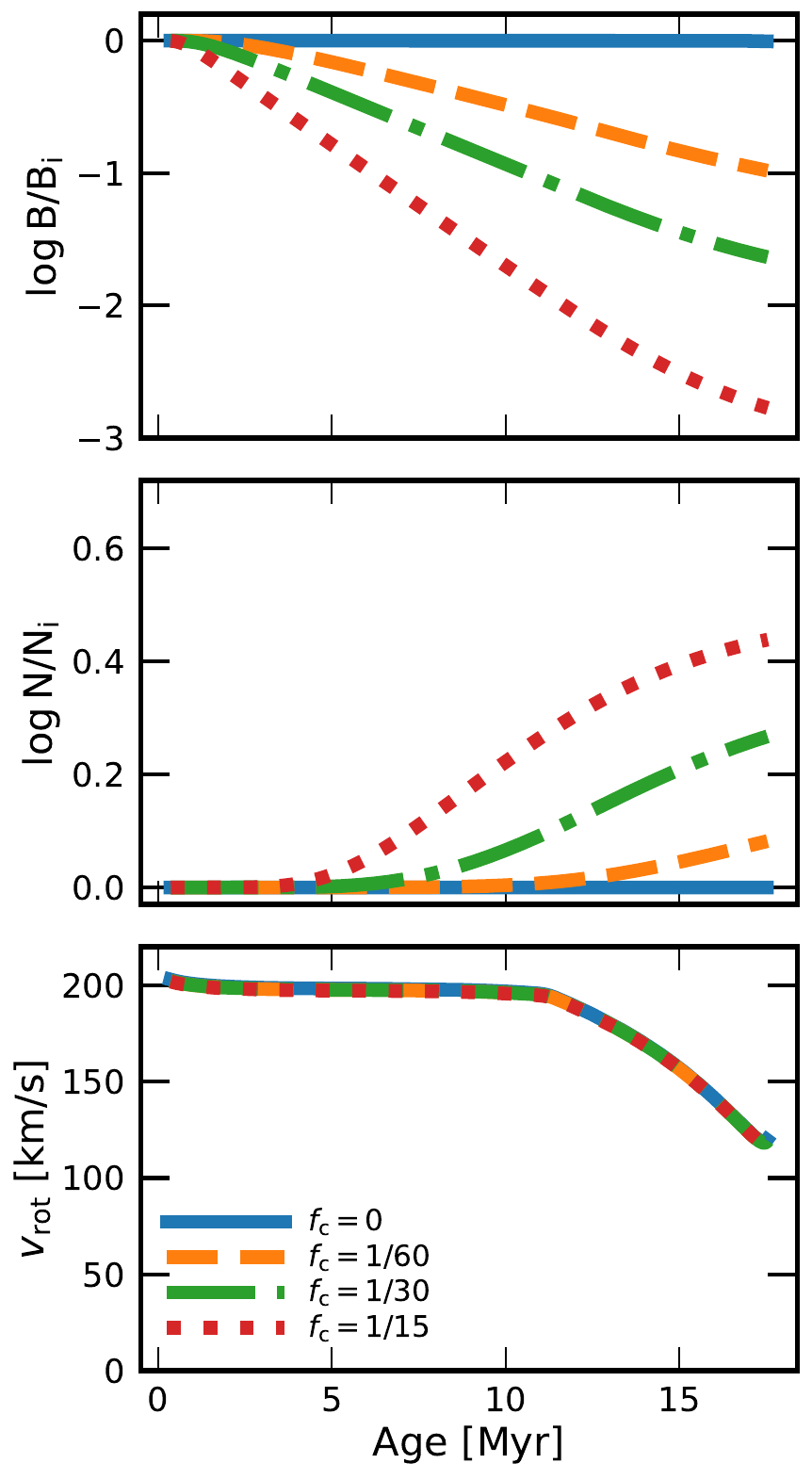}
	\includegraphics[width=0.30\linewidth]{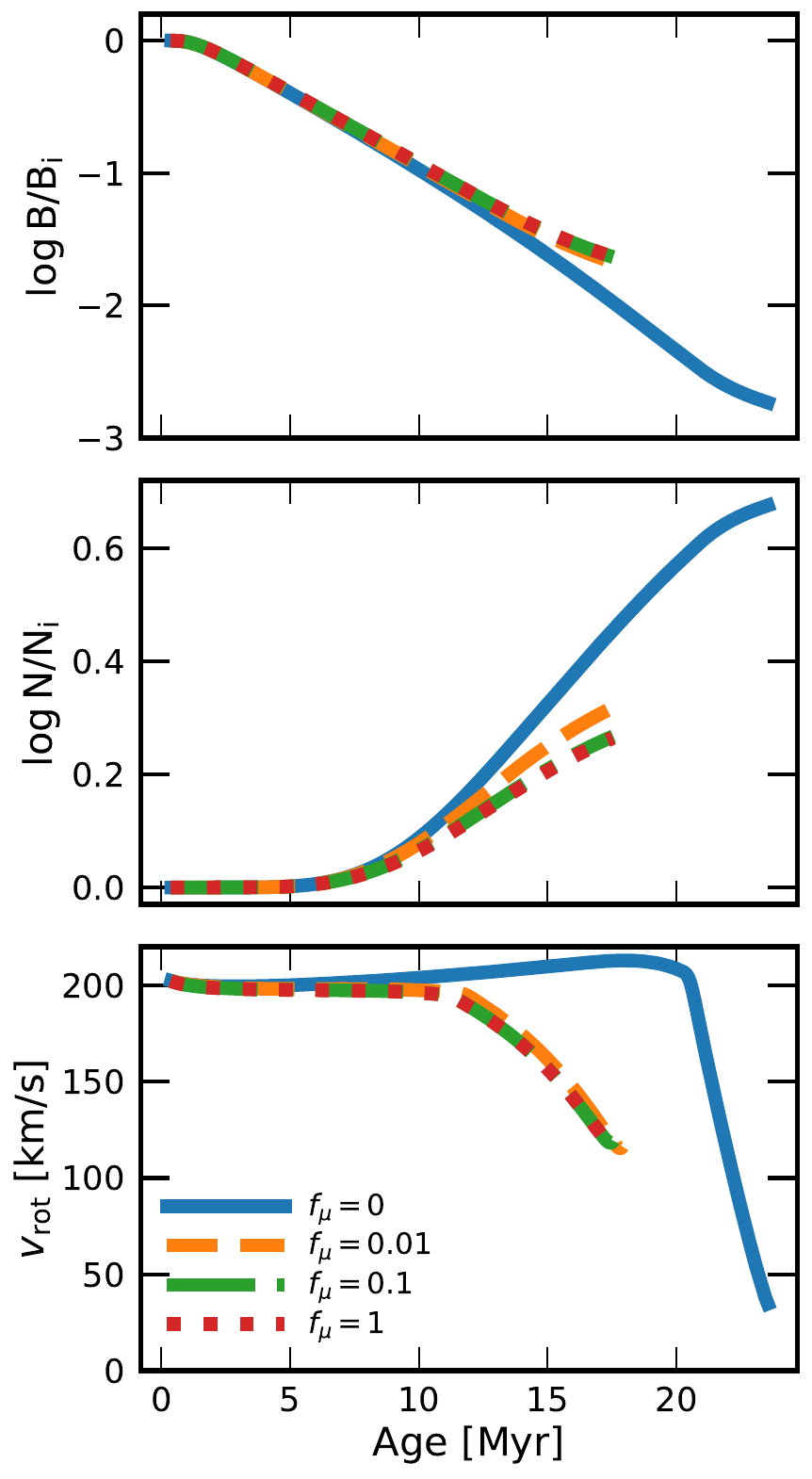}
	\includegraphics[width=0.30\linewidth]{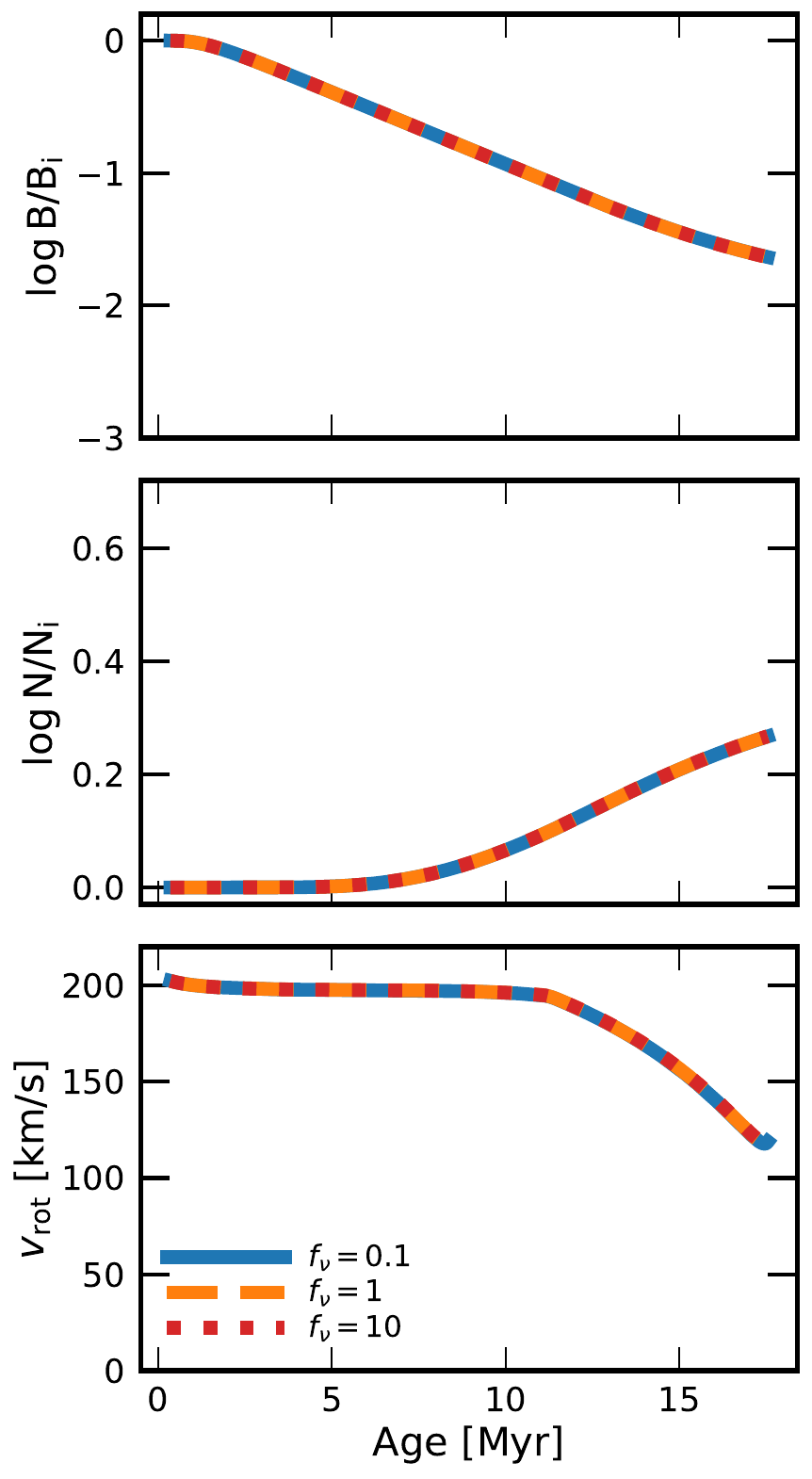}
	\caption{Evolution of surface abundances of boron and nitrogen and rotational velocity in a 12$\mso$ stellar model with $\vini = 200 \kms$ during the main sequence evolution. Three panels show the effects of varying rotational mixing efficiency $\fc$ (left), degrees of inhibition by chemical gradient $\fmu$ (middle), and degrees of angular momentum transport $\fnu$ (right).}
	\label{fig_fcfmufnu}
\end{figure*}

The implementation of the parameter $\fc$ in MESA is such that it acts as a scaling factor directly multiplied to the diffusion coefficient originating from rotationally-induced instabilities. Therefore, it directly influences the strength of rotational mixing. The left panel of \Fig{fig_fcfmufnu} shows how $\fc$ affects the evolution of surface abundances and rotational velocity. While it has no effect on the rotational velocity, the degrees of boron depletion and nitrogen enhancement increase significantly as $\fc$ increases.

In MESA, $\fmu$ is multiplied by the mean molecular weight term in the diffusion equations. The value of $\fmu=0$ corresponds to the case where the chemical gradient does not inhibit rotational mixing at all. 
The middle panel of \Fig{fig_fcfmufnu} shows that varying $\fmu$ by orders of magnitude does not yield significant changes in surface abundances of boron and nitrogen. The sole exception is when $\fmu=0$, which is physically unrealistic and results in quasi-chemically homogeneous evolution, showing distinctively different behavior. Furthermore, boron is a trace element,
and the efficiency of boron depletion is independent of $\fmu=0$ because
mean molecular weight gradients are extremely shallow in the boron depletion region.

$\fnu$ is the factor that scales the diffusion coefficient for angular momentum transport in MESA. \citet{Yoon2006} found that $\fnu$ has a self-regulating characteristic. As $\fnu$ increases, i.e., angular momentum transport becomes more efficient, it weakens differential rotation, which in turn makes the angular momentum transport less efficient. The right panel of \Fig{fig_fcfmufnu} also confirms their finding, as models do not show any difference in surface abundances and rotational velocity.

The value of $\fc$ is therefore the primary parameter that significantly impacts the surface abundances. Thus, we will focus on this parameter, speaking of it as describing the rotational mixing efficiency, in the subsequent analysis. 

\citet{Chaboyer1992} suggested $\fc=1/30$ based on their theoretical work. This value has been widely adopted in various stellar model grids \citep{Heger2000b,Marchant2017,Wang2020,Menon2021,Keszthelyi2022}. We use the same value for our work. Notably, \citet{Brott2011a} used a lower value, i.e., $\fc=0.023$ for their models calculated by STERN, based on a comparison of their models with observational data from the VLT-FLAMES survey \citep{Hunter2009}. They concluded that $\fc=0.023$ can best reproduce the observed nitrogen enrichments in both, slow and fast rotators.

\subsection{Varying the mixing efficiency}\label{sec_fc}
Unless the rotational mixing of hydrogen and helium is significant, which can occur when a star rotates very fast or when $\fc$ is very large, the stellar luminosity, temperature, gravity, and rotational velocity are approximately independent of $\fc$ in our models, while the surface abundances of boron and nitrogen depend sensitively on $\fc$. This is because the mixing of boron and nitrogen, which are of low abundance, is insufficient to affect the global properties of the star, as illustrated in \Fig{fig_fcfmufnu}. If $\fc$ is very large, the development of a steep mean molecular weight gradient may be prevented by the transport of a significant amount of helium into the envelope. This quasi-chemically homogeneous evolution alters the evolutionary track \citep[see also][]{Brott2011a}. Since none of our stars shows the very fast rotation that is required for this channel, we only consider the change of surface abundances for different mixing efficiencies, but assume that the other stellar properties are not affected.

To check the quantitative dependency of the change of surface abundances on the mixing efficiency, we compute small grids of models with different $\fc$ values of $\fc \times 30 = 0\dots2$. I.e., we investigate rotational mixing efficiency ranging from 0\% to 200\% relative to the widely adopted value of $\fc \times 30 = 1$. We do not consider mixing efficiency parameters above $\fc \times 30 =2$, since those have been ruled out by the analysis of
the nitrogen surface abundances in massive main sequence stars \citep{Hunter2008b, Brott2011a}. Each of these grids is composed of $\sim 100$ models with $\sim 10$ different initial masses ($\Mini$) and $\sim 10$ different initial rotational velocities ($\vini$), significantly sparser compared to the original grid discussed in Section~\ref{sec_grid}.

\Figure{fig_fcfmufnu} shows that the logarithm of the boron abundance drops
nearly linearly as a function of time in our models (see also Fig.\,\ref{fig_com}). This can be understood as follows.
In a first-order approximation, we can assume that
${d \mathrm{B} \over dt}  \propto \mathrm{B} / \tau_{\rm mix}$, and using
$\tau_{\rm mix} \simeq R^2 / D$  yields
\begin{equation}
{d\log \mathrm{B} \over dt} \propto {D \over R^2} \propto \fc 
\end{equation}
which, for a constant diffusion coefficient $D$ and neglecting the stellar radius change during the main sequence evolution, implies that $\log \mathrm{B}$ should decline linearly with time, with a slope that is proportional to $\fc$.

\begin{figure} 
	\centering
	\includegraphics[width=\linewidth]{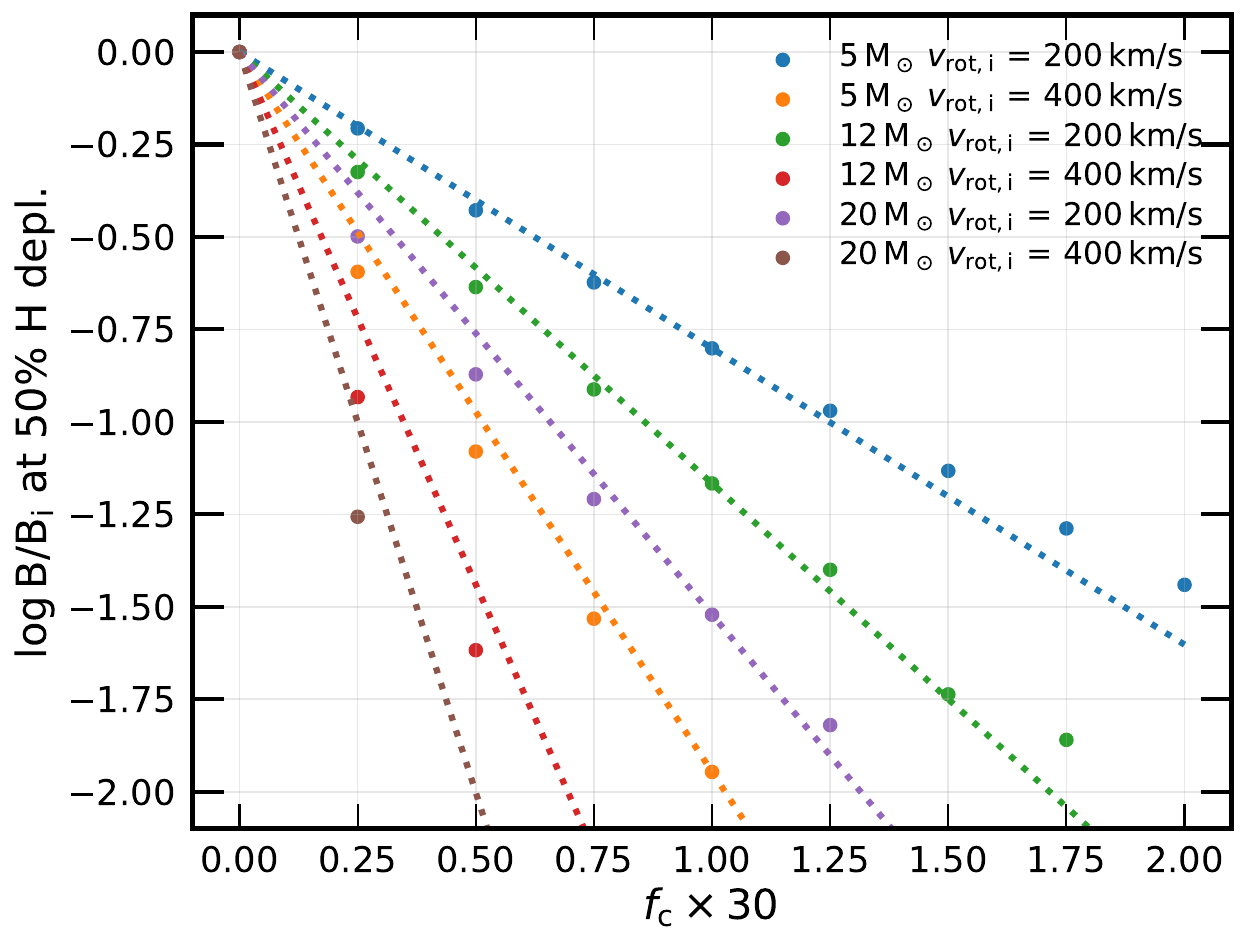}
	\includegraphics[width=\linewidth]{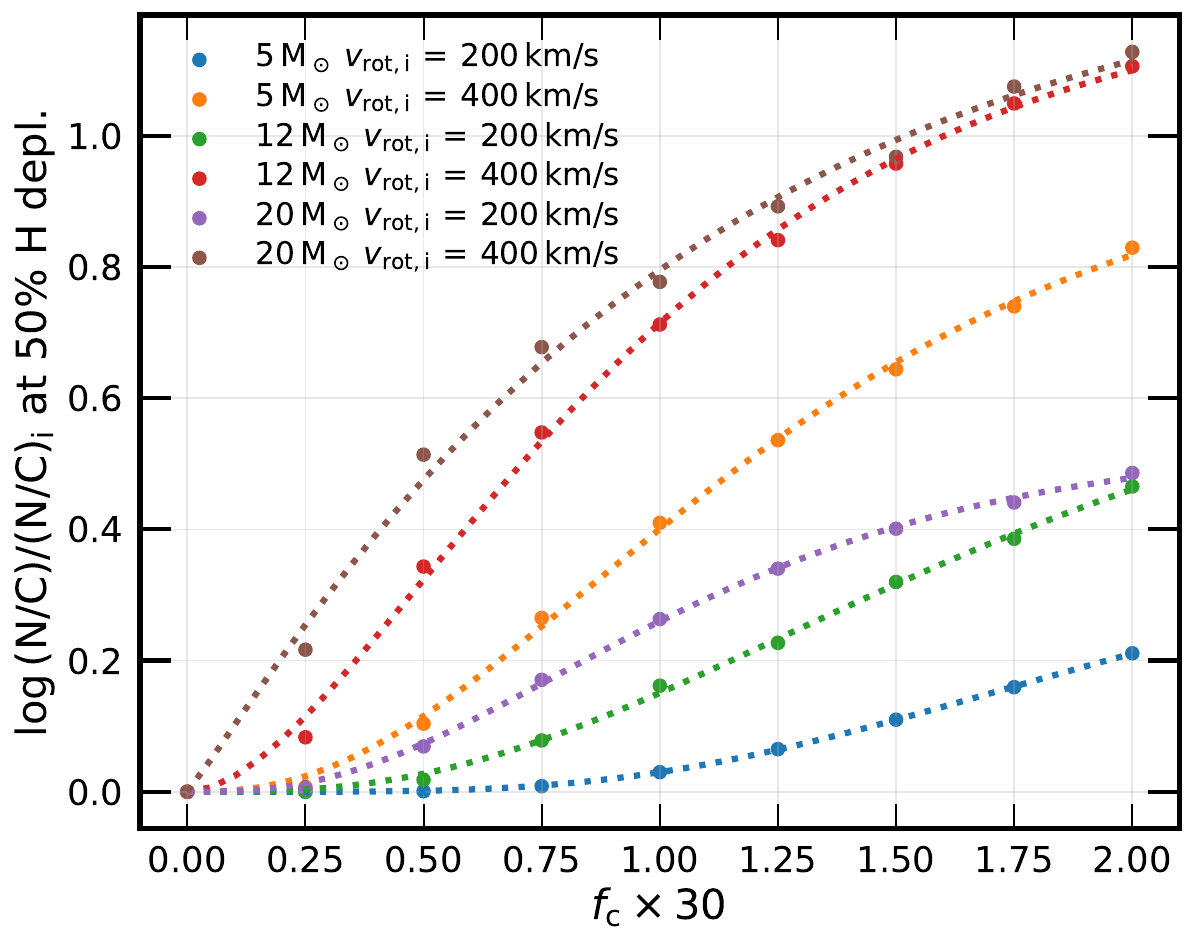}
	\caption{Boron depletion factor (top) and nitrogen enhancement factor (bottom) at 50\% central hydrogen depletion for six evolutionary models (filled circles; see legend), as a function of the adopted rotational mixing efficiency parameter $\fc$ . The dotted lines show our analytic fits for the boron depletion and the nitrogen enhancement factors (see text).}
	\label{fig_logX_fc}
\end{figure}

Thus, for a given $\Delta t$, $\Delta \log \mathrm{B}$ is directly proportional to $\fc$. \Figure{fig_logX_fc} shows that model data points from the small grids align well along the lines connecting points at $\fc \times 30 = 0$ (no rotational mixing) and points at $\fc \times 30 = 1$ (the conventional value). This relationship can be expressed as a simple scaling equation as
\begin{equation}\label{eq_logBBi}
    \log \left(\mathrm{\frac{B}{B_i}}\right)(f_{c}) = \log \left(\mathrm{\frac{B}{B_i}}\right) \left(f_{c} =\frac{1}{30} \right)\, \frac{f_{c}}{1/30} ,
\end{equation}
which we adopt to explore the effect of varying $\fc$ on the boron depletion factor
(cf., top panel of \Fig{fig_logX_fc}).

The change of the logarithm of the surface nitrogen abundance $N$ with time is not
well described by a linear function. The reason is that nitrogen is mixed up from 
the stellar core, such that it may take some time before an enhanced nitrogen abundance appers at the surface. Furthermore, the nitrogen enhancement may saturate at CN- or CNO-equilibrium values. \cite{Kohler2012} found that 
the dependence of the nitrogen enhancement factor on the fractional main sequence 
time $\tau=t/\tau_{\rm H}$, where $\tau_{\rm H}$ is the core hydrogen burning lifetime, is well described by the incomplete gamma function. 
We use their ansatz here, where, for a given evolutionary model, we assume that
$\logN (\tau, f_\mathrm{c0}) = \logN (\tau a, f_\mathrm{c0}/a)$,
where $f_\mathrm{c0}$ is our canonical mixing efficiency.
We explore 11 values of $\tau=0, 0.1, \dots 1$. Based on the models from our small model grids, for each of the $\sim 1000$ different triples of ($\Mini$, $\vini$, $\tau$) we obtain the best-fit incomplete gamma function, 
and we can obtain $\logNCNCi$ for any ($\Mini$, $\vini$, $\tau$, $\fc$) through multi-linear interpolation (cf., bottom panel of \Fig{fig_logX_fc}). We use this method to explore the effect of $\fc$ on the nitrogen enhancement factor, on $\logNCNCi$, and on $\logNONOi$.

\subsection{Bayesian analysis}\label{sec_app}
\textsc{Bonnsai} is a Bayesian tool for comparing observationally derived stellar parameters with those of stellar models \citep{Schneider2014}. We adopt its framework here to quantify the compatibility between our stellar models and the stars in our data set. We use all the observationally derived stellar parameters for each star. All the stars in our sample have measurements of $\Teff$, $\logg$, $\logL$, and $\vsini$. About two-thirds of them have a measurement of $\logB$, while one-third of them only have an upper limit to $\logB$. Additionally, the majority of the stars have nitrogen enhancement factors determined ($\logNCNCi$, $\logNONOi$), and seven stars have equatorial rotational velocity ($\veq$) determined. See Table~\ref{tab_obs} for their values and references.

In our analysis, we generally use the uncertainties reported for the measured quantities in the original literature.
Since the uncertainties of $\vsini$ are often not reported 
we adopt an uncertainty of $\max(0.1 \, \vsini, 10) \kms$ for each star. For $\veq$, we adopt an uncertainty of $\max(0.5 \, \veq, 10) \kms$ \citep[e.g., $\veq = 55 \pm 28 \kms$ for HD\,3360;][]{Neiner2003}. 
We adopt an uncertainty of the maximum of 0.4\,dex or the original $\Delta \logNCNCi$ for $\logNCNCi$, and do the same for the uncertainty of $\logNONOi$ (see \App{app_cno}).

For our Bayesian analysis, we interpolate our original model grid described in Section~\ref{sec_mod} in order to make a denser grid of main sequence stars with a resolution of $\Delta M_\textrm{i} = 0.2 \mso$, $\Delta \vini = 10 \kms$, and $\Delta \, \mathrm{age} = 0.02 \Myr$ as was done in \citet{Schneider2014}. For interpolation, we follow the approach of \textsc{Starmaker} \citep{Brott2011b}. This increases the number of initial masses of the stellar models from 46 to 175, and the number of model sequences from $\sim 2600$ to $\sim 10\,000$. For the prior functions for the model parameters, we use \citet{Salpeter1955} initial mass function, rotational velocity distribution from \citet{Dufton2013}, and flat age distribution.

\begin{figure*} 
	\centering
	\includegraphics[width=\linewidth]{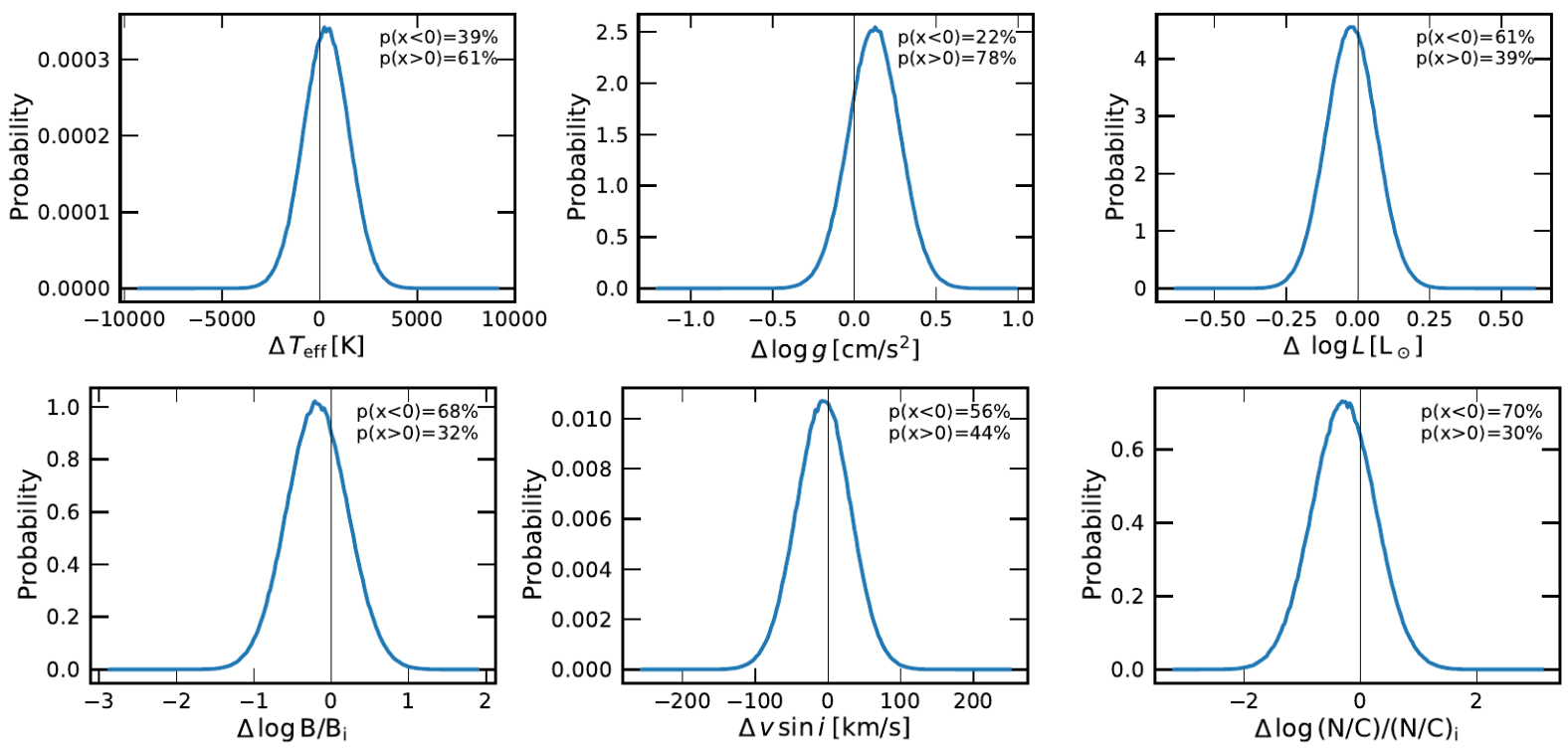}
	\caption{The result of posterior predictive check for $\fc=1$ for Star\,51, namely, probability distributions of the differences between the predicted values from the stellar models and the observables. The integrals of both sides of the distributions with respect to zero, p(x$<$0) or p(x$>$0), are also presented.}
	\label{fig_ppt}
\end{figure*}

A crucial goodness-of-fit test conducted in the Bayesian framework is posterior predictive check \citep{Schneider2014}. This is done by comparing the observables with the predicted values from stellar models. The probability distribution of the difference between the predicted value and the observable is calculated. If this distribution is significantly skewed away from zero, we conclude that the stellar models cannot reproduce the observable. \Figure{fig_ppt} presents the result of the posterior predictive check for Star\,51 (3293-031) when using all the observationally derived parameters for this star. We select the minimum of p(x$<$0) or p(x$>$0) for each observable $\mathrm{x}$, divide it by 50\%, and refer to it as $\pfc^\mathrm{x}$. E.g., $P(1/30)^\mathrm{\Teff}=38\%/50\%=0.76$ for Star\,51.

Assuming that our sample stars are main sequence stars, we generally expect a high probability for recovering a given star with our models when restricting the Bayesian analysis to only $\Teff$ and $\logL$, because any point in the main sequence band is reached by a star of our synthetic population for a specific mass and time. Even binary products, i.e., merger products or mass gainers, are expected to be recovered as they are thought to evolve similarly to single stars \citep[e.g.,][]{Hellings1983,Hellings1984,Schneider2020}. Notably, as our evolutionary tracks are rather independent of the rotation rate in the analyzed range of rotational velocities, we do not need to consider the rotational mixing efficiency parameter $\fc$ here. However, stars that are located close to one of the two edges of the main sequence band in the HRD may still obtain a rather low probability. The reason is that their error ellipse reaches into a part of the parameter space which contains no (leftward of the ZAMS) or few (rightward of the TAMS) models. For example, a star located on the ZAMS will obtain a probability of $\ptl=\pfc^{\Teff} \pfc^{\logL}$ of only $\sim 0.5$. A similar consideration holds for the spectroscopic Hertzsprung-Russell diagram, or $\ptg=\pfc^{\Teff} \pfc^{\logg}$. We list the values of $\ptl$ and $\ptg$ in Table~\ref{tab_obs}.

A second issue that may lead to a low probability of a given star being recovered by our models is referred to as mass discrepancy. For example, Star\,82 is located near the $15 \mso$ evolutionary track in the HRD, while it is near the $20 \mso$ evolutionary track in the sHRD. While at high mass, the mass discrepancy has been related to helium enrichment and envelope stripping \citep[e.g.,][]{Langer1992, Langer2014}, the cause for this in B-type stars is unclear. As several of our mass-discrepant stars show boron at their surface, it is unlikely that mixing or stripping is the cause here. We therefore assume in the following that the cause lies in the determination of the stellar surface properties  
rather than in the star itself.
 
In order to be able to still use all our sample stars in our investigation, we perform a Bayesian analysis for all stars which uses only $\Teff$, $\logL$, and $\logg$. From this, we define the quantity $\pev$ as $\pev = \pfc^{\Teff} \, \pfc^{\logg} \, \pfc^{\logL}$. The value of $\pev$ can be low for a given star for two reasons. One is that our models can not reproduce the three essential surface properties of the star simultaneously, which is so for the mass-discrepant cases. The other is the location of the star near an edge of the main sequence band, which would be reflected in low values of $\ptl$ and $\ptg$ defined above.

We perform another Bayesian analysis for all stars in our sample, but in this case, we consider all the observationally derived parameters for each star. Then we define the quantity $\pfcproduct$ as the probability that the stellar models can reproduce the observed star's $\Teff$, $\logg$, $\logL$, $\logBBi$, and $\vsini$ simultaneously, i.e., the parameters that are derived for all the stars in our sample, as
\begin{equation}
    \pfcproduct = \frac{\pfc^{\Teff} \, \pfc^{\logg} \, \pfc^{\logL} \, \pfc^{\logBBi} \, \pfc^{\vsini}}{\pev}.
    \label{e3}
\end{equation} 
Mass, age, and rotation are the stellar model parameters that determine boron depletion. Thus, comparing boron depletion between the models and the star can only be made when these parameters are considered simultaneously. This is implicitly done by considering $\logL$ (mass), $\Teff$, $\logg$ (age), and $\vsini$ (rotation) through the Bayesian analysis.

Since our focus lies on analyzing the internal mixing efficiency, and the deviation of $\pev$ from one reflects a seeming discrepancy, which would not exist had we infinitely accurate data, we correct for this by normalizing to $\pev$ in Eq.\,\ref{e3}. We show in \App{app_pev} that this normalization does not lead to notable bias effects. 
Mind that $\pfc^{\Teff} \, \pfc^{\logg} \, \pfc^{\logL}$ appearing in Eq.\,\ref{e3} and $\pev$ are different since the former is the result of using the observed star's boron depletion as a constraint, while the latter is not.

\begin{figure} 
	\centering
	\includegraphics[width=0.95\linewidth]{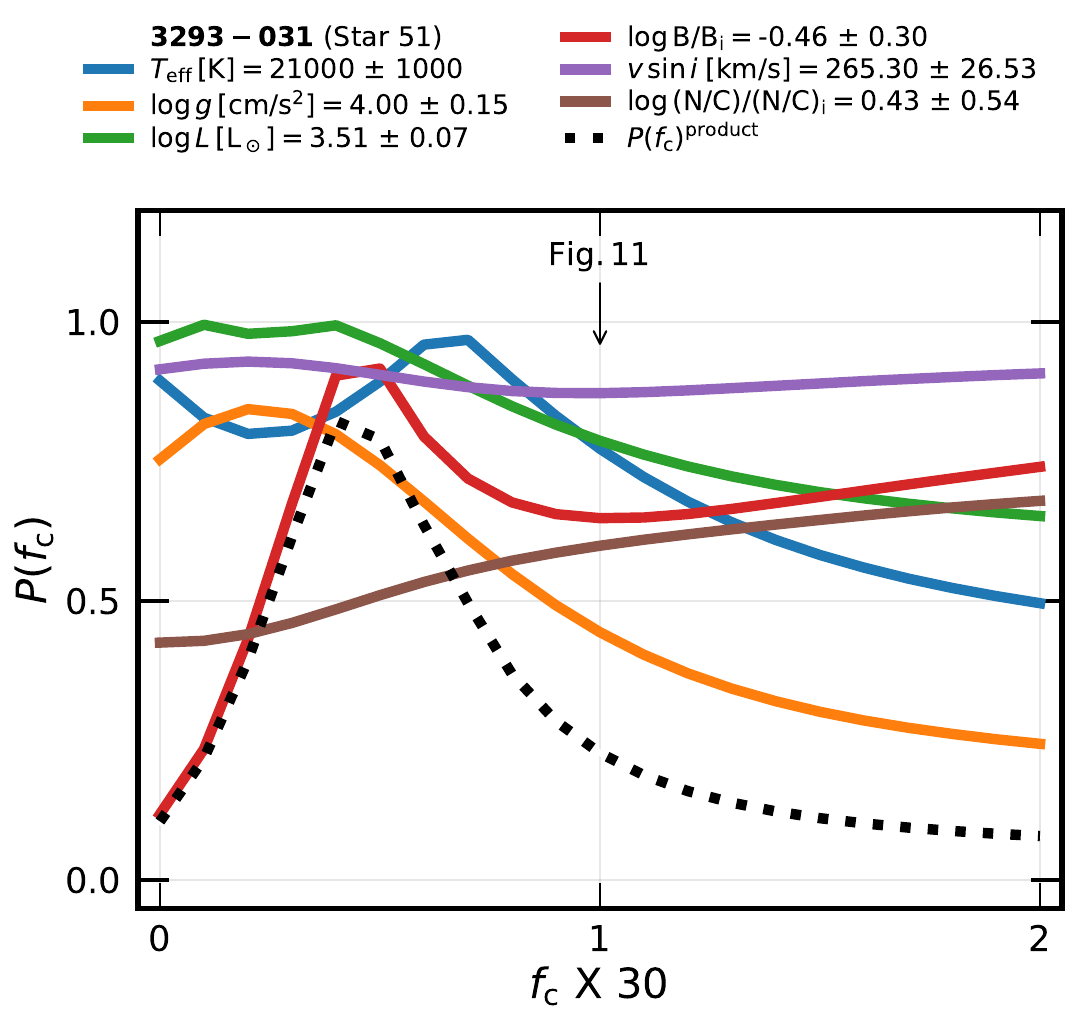}
	\caption{The distributions of $\pfc$ for each observable (solid lines) and the distribution of $\pfcproduct$ (dotted line) for Star\,51. The arrow marks the $\fc$ value corresponding to \Fig{fig_ppt}.} 
	\label{fig_pfc}
\end{figure}

\Figure{fig_pfc} displays the effects of varying $\fc$ on the values of $\pfc$ and $\pfcproduct$ for the boron depleted Star\,51. The value of $\pfc^{\logBBi}$ at $\fc \times 30 = 0$ is low because our models without rotational mixing do not entail boron depletion unless there is a significant mass stripping (see the lower left panel of \Fig{fig_grid}). The mass of Star\,51 estimated from the evolutionary tracks is $\sim 8 \mso$, where mass stripping is insignificant. The distribution of $\pfc^{\logBBi}$ shows a peak at $\fc \times 30 = 0.5$. The distributions of $\pfc$ corresponding to other parameters are not easy to understand because all $\pfc$ values are determined simultaneously. For example, the value of $\pfc^{\logg}$ decreases for $\fc \times 30 \gtrapprox 0.2$. This is because the boron depletion is not too strong such that models with smaller stellar ages (i.e., higher $\logg$) are preferred. This preference for smaller age becomes stronger for larger rotational mixing efficiency because, for stronger rotational mixing, the star has to be younger to avoid strong boron depletion. We define the maximum of $\pfcproduct(\fc)$ as the probability $\pmax$, which describes the highest possible probability of our stellar models to represent a given sample star for any rotational mixing efficiency, and
which defines the value of the mixing efficiency for which this is achieved.

If a star only has an upper limit on the surface boron abundance, $\logB_\mathrm{upper}$, we probe surface boron abundances of $\logB_\mathrm{upper}-0.0\dots2.0$ with a typical uncertainty of 0.3\,dex. When the boron abundance is below $\logB_\mathrm{upper}-2.0$, the value of $\pfcproduct$ is inherently low since it is too much boron depletion given our sample stars' mass and rotation (i.e., low $\pfc^{\logB}$). 
See the lower left panel of \Fig{fig_grid}, in particular, see $M_\mathrm{i} < 20 \mso$ and $\vini < 300 \kms$, where most of our stars would lie. We thereby obtain an upper limit of $\pmax$ for the star.


\section{Results}\label{sec_res}

\subsection{Two groups of stars}
\label{sec_two}
The result of our Bayesian analysis is presented in Table~\ref{tab_obs}. We find that about two-thirds of the stars in our sample show values of $\pmax$ in the range $0.6\dots1$, which implies that they can be well represented by our single star models if the mixing efficiency is tuned accordingly. This includes most of the relatively fast rotators ($\vsini > 50 \kms$) regardless of their boron abundance (e.g., Stars\,4, 63). Slow rotators with high boron abundance also belong to this group (e.g., Stars\,1, 3).
One-third of the stars in our sample show low $\pmax$ values of $0.55\dots 0$. 
%
While we can not define a minimum $\pmax$ value for a star to be considered to fit well to our models, the distribution of the $\pmax$ values (top panel of \Fig{fig_pvsini}) suggests a division into two distinct subsamples, with the majority of our stars clustering around $\pmax\simeq 0.85$, and the low $\pmax$ group showing a rather flat distribution in the range $\pmax \simeq 0.55\dots 0$.  

As our analysis is based on a grid of single star models, one possible interpretation of this finding is as follows.
If a star has evolved as a single star, it may show a good agreement with the stellar models, leading to a high $\pmax$ value. On the contrary, if a star has experienced a non-canonical evolution (e.g., binary interaction), it likely evolved in a way that is not predicted by the single star models, leading to a low $\pmax$ value. Therefore, to find one group of stars at high, and another at low $\pmax$ might actually be expected.

That the considered split of the sample is meaningful is emphasized by the main panel of \Fig{fig_pvsini}, which shows the $\pmax$-values of our sample stars against their apparent rotation rate. It turns out that most of the stars in the low $\pmax$ group are slow rotators. While for most of them, only projected rotational velocities are known, 24 out of the 29 stars in this group show a $v\sin i$ of less than $50\,$km/s. We consider these as slow rotators.  
Equatorial velocities are derived from measured rotation periods for seven stars in our sample. All of them are slow rotators except for Star\,57 with $\vsini=62\kms$. One of them (Star\,1) belongs to the high $\pmax$ group because this star has the highest surface boron abundance amongst our stars, which is even compatible with the initial boron abundance of our models. The remaining six stars belong to the low $\pmax$ group since their boron depletion is too strong to be compatible with their equatorial rotational velocities.

The distribution of $\vsini$ in the lower right panel of \Fig{fig_pvsini} reveals that 60\% of slow rotators belong to the low $\pmax$ group. This increases by up to 10\% when we include the possible contribution from the stars with an upper limit on the boron abundance. The bimodality in rotational velocity distribution of early B-type stars from the VLT-FLAMES Tarantula Survey \citep{Dufton2013} brought interest to the origin of the slowly rotating B-type stars \citep{Bastian2020,Wang2022}. Our results support the idea of a non-canonical origin of the slow rotators, which were suggested to originate from binary mergers \citep{Schneider2019,Wang2022} or magnetic braking \citep{Meynet2011,Keszthelyi2022}.

Stars\,64, 82, 88 are noticeable outliers in the low $\pmax$ group since they have high $\vsini$ of $\gtrsim 90 \kms$. These are highly evolved fast rotators with high luminosity, which are expected to spin down due to angular momentum loss accompanying wind mass loss. These stars show low $\pmax$ not because of too much boron depletion for their age and rotation, but because of their rotation being too fast compared to our highly evolved models at the highest luminosity regime (see \App{app_out}).

An unambiguous nitrogen enhancement ($\logNCNCi>0.3$ and $\logNONOi>0.3$ in \Fig{fig_cno}) is present in nine stars. Six of them (Stars\,65, 66, 85, 86, 88, 90)  belong to the low $\pmax$ group, while three (Stars\,80, 81, 89) of them belong to the high $\pmax$ group. These three stars are consistent with having undergone canonical stellar evolution wherein rotational mixing has enhanced their surface nitrogen abundances, even though we cannot conclude this definitely since they only have an upper limit on $\pmax$.

\subsection{Rotational mixing efficiency}
\begin{figure} 
	\centering
	\includegraphics[width=\linewidth]{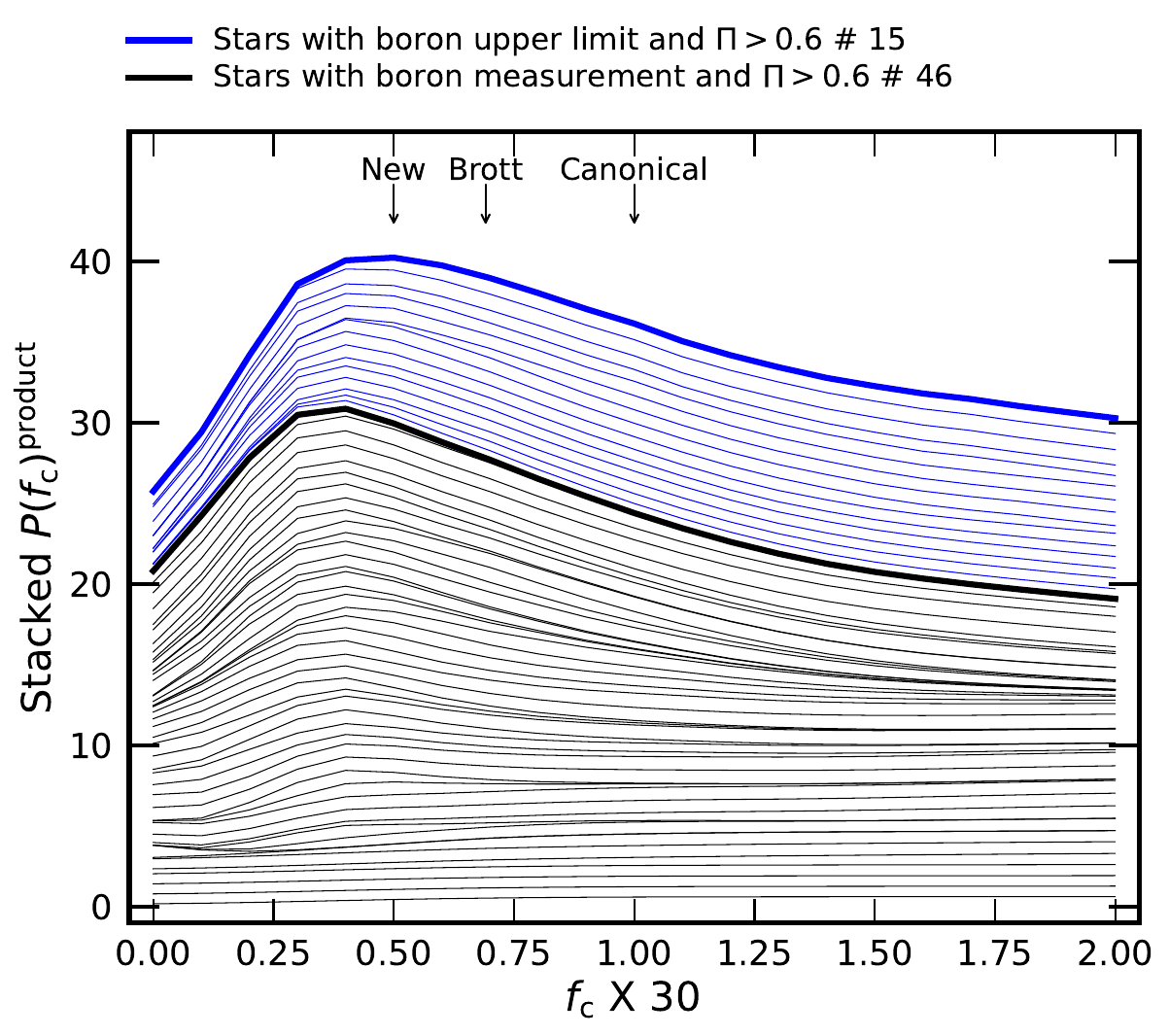}
	\caption{Stacked $\pfcproduct$ distributions for the stars belonging to high $\pmax$ group ($\pmax>0.6$). The distributions of stars with boron measurement (upper limit) are shown in the black (blue) lines. The arrows at the top indicate the canonical value from \citet{Chaboyer1992}, the preferred value by \citet{Brott2011a}, and the new preferred value by this work.}
	\label{fig_stacked}
\end{figure}

For the stars in the high $\pmax$ group, i.e., showing compatibility with stellar models, we present a stacked $\pfcproduct$ distribution in \Fig{fig_stacked}. The stacked distribution shows a peak at $\fc \times 30 = 0.5$. This implies that, if rotational mixing were the only mechanism to modify the surface boron abundance, its strength would be 50\% weaker ($\fc=0.017$) than the widely adopted value $\fc=0.033$. This agrees with the result of \citet{Proffitt2024} obtained from the comparison of the subset of our sample stars (the stars in NGC\,3293) with the same stellar models as us. This is also in qualitative agreement with the analysis of the nitrogen abundances in early B-type main sequence stars by \citet{Brott2011a}, who found 68\% of the conventional value, $\fc=0.023$, is preferred.

One of the key uncertainties that come into play in the analysis is the initial boron abundance, $\logBi$. Higher (lower) initial boron abundance translates into stronger (weaker) boron depletion, so higher (lower) $\fc$ would be preferred than obtained from our fiducial value. To first order, one can estimate the effect of varying $\logBi$ on the preferred $\fc$ through the dependency of $\logBBi$ on $\fc$ (the slopes of the lines in \Fig{fig_logX_fc}). A 12$\mso$ star with $\vini=200\kms$ (approximately the average initial mass and initial rotational velocity predicted by our Bayesian analysis; see Table~\ref{tab_obs}) in the middle of the main sequence evolution (approximately the average evolutionary stage; see \Fig{fig_HRD}) predicts $\log B/B_{\mathrm{i}}(\fc) \sim - \fc \times 30$. Thus, if $\logBi$ was 0.1\,dex higher, $\logBBi$ would be 0.1\,dex lower, and the stacked $\pfcproduct$ distribution would show 0.1 higher value of $\fc \times 30$ at the peak. The uncertainty of 0.1\,dex in $\logBi$ translates into the shift of the peak $\fc \times 30$ by 0.1, i.e., 40-60\% of the conventional rotational mixing efficiency. The preference for the weaker rotational mixing efficiency holds unless we underestimated $\logBi$ by more than $\sim 0.5\, \rm dex$; if the real $\logBi$ was $\sim 0.5\, \rm dex$ higher, the conventional value of $\fc \times 30 = 1$ would be preferred. This is unlikely since the stars with the highest boron abundance in the solar neighborhood and NGC\,3293, Stars\,1 and 10, show $\logBBi \approx 0$ (see Table~\ref{tab_obs}).

\subsection{Stars in NGC\,3293}
\citet{Proffitt2024} investigated 18\,stars with boron abundance data in the young NGC\,3293, and compared them to our single star models which use the standard value for the rotational mixing efficiency (cf., Sect.\,\ref{sec_mod}). They found that Stars\,14, 20, 39, 45, 47, 51, which belong to their low-luminosity group ($3.2 < \logL [\lso] < 3.8$), follow the predictions of rotating single star models well. Our analysis identifies these stars to be highly compatible with our single star models based on their high $\pmax$ value. Also, the stars in their mid-luminosity group ($3.8 < \logL [\lso] < 4.5$) and three stars in their high luminosity group ($\logL [\lso] > 4.5$) belong to our high $\pmax$ group. Notably, the result for Star\,10, which belongs to their mid-luminosity group, is uncertain due to its line profile variations. Only the three rather rapidly rotating stars in the high-luminosity group, Stars\,64, 82, 88, belong to our low $\pmax$ group, as discussed in Sect.\,\ref{sec_two}. 

Stars\,63 (3293-010), 76 (3293-016), 79 (3293-007), 87 (3293-003), 88 (3293-002) are the five rather slowly rotating stars in the sample of \citet{Proffitt2024}, which appeared to have a poor fit to their synthetic single star population with an age of 12\,Myr in the Hunter diagram for boron (their fig.\,7). In our analysis, one of these five stars (Star\,76) ends up with a high $\pmax$ of 0.96, three with $\pmax$-values near 0.7 but still in our high-$\pmax$ group, and only one (Star\,88) obtained a low $\pmax$.

The reason for the seemingly better fit of these sources to the single star models according to our analysis is that we have opened up two parameters which were fixed in \citet{Proffitt2024}, namely the rotational mixing efficiency and the stellar age. E.g., Star\,63 may be compatible with single star models if a larger age (cf., fig.\,7 of Proffitt et al. for 15\,Myr) and/or a mixing efficiency below the standard value is adopted. Furthermore, according to our Bayesian analysis, Stars\,76, 79, and\,87 can be more or less reproduced by our single star models, but only if --- despite their low $\vsini$ values --- fast rotation and a low inclination are assumed. While for one source, this is always possible, it appears unlikely for all of them, which, however, is not accounted for in the Bayesian analysis.

\subsection{Stars in binaries}
Out of 90 stars in our sample, 27 show indications of binarity. Of these, 19 (70\%) belong to the high probability group ($\Pi > 0.6$), with an average of $\Pi = 0.84$. These comply well with the expectation that the majority of the binaries are pre-interaction systems, where the primaries are the ``best single stars'' \citep{Mink2011}, in the sense that we can assume more safely for their primaries that no binary interaction took place than for current single stars.  

For 19 sources, orbital period and/or eccentricity estimates are available (see Table~\ref{tab_obs}). However, this data is not homogeneously derived, and often obtained many decades ago, such that we refrain from assessing the corresponding uncertainties. Here, we use the orbital parameters to estimate the Roche lobe filling factor of our corresponding sample stars, which can indicate whether tidal interaction or even mass transfer could have played a role. From Kepler's third law and the fit formula for the Roche lobe radius from \citet{Eggleton1983}, the Roche lobe radius of a star at periastron can be obtained as
\begin{equation}
    R_\mathrm{RL} = \left( \frac{P_\mathrm{orb}^2 G M (1+1/q)}{4 \pi^2} \right)^{1/3} \frac{0.49 q^{2/3}}{0.6 q^{2/3} + \ln(1+q^{1/3})} (1-e),
\end{equation}
where $M$ is the mass of the star, $q$ is the mass of the star over that of the companion, and $e$ is the orbital eccentricity. 
For given $P_\mathrm{orb}$, $e$, and primary mass, $R_\mathrm{RL}$
varies by less than 10\% as function of $q$. We adopt its minimum value, which is obtained near $q = 0.2$, to estimate the maximum Roche-lobe filling factor $(R/R_\mathrm{RL})_\mathrm{max}$ which we list for the 19 evident binaries in Tab.\,\ref{tab_obs}. The eccentricity is not known for two short-period systems, for which we assume a circular orbit. The radii of stars are calculated from the Stefan-Boltzmann law based on their luminosity and effective temperature. 

The Roche-lobe filling factors of all but one of the 19 binaries with known orbital periods are well below one, indicating that these systems are currently not affected by mass transfer. One object, Star\,36, has a formal Roche-lobe filling factor of 1.8. This star belongs to the $\gamma^1$ Velorum binary system, which comprises a quadruple star system along with the $\gamma^2$ Velorum binary system (Wolf-Rayet+blue supergiant). This star might have undergone a complex interaction with its companions, or its observation might have been influenced by them. 

Stars\, 4, 36, 41, and\,70 have orbital periods in the range 1.5 to 4\,d, and Rolche-lobe filling factor equal or above 0.6. These stars are in relatively close binaries in that they will likely experience Roche lobe overflow as they expand more. They all belong to the high $\pmax$ group. While Stars\,41 and\,70 are near ZAMS, Stars\,4 and\,36 are somewhat evolved (see \Fig{fig_HRD}) and do not show too much boron depletion. As these stars correspond well to the single star models, they do not provide evidence for significant tidal mixing.  

From the 19 binaries with measured periods, 13 have $\pmax\simgr 0.65$. The remaining 6 binary systems show a low probability to fit to our single star models, with an average of $\pmax\sim 0.19$. Of these, three (Stars\,24, 44, and 48) have rather large orbital periods, which does not exclude that these systems were previously in a triple configuration and the primary star has merged with a close companion. The other three, however, have orbital periods in the range 7 to 12\,d. Unless the orbital period has drastically decreased, the merger scenario seems unlikely for them.  

Nevertheless, the evidence that these three stars (Stars\,40, 42, 85) underwent non-canonical evolution is strong. They combine two attributes which are mutually exclusive in our single star models. They are boron depleted, by factors of 2.5 to 20, and they show almost no rotation. We conclude the latter from their extremely low $\vsini$ values, 10, 15, and 32$\kms$. One of them, Star\,42, also known as 16\,Lac, is an eclipsing binary with an orbital period of $12\,$d, which excludes that we observe the system pole-on.
Stars\,40 and\,85 are also magnetic (see below).
We return to these sources in the next section.


\begin{figure*} 
	\centering
	\includegraphics[width=\linewidth]{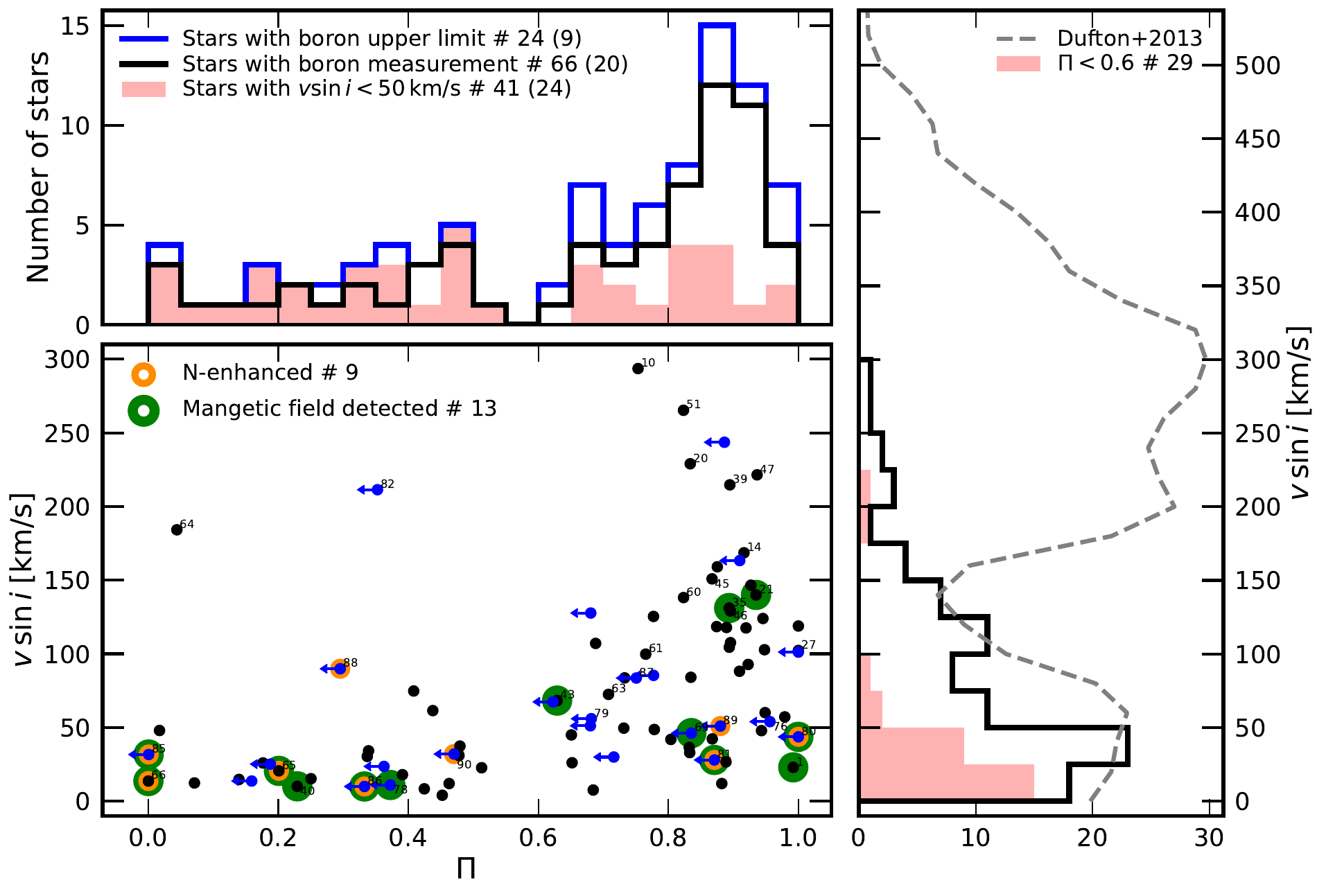}
	\caption{\textit{Lower left}: the stars with boron upper limit (blue circles with leftward arrows) and the stars with boron measurement (black circles) in the $\vsini$ versus $\pmax$ diagram. Stars with a clear sign of nitrogen enhancement are marked with a yellow edge, and stars with magnetic field detection are marked with a green edge (see text). Stellar numbers are shown if a star is nitrogen enhanced, magnetic field detected, or a member of NGC\,3293. Four stars (Star\,15, 27, 70, 80) showing $\pmax$ or the upper limit on $\pmax$ larger than 1.0 are placed at $\pmax=1.0$ for better visibility. \textit{Upper left}: the distribution of $\pmax$ of stars with boron upper limit (blue) and boron measurement (black). The results are stacked. The distribution corresponding to slow rotators ($\vsini < 50 \kms$) is shown in red. The numbers corresponding to each group are presented in the legend, and the numbers in parentheses are the number of stars with $\pmax<0.6$ in each group. \textit{Lower right}: the distribution of $\vsini$ of stars. The distribution corresponding to the low $\pmax$ group ($\pmax<0.6$) is shown in red. The number of corresponding stars is shown in the legend.}
	\label{fig_pvsini}
\end{figure*} 

\subsection{Magnetic stars}
Of the 90 stars in our sample, 44 have spectropolarimetric data. Of these, magnetic field measurements have been performed for 27 stars. In 13 of these stars, magnetic fields have been detected, although with various levels of significance. No field has been found in the other 14 stars. In Table~\ref{tab_obs}, we highlight those five stars for which a magnetic field has been detected through more than one measurement; for the other eight stars with field detections, also non-detections are published. Below, we do not distinguish between the significance levels of the field detections in these 13 stars. Unfortunately, the magnetic field analysis for some stars is based on very sparse spectropolarimetric data, and the data for most stars are not obtained from one homogenous survey, which hinders a statistically sound analysis. Also, not all the 13 magnetic stars have been analyzed via the dipole oblique rotator model \citep[see][for the case of Star\,65]{Briquet2016}, sich that the dipolar field strength is not available for all of them. We therefore present longitudinal field strengths in Table~\ref{tab_obs}.

Six of the 13 magnetic stars belong to the low $\pmax$ group (Stars\,40, 65, 66, 78, 85, 86). Of the remaining seven, three have a high $\pmax$ value (Stars\,1, 21, 35), Star\,43 is at the borderline between high and low ($\pmax =0.63$), and for the other three (stars 69, 80, 81) we only have upper limits on $\pmax$ owing to the upper limits of their boron abundance. 
This leads to incidence fractions of magnetic stars in the range 7$\dots$11\% in the high $\pmax$ group, and 21$\dots$28\% for low $\pmax$. While the total magnetic fraction of $\sim 14$\% is larger than the general incidence fraction of magnetic stars in OB dwarfs of $\sim 7$\% \citep{Wade2014, Fossati2015}, this may just reflect our sample bias, which favors slow rotators. We do not find a significant difference in the measured field strengths between the two groups. Notably, the boron abundances in the three magnetic stars with the unambiguously high $\pmax$-values are compatible with no boron depletion, within the error bars. 

We find that from the nine stars in our sample that show a nitrogen surface enhancement, six stars are magnetic. While this may suggest a correlation between nitrogen enhancement and the presence of a magnetic field, the remaining seven magnetic stars are not significantly nitrogen enhanced. This result resembles the finding of \cite{Morel2008} that nitrogen enriched stars have a higher chance to be magnetic than nitrogen-normal ones,
but there is no one-to-one correlation between both attributes.

Three of the magnetic stars are members of binary systems, with orbital periods of 19\,d (Star\,21), 7\,d (Star\,40), and 11\,d (Star\,85), with the latter two showing low $\pmax$ values.



\section{Evolutionary scenarios\label{sec_dis}}
In this section, we first discuss which potentially important processes are neglected in our single star models. With this in mind, we then draw tentative conclusions on the evolutionary status of the fraction of our sample stars that do not correspond to our single star models.

\subsection{Neglected physical processes}
In principle, the boron surface abundance and the surface rotation rate of some of our sample stars may be influenced by physical processes intrinsic to single stars which are not considered in our evolutionary models.  In particular, non-radial g-mode pulsations as found in the slowly-pulsating B-type (SPB) stars are thought to be able to induce a significant radial transport of angular momentum \citep[e.g.,][]{Schatzman1993,Kumar1997,Townsend2018}
and of chemical species
\citep{Montalban1994,Talon2005,Rogers2017}

In our sample, Star\,65 (=HD\,3360) is the only identified SPB star \citep{Briquet2007}, which shows a clear nitrogen enhancement. \citet{Briquet2016} found that this star exhibits solid body rotation, and that it has a polar magnetic field of $100-150 \, \rm G$. They argued that rotational mixing can not be the underlying mechanism for its nitrogen enhancement, 
which is in line with our result, since this star shows a value of $\Pi$ of 0.20. Since they found no indication of a companion, they argued against binary interaction as a possible explanation for the nitrogen enhancement (however, see Sect.\,\ref{S62}), and that chemical mixing induced by internal gravity waves is the most plausible mechanism.

While we are not aware of the SPB nature of our targets except for the aforementioned star, we can not exclude having some more such stars in our sample. 
However, the vast majority of SPB stars are found at luminosities below $\sim 1000\,\lso$
\citep{Waelkens1998}, such that most likely they do not affect our results too much. 
The luminosity range of our sample has a large overlap with the $\beta\,$Cephei-pulsators, and 
indeed many stars in our sample belong to that class: Stars\,1, 24, 27, 31, 40, 42, 46, 57, 63, 66, 80, 85, 86 \citep{Morel2006,Morel2008,Proffitt2016,Southworth2022}.
However, the variability in these stars is dominated by p-modes, in which transport processes are less likely to occur.

Furthermore, gravitational settling and radiative levitation could affect the boron surface abundance of some of our stars. Indeed,
\citet{Richard2001} find that in models of metal-poor stars of up to 1.2$\mso$, the boron abundance may increse or decrease by up to one order of magnitude due to these effects. It is well established that in chemically peculiar (CP) A stars, settling and levitation can even have much stronger effects, in particular in the slowest rotators \citep{Smith1996}. However, as these processes operate on very long timescales, and the vast majority of the CP stars are again found below $1000\lso$, we assume here that they do not affect our results.

Also, the presence of strong magnetic fields is known to be able to induce chemical peculiarity, at least partly in combination with settling and levitation. This is so for the classical Ap stars. However, it may also hold for the so-called He-weak and He-strong stars \citep{Smith1996}, and could potentially overlap with the nitrogen enhancement found in many of the magnetic stars in our sample. This could affect a significant fraction of our sample stars because we do not know the magnetic status of most of them. \citet{Meynet2011, Keszthelyi2022} investigate magnetic stellar evolution models, and they find that the interplay between magnetic braking and internal mixing in stars with fossil magnetic fields can reproduce nitrogen-enhanced slow rotators.

With these caveats in mind, we discuss the evolutionary implications of our findings in the following, assuming that rotationally induced mixing is the only process that affects the surface abundances of single stars.

\subsection{Evolutionary considerations}
\label{S62}
Our analysis showed that $\sim 70$\% of our sample stars are in good agreement with our single star models. While this involved the adjustment of the rotational mixing efficiency parameter, this adjustment was moderate ($f_{\rm c} \times 30=1 ~ \rightarrow f_{\rm c} \times 30 \simeq 0.5$) and in line with a previously suggested adjustment to $f_{\rm c} \times 30\simeq 0.7$ based  on nitrogen surface abundances (\citet{Brott2011a}. This result supports the paradigm that ordinary single star models are suitable to describe the majority of the massive main sequence stars, and that their surface boron abundance is determined by rotational mixing.

 On the other hand, about one-third of our sample stars are clearly not compatible with our single star models. The fraction of stars in our sample with a peculiar evolution could be higher, because in our high-$\pmax$ group (61 stars) we have stars with peculiarities, namely 16 slow rotators ($\vsini < 50\kms$) and seven magnetic stars. 

With the evidence of a high binary fraction of massive stars, it appears natural to pursue the path to identify some or all of our peculiar stars as binary interaction products. When leaving out tidal interaction, for which there is little evidence, we need to focus on mass transfer effects. However, the donor stars in mass transfer systems lose all their envelope, and live only a very short amount of time in the surface temperature range compatible with main sequence stars \citep[cf.,][]{Schootemeijer2018,Schurmann2022}. Examples have been found in recent years in the context of searching for X-ray quiet black-hole binaries \citep[e.g.,][]{Shenar2020,Lennon2021,Badry2022}, but these objects are too rare to affect our conclusions.

The mass gainers in mass transfer systems, on the other hand, are expected to be spun-up to rapid rotation by the accretion process \citep{Packet1981,Petrovic2005}, presumably feeding in the large population of Be stars and Be/X-ray binaries \citep{Heuvel1987,Pols1991,Hastings2021}. 
Notably, a spin-down of such fast rotators is expected only in stars with high stellar wind mass loss rates, which could only have occurred in the few most massive stars of our sample. Otherwise, the stars might rather be expected to spin up rather than spin down due to their internal evolution \citep{Ekstrom2008,Hastings2020b}. 
 
In our sample, three stars (Star\,33, 56, 68) are identified as Be stars \citep{Merrill1925,Mendoza1958,Cote1993}. Interestingly, none of these stars exhibit strong boron depletion. They are fairly evolved stars with fractional main sequence ages (the current stellar age over the respective terminal main sequence age) of $\sim 0.8$, $\sim 0.8$, $\sim 0.3$, respectively. If they are rotating near-critically, they are expected to show more boron depletion than reported (see the depletion factors for $400 \kms$ models in \Fig{fig_logX_fc}). This might support the idea that Be stars got spun up via mass and angular momentum transfer from companions \citep[e.g.,][]{Bodensteiner2020,Langer2020}. The accretion of only a few percent of the stellar mass is sufficient to spin up a star to near critical \citep{Packet1981}, so mass gainer may become a Be star via highly non-conservative mass transfer. In 
corresponding detailed binary evolution models \citep{Langer2020,Wang2023}, many mass gainers accrete only material from the outermost part of the donor, which is not contaminated with nuclear-processed material. If the accretor was slowly rotating before the mass transfer event such that it retained boron at its surface, the resultant spun-up Be star can still contain boron in its outer layers \citep[see][for a Be stars exhibiting no nitrogen enhancement]{Lennon2005,Dufton2024}. However, such fast rotators will soon deplete boron at the surface due to strong rotational mixing. This channel will be further investigated in our future work involving binary evolution models.

Only systems that undergo mass transfer during core hydrogen burning may avoid the near-complete stripping of the donor and the spin-up of the mass gainer. However, in these so-called Algol-systems, nuclear-timescale mass transfer prevails \citep{Sen2022} and gives prominent features in the spectra of such sources \citep[e.g.,][]{Howarth2015,Mahy2020}. Thus, we do not expect any Algol binary in our sample. 

This leaves only one type of binary interaction product which can be expected to occur in a significant number in our sample, namely merger stars. Binary merger products could also be spun up to rapid rotation, as the pre-merger orbital angular momentum exceeds the value that the merger product could host. However, recent hydrodynamic calculations show that such a spin-up may occur initially, but that mass loss in the puffed-up merger product drains so much angular momentum that the relaxed merger product is in fact slowly rotating \citep{Schneider2019}. Furthermore, the population synthesis calculations of \citet{Mink2014} show that  $8^{+9}_{-4}$\% of the main sequence stars are expected to be merger products. The order of magnitude of this number could fit the fraction of stars in our low-$\pmax$ group, given the sample bias towards slow rotators.

Could we understand the low $\pmax$-group, and possibly the unconventional stars (slow rotators and/or magnetic stars) in the high $\pmax$-group as mergers? We think yes, but with significant constraints. First, it would be necessary to assume that most binary mergers produce slow rotators, while we can argue for that so far only based on one merger model. If so, it is useful to consider other expected properties of merger stars.

\citet{Schneider2019} find that a strong magnetic field is produced in their model merger product. This supports our attempt to understand the unconventional stars as merger products because they have a larger fraction of magnetic stars than the stars in the high $\pmax$-group. However, it would also imply that either not every merger product is magnetic, or that the magnetic field decays on the nuclear timescale or faster \citep[cf.][]{Fossati2016}, as for seven stars in our low $\pmax$-group, no magnetic field was detected at the $3-5 \, \sigma$ level.

It would further imply that binary mergers would undergo various degrees of envelope mixing. While the majority of stars in the low $\pmax$-group are boron-depleted slow rotators, the degree of depletion can be rather low 
(e.g., Stars\, 12, 18, 22). 
Also, unless one wants to consider alternative mechanisms of magnetic field generation, the three nearly boron-undepleted magnetic stars (Stars 1, 21, 35) imply very little mixing in their merger process. On the other side, we already noted that the fraction of nitrogen-enriched stars in our magnetic sub-sample is high, and on a wider scope, the nitrogen-enhancements in slowly-rotating B\,main sequence stars can be very high \citep{Morel2006,Hunter2008b}. 

The strongest constraints on the merger scenario emerge from the three close binaries with primary stars (Stars 40, 42, 85) which are incompatible with our single star models, of which Stars 40 and 85 are magnetic. At first glance, their orbital periods (7\,d, 12\,d, and 11\,d, respectively) appear too short to allow for a scenario in which the primary star is a merger product. However, a solution could be a triple star scenario, in which the inner binary merged while the outer companion was still farther away. The required orbital shrinkage could be induced by circum-binary material, thereby linking this phase to the pre-main sequence evolution of the stars \citep{Bate2002,Tokovinin2020}.

While speculative here, pre-main sequence mergers have also been suggested by \citet{Wang2022} to explain the blue main sequence stars in young open star clusters \citep{Milone2018}, and are predicted to occur in as many as $30$\% of the massive pre-main sequence binaries \citep{Tokovinin2020}. Notably, while avoiding any boron depletion in a merger of two main sequence stars may be difficult, a merger during the pre-main sequence evolution could happen at a time when the temperature is too low for boron destruction even in the centers of both stars, such that no boron depletion would be expected.  

Thus it appears that the merger scenario may explain the properties of our unconventional stars, if mergers produce slow rotators, if they sometimes produce magnetic fields or fields that decay on a long timescale, if they produce a variety of mixing degrees, and if they occur partly already during the pre-main sequence evolution. This involves many if-conditions, which would need to be explored by corresponding theoretical experiments in the future.

\section{Conclusions}\label{sec_con}

 We have computed a new grid of rotating single star evolution models with MESA which includes the physics of rotational mixing as well as a nuclear network that follows the abundance changes of the relevant isotopes of lithium, beryllium, and boron in detail. We have extended this grid semi-analytically to represent a large range of rotational mixing efficiencies. We also compiled the observationally derived physical parameters for all the Galactic early B-type stars for which their boron surface abundance is either measured or constrained.  Furthermore, we tested the compatibility of the stellar models with the observed stars quantitatively by constructing a Bayesian framework similar to the BONNSAI method \citep{Schneider2014}, in which we simultaneously considered all the observed stellar properties which are predicted by the models.

We found that the stars in our sample can be divided into two groups. The first group, in which the probability of the models representing the stars is on average about 80\%, comprises about 68\% of our sample. These stars are well explained by our single star models, and as such support the concept of rotationally induced mixing.  At the same time, the overall comparison between the models and the stars reveals that a 50\% weaker rotational mixing efficiency ($\fc=0.017$) than the conventional value ($\fc=0.033$) is favored. This has implications for the predictions of rotating single star models. 

For the second group of stars in our sample (32\%), we find that the probability that our single star models can explain them is, on average, below 29\%. In many of them, boron is depleted too much given their ages and rotation, implying that rotational mixing is not the main mechanism to deplete boron. These boron-depleted slow rotators may be the counterparts of nitrogen-enhanced slow rotators in which non-canonical single star evolution is expected to have occurred \citep{Morel2006,Hunter2008b,Hunter2009,Rivero2012,Grin2017}. That most of our sample stars show mild or no nitrogen enhancement demonstrates that boron is a more sensitive indicator for rotational mixing than nitrogen. 

About half of the slow rotators ($< 50 \kms$) in our sample belong to the second group. This implies that the origin of a significant fraction of slow rotators can be different from that of fast rotators. This idea is strengthened by the high incidence ($\gtrsim 21 \%$) of magnetic stars in this group. The presence of some of them in close binary systems may give the strongest constraints on their past evolution. We speculate that mergers due to tidally induced orbit decay in pre-main sequence binaries may play a role in explaining these stars. In any case, this or any other model for their formation needs to be able to account for the large spread in the observed boron surface abundances, as well as for the presence of large scale magnetic fields, in these many of these slowly rotating stars.


\begin{acknowledgements}
HJ received financial support for this research from the International Max Planck Research School (IMPRS) for Astronomy and Astrophysics at the Universities of Bonn and Cologne. The authors gratefully acknowledge the granted access to the Bonna cluster hosted by the University of Bonn.
\end{acknowledgements}

\bibliographystyle{aa}
\bibliography{Reference_list}

\begin{appendix}
\section{STERN network}\label{app_com}
\begin{figure} 
	\centering
	\includegraphics[width=\linewidth]{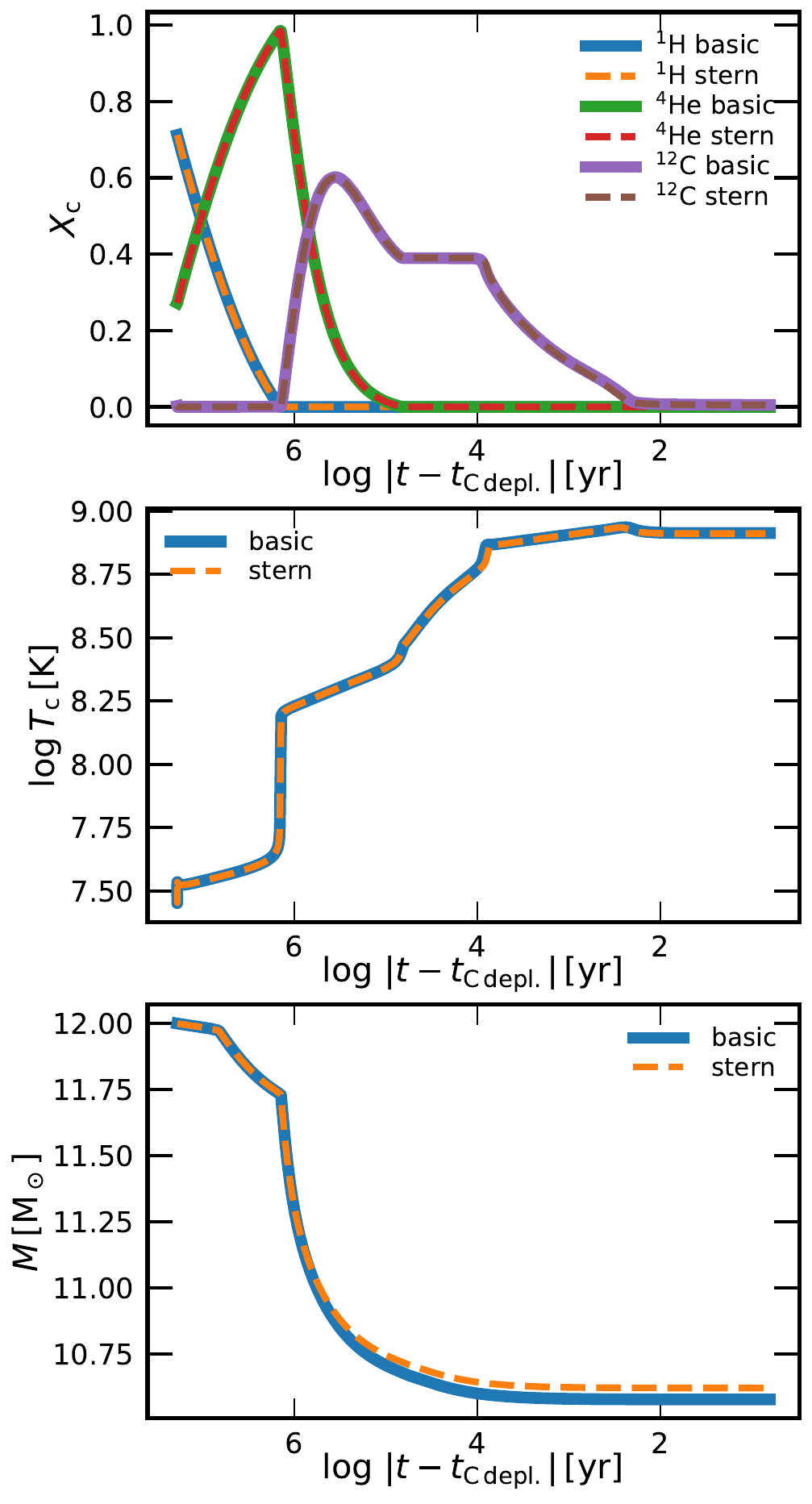}
	\caption{Evolution of physical quantities in 12$\mso$ single star models computed with \texttt{basic} network (solid lines) and STERN network (dashed lines). The evolution of central mass fractions of hydrogen, helium, and carbon (top), central temperature (middle), and total mass (bottom) as a function of remaining time until the core carbon depletion.}
	\label{fig_basic}
\end{figure}

\begin{figure} 
	\centering
	\includegraphics[width=\linewidth]{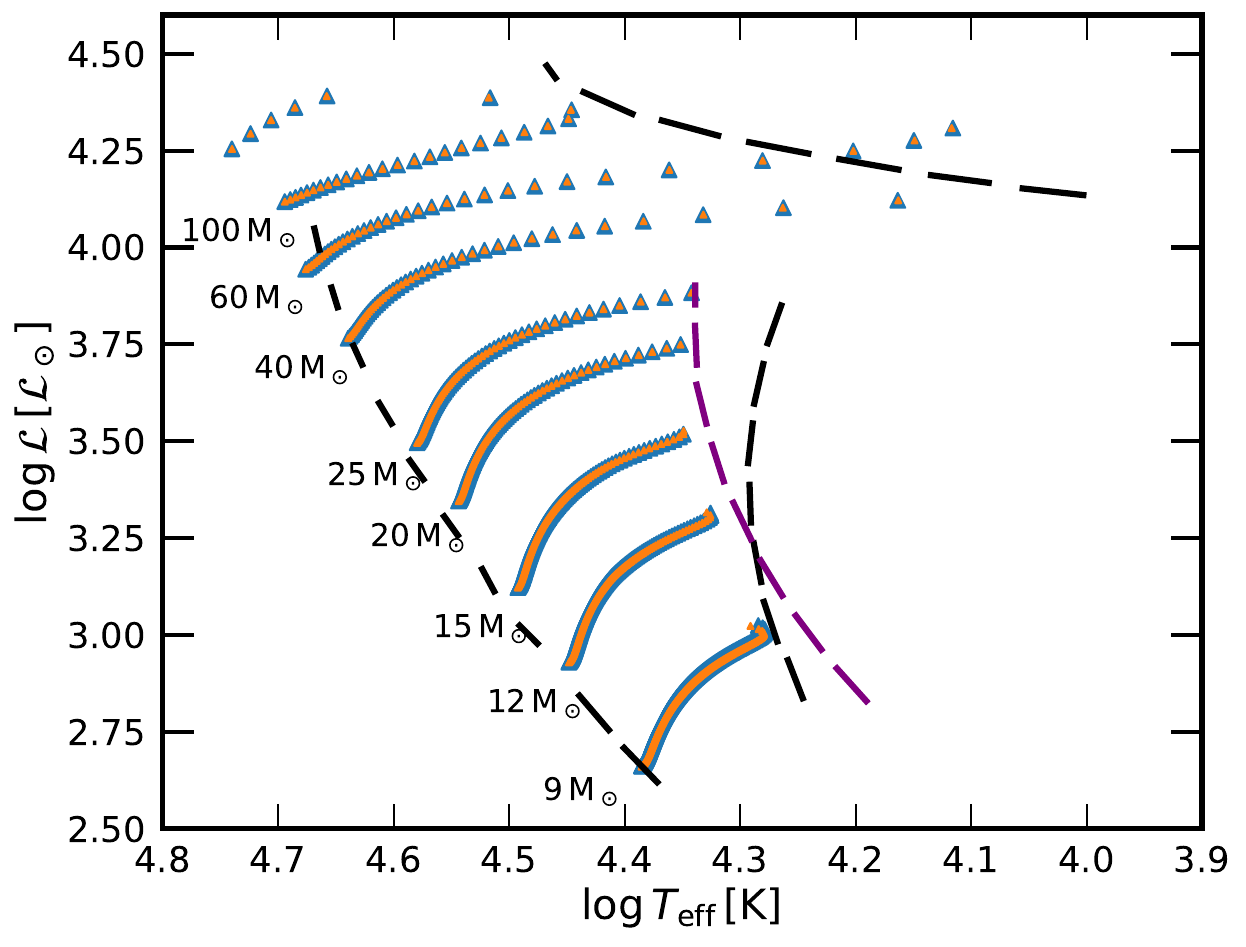}
	\caption{main sequence evolutionary tracks of single star models computed with \texttt{basic} network (blue triangles) and STERN network (yellow triangles) on the spectroscopic Hertzsprung-Russel diagram. Along each track, data points are separated by a time interval of 0.1$\Myr$. The black dashed lines show the boundaries of a region that is densely populated by Galactic stars, and the purple dashed line corresponds to the terminal age main sequence (terminal-age main sequence) line of the Galactic models of \citet{Brott2011a} (see fig.\,1 in \citet{Castro2014}).}
	\label{fig_basic2}
\end{figure}

The default nuclear network in MESA, \texttt{basic} network, contains reactions involving eight isotopes and was tested to follow hydrogen and helium burning phases with sufficient accuracy in mass fractions and energy generation \citep{Paxton2011}. STERN network, on the other hand, includes more detailed reactions (Table~\ref{tab_reac}) that were approximated in \texttt{basic}. To check if the STERN network performs well in the MESA calculation, we compute two single star models with \texttt{basic} network and STERN network. 
\Figure{fig_basic} compares the time-evolution of physical parameters of the two models, central abundances of hydrogen, helium, carbon, central temperature, and total mass until core carbon depletion, which show negligible differences. The respective lifetimes of these models differ by less than 1\%. \Figure{fig_basic2} compares the main sequence evolutionary tracks of the two models. Once again, these tracks do not display any discernible differences. 

\begin{table*}[]
\caption{List of reactions in STERN nuclear network and corresponding handles in MESA}
\label{tab_reac}
\resizebox{\linewidth}{!}{%
\begin{tabular}{lll|lll}
\hline \hline
No.\,& Reaction & MESA & No.\,& Reaction & MESA \\ \hline
1 & ${}^{1}\mathrm{H}\left(\mathrm{p},\beta^{+}\nu_{\mathrm{e}}\right){}^{2}\mathrm{H}(\mathrm{p},\gamma){}^{3}\mathrm{He}$ & rpp\_to\_he3 & 41 & ${ }^{17} \rm O (\alpha, n) { }^{20} \mathrm{Ne}$ & r\_o17\_an\_ne20 \\
2 & ${}^{3}\mathrm{He}\left({}^{3}\mathrm{He},2\mathrm{p}\right){}^{4}\mathrm{He}$ & r\_he3\_he3\_to\_h1\_h1\_he4 & 42 & ${ }^{18} \mathrm{O} (p, \gamma){ }^{19} \mathrm{F}$ & r\_o18\_pg\_f19 \\
3 & ${}^{3}\mathrm{He}(\alpha,\gamma){}^{7}\mathrm{Be}\left(\beta^{+}\nu_{\mathrm{e}}\right){}^{7}\mathrm{Li}$ & r\_he3\_ag\_be7, r\_be7\_wk\_li7 & 43 & ${ }^{18} \rm O (p, \alpha)  { }^{15}  \mathrm{N}$ & r\_o18\_pa\_n15 \\
4 & ${}^{4}\mathrm{He}(\alpha,\gamma){}^{8}\mathrm{Be}(\alpha,\gamma){}^{12}\mathrm{C}$ & r\_he4\_he4\_he4\_to\_c12 & 44 & ${ }^{18} \rm O (\alpha, \gamma){ }^{22} \mathrm{Ne}$ & r\_o18\_ag\_ne22 \\
5 & ${}^{6}\mathrm{Li}(\mathrm{p},\alpha){}^{3}\mathrm{He}$ & r\_li6\_pa\_he3 & 45 & ${ }^{19} \mathrm{F} (p, \gamma){ }^{20} \mathrm{Ne}$ & r\_f19\_pg\_ne20 \\
6 & ${}^{7}\mathrm{Li}(\mathrm{p},\alpha){}^{4}\mathrm{He}$ & r\_li7\_pa\_he4 & 46 & ${ }^{19} \mathrm{F} (p, \alpha){ }^{16} \mathrm{O}$ & r\_f19\_pa\_o16 \\
7 & ${}^{7}\mathrm{Li}(\alpha,\gamma){}^{11}\mathrm{B}$ & r\_li7\_ag\_b11 & 47 & ${ }^{20} \mathrm{Ne} \left(\gamma,\alpha\right)^{16} \mathrm{O}$ & r\_ne20\_ga\_o16 \\
8 & ${}^{7}\mathrm{Be}\left(\beta^{+}\nu_{\mathrm{e}}\right){}^{7}\mathrm{Li}$ & r\_be7\_wk\_li7 & 48 & ${ }^{20} \mathrm{Ne} \left(n,\gamma\right)^{21} \mathrm{Ne}$ & r\_ne20\_ng\_ne21 \\
9 & ${}^{7}\mathrm{Be}\left({}^{3}\mathrm{He},2\mathrm{p}\right){}^{4}\mathrm{He}+{}^{4}\mathrm{He}$ & r\_he3\_be7\_to\_h1\_h1\_he4\_he4 & 49 & ${ }^{20} \mathrm{Ne}(p, \gamma){ }^{21} \mathrm{Na}\left(\beta^{+} \nu_{e}\right)^{21} \mathrm{Ne}$ & r\_ne20\_pg\_na21, r\_na21\_wk\_ne21 \\
10 & ${}^{7}\mathrm{Be}(\mathrm{p},\gamma){}^{8}\mathrm{B}$ & r\_be7\_pg\_b8 & 50 & ${ }^{20} \mathrm{Ne}(\alpha, \gamma){ }^{24} \mathrm{Mg}$ & r\_ne20\_ag\_mg24 \\
11 & ${}^{7}\mathrm{Be}(\alpha,\gamma){}^{11}\mathrm{C}$ & r\_be7\_ag\_c11 & 51 & ${ }^{21} \mathrm{Ne}(p, \gamma){ }^{22} \mathrm{Na}\left(\beta^{+} \nu_{e}\right)^{22} \mathrm{Ne}$ & r\_ne21\_pg\_na22, r\_na21\_wk\_ne21 \\
12 & ${}^{9}\mathrm{Be}\left(p,{}^{2}\mathrm{H}\right){}^{4}\mathrm{He}+{}^{4}\mathrm{He}$ & r\_h1\_be9\_to\_h2\_he4\_he4 & 52 & ${ }^{21} \mathrm{Ne} (n, \gamma){ }^{22} \mathrm{Ne}$ & r\_ne21\_ng\_ne22 \\
13 & ${}^{9}\mathrm{Be}(\mathrm{p},\alpha){}^{6}\mathrm{Li}$ & r\_be9\_pa\_li6 & 53 & ${ }^{21} \mathrm{Ne} (\alpha, \gamma){ }^{25} \mathrm{Mg}$ & r\_ne21\_ag\_mg25 \\
14 & ${}^{8}\mathrm{B}\left(\beta^{+}\nu_{\mathrm{e}}\right){}^{4}\mathrm{He}+{}^{4}\mathrm{He}$ & r\_b8\_wk\_he4\_he4 & 54 & ${ }^{21} \mathrm{Ne}(\alpha, n){ }^{24} \mathrm{Mg}$ & r\_ne21\_an\_mg24 \\
15 & ${}^{8}\mathrm{B}(\alpha,\mathrm{p}){}^{11}\mathrm{C}$ & r\_b8\_ap\_c11 & 55 & ${ }^{22} \mathrm{Ne} (p, \gamma){ }^{23} \mathrm{Na}$ & r\_ne22\_pg\_na23 \\
16 & ${}^{8}\mathrm{B}(\gamma,\mathrm{p}){}^{7}\mathrm{Be}$ & r\_b8\_gp\_be7 & 56 & ${ }^{22} \mathrm{Ne} (p, \alpha)^{19} \mathrm{F}$ & r\_ne22\_pa\_f19 \\
17 & ${}^{10}\mathrm{B}(\mathrm{p},\alpha){}^{7}\mathrm{Be}$ & r\_b10\_pa\_be7 & 57 & ${ }^{22} \mathrm{Ne} (\alpha, \gamma){ }^{26} \mathrm{Mg}$ & r\_ne22\_ag\_mg26 \\
18 & ${}^{11}\mathrm{B}(p,\gamma){}^{12}\mathrm{C}$ & r\_b11\_pg\_c12 & 58 & ${ }^{22} \mathrm{Ne} (\alpha, n){ }^{25} \mathrm{Mg}$ & r\_ne22\_an\_mg25 \\
19 & ${}^{11}\mathrm{B}(\mathrm{p},\alpha){}^{4}\mathrm{He}+{}^{4}\mathrm{He}$ & r\_h1\_b11\_to\_he4\_he4\_he4 & 59 & ${ }^{23} \mathrm{Na} (p, \gamma){ }^{24} \mathrm{Mg}$ & r\_na23\_pg\_mg24 \\
20 & ${}^{11}\mathrm{C}\left(\beta^{+}\nu_{\mathrm{e}}\right){}^{11}\mathrm{B}$ & r\_c11\_wk\_b11 & 60 & ${ }^{23} \mathrm{Na} (p, \alpha){ }^{20} \mathrm{Ne}$ & r\_na23\_pa\_ne20 \\
21 & ${}^{11}\mathrm{C}(p,\gamma){}^{12}\mathrm{N}\left(\beta^{+}\nu_{e}\right)^{12}\mathrm{C}$ & r\_c11\_pg\_n12, r\_n12\_wk\_c12 & 61 & ${ }^{24} \mathrm{Mg}(p, \gamma){ }^{25} \mathrm{Al}\left(\beta^{+}\nu_{e}\right)^{25} \mathrm{Mg}$ & r\_mg24\_pg\_al25, r\_al25\_wk\_mg25 \\
22 & ${}^{12}\mathrm{C}(\mathrm{p},\gamma){}^{13}\mathrm{N}\left(\beta^{+}\nu_{\mathrm{e}}\right)^{13}\mathrm{C}$ & rc12\_to\_c13 & 62 & ${ }^{24} \mathrm{Mg} (n, \gamma){ }^{25} \mathrm{Mg}$ & r\_mg24\_ng\_mg25 \\
23 & ${}^{12}\mathrm{C}(\alpha,\gamma)^{16}\mathrm{O}$ & r\_c12\_ag\_o16 & 63 & ${ }^{24} \mathrm{Mg}(\alpha, \gamma){ }^{28} \mathrm{Si}$ & r\_mg24\_ag\_si28 \\
24 & ${}^{12}\mathrm{C}\left({}^{12}\mathrm{C},\mathrm{p}\right){}^{23}\mathrm{Na}$ & r\_c12\_c12\_to\_h1\_na23 & 64 & ${ }^{25} \mathrm{Mg} (p, \gamma){ }^{26} \mathrm{Al}$ & r\_mg25\_pg\_al26-1 \\
25 & ${}^{12}\mathrm{C}\left({}^{12}\mathrm{C},\alpha\right){}^{20}\mathrm{Ne}$ & r\_c12\_c12\_to\_he4\_ne20 & 65 & ${ }^{25} \mathrm{Mg}(p, \gamma){ }^{26} \mathrm{Al}^{*}\left(\beta^{+} \nu_{e}\right)^{26} \mathrm{Mg}$ & r\_mg25\_pg\_al26-2, r\_al26-2\_wk\_mg26 \\
26 & ${}^{12}\mathrm{C}(\mathrm{n},\gamma)^{13}\mathrm{C}$ & r\_c12\_ng\_c13 & 66 & ${ }^{25} \mathrm{Mg} (n, \gamma){ }^{26} \mathrm{Mg}$ & r\_mg25\_ng\_mg26 \\
27 & ${}^{13}\mathrm{C}(\mathrm{p},\gamma){}^{14}\mathrm{N}$ & r\_c13\_pg\_n14 & 67 & ${ }^{25} \mathrm{Mg}(\alpha, \gamma){ }^{29} \mathrm{Si}$ & r\_mg25\_ag\_si29 \\
28 & ${}^{13}\mathrm{C}(\alpha,\mathrm{n}){}^{16}\mathrm{O}$ & r\_c13\_an\_o16 & 68 & ${ }^{25} \mathrm{Mg}(\alpha, n){ }^{28} \mathrm{Si}$ & r\_mg25\_an\_si28 \\
29 & ${}^{12}\mathrm{N}\left(\beta^{+}\nu_{\mathrm{e}}\right)^{12}\mathrm{C}$ & r\_n12\_wk\_c12 & 69 & ${ }^{26} \mathrm{Mg}(p, \gamma){ }^{27} \mathrm{Al}$ & r\_mg26\_pg\_al27 \\
30 & ${}^{14}\mathrm{N}(\mathrm{p},\gamma){}^{15}\mathrm{O}\left(\beta^{+}\nu_{\mathrm{e}}\right)^{15}\mathrm{N}$ & rn14\_to\_n15 & 70 & ${ }^{26} \mathrm{Mg}(\alpha, \gamma){ }^{30} \mathrm{Si}$ & r\_mg26\_ag\_si30 \\
31 & ${}^{14}\mathrm{N}(\alpha,\gamma){}^{18}\mathrm{F}\left(\beta^{+}\nu_{e}\right)^{18}\mathrm{O}$ & rn14ag\_to\_o18 & 71 & ${ }^{26} \mathrm{Mg}(\alpha, n){ }^{29} \mathrm{Si}$ & r\_mg26\_an\_si29 \\
32 & ${}^{15}\mathrm{N}(\mathrm{p},\alpha){}^{12}\mathrm{C}$ & r\_n15\_pa\_c12 & 72 & ${ }^{26} \mathrm{Al}\left(\beta^{+} \nu_{e}\right)^{26} \mathrm{Mg}$ & r\_al26-1\_wk\_mg26 \\
33 & ${}^{15}\mathrm{N}(p,\gamma){}^{16}\mathrm{O}$ & r\_n15\_pg\_o16 & 73 & ${ }^{26} \mathrm{Al} (p, \gamma){ }^{27} \mathrm{Si}\left(\beta^{+} \nu_{e}\right)^{27} \mathrm{Al}$ & r\_al26-1\_pg\_si27, r\_si27\_wk\_al27 \\
34 & ${}^{15}\mathrm{N}(\alpha,\gamma){}^{19}\mathrm{F}(\alpha,\gamma)^{23}\mathrm{Na}$ & r\_n15\_ag\_f19, r\_f19\_ag\_na23 & 74 & ${ }^{27} \mathrm{Al} (p, \gamma){ }^{28} \mathrm{Si}$ & r\_al27\_pg\_si28 \\
35 & ${}^{16}\mathrm{O}(\mathrm{p},\gamma){}^{17}\mathrm{F}\left(\beta^{+}\nu_{\mathrm{e}}\right)^{17}\mathrm{O}$ & ro16\_to\_o17 & 75 & ${ }^{27} \mathrm{Al} (p, \alpha){ }^{24} \mathrm{Mg}$ & r\_al27\_pa\_mg24 \\
36 & ${}^{16}\mathrm{O}(\alpha,\gamma){}^{20}\mathrm{Ne}$ & r\_o16\_ag\_ne20 & 76 & ${ }^{28} \mathrm{Si} (n, \gamma){ }^{29} \mathrm{Si}$ & r\_si28\_ng\_si29 \\
37 & ${}^{16}\mathrm{O}\left({}^{12}\mathrm{C},\alpha\right){}^{24}\mathrm{Mg}$ & r\_c12\_o16\_to\_he4\_mg24 & 77 & ${ }^{29} \mathrm{Si} (n, \gamma)^{30} \mathrm{Si}$ & r\_si29\_ng\_si30 \\
38 & ${}^{16}\mathrm{O}\left({}^{16}\mathrm{O},\alpha\right)^{28}\mathrm{Si}$ & r\_o16\_o16\_to\_he4\_si28 &  &  &  \\
39 & ${}^{17}\mathrm{O}(\mathrm{p},\gamma){}^{18}\mathrm{F}\left(\beta^{+}\nu_{e}\right)^{18}\mathrm{O}$ & ro17\_to\_o18 &  &  &  \\
40 & ${}^{17}\mathrm{O}(\mathrm{p},\alpha){}^{14}\mathrm{N}$ & r\_o17\_pa\_n14 &  &  &  \\ \hline
\end{tabular}}
\end{table*}


\section{Stellar distances and luminosities}\label{app_field}
All parameters for sources in NGC\,3293 are taken directly from \cite{Proffitt2024}, while effective temperature and surface gravity estimates are taken from a variety of sources, as listed in Table \ref{tab_obs}. Concerning distances for this latter group of sources we used the parallax estimates from $Gaia$ DR3 \citep{GaiaEDR3} or, for the brightest stars, Hipparcos \citep{Perryman1997}.
Typical uncertainties in parallax are $\sim$5\% , with only six stars having uncertainties above 10\%, a consequence of the bulk of the sample being closer than 500\,pc.
Note that for a small number of sources there are alternate estimates of distances, though in general the differences are small, for example the dynamical distance to the Trapezium stars HD\,37020, 37023 and 37042 is 410$\pm$20\,pc \citep{Kraus2009}.
Also, we did not correct $Gaia$ parallax measurements for potential bias, as discussed by \cite{Lindegren2021} as such corrections are of order a few tens of micro-arcseconds and hence small compared to the parallaxes of our sources. 
Furthermore, the error budget for the stellar luminosities is dominated by the uncertainties in stellar parameters and extinction. 
Adopted distance estimates are listed in Table \ref{tab_field} along with adopted visual magnitudes and extinction, that were used to estimate the associated absolute visual magnitudes ($M_{\rm V}$). 
Bolometric corrections (BC) were by obtained by interpolation in the TLUSTY model atmosphere BC tables \citep{Lanz2007} with which the stellar luminosity and radius were determined.
Uncertainties in luminosity were derived by propagating observational errors of the stellar parameters, photometry and parallaxes. Errors in $\Teff$ and $\logg$ are listed in \Table{tab_obs}, and where these were not available from the literature we adopted an error of $\pm$1000\,K in $\Teff$ and $\pm$0.1 in $\logg$. Photometric errors of 0.03\,mag were adopted for all sources.  As detailed knowledge is lacking for the specific extinction law towards most of our sources we adopted $R_V = 3.1$, exceptions being the stars HD\,34078, 36629, 37020, 37023, 37042, 37356 and BD+56\,576 where we used explicit estimates from the literature of 3.40, 3.36, 6.33, 4.82, 5.50, 3.83 and 2.80 respectively \citep{Fitzpatrick2007,Simondiaz2006,Kraus2009}.

\begin{table*}[]
\begin{center}
\caption{List of stellar parameters of the sample related to the luminosity determination. Here, the stars are ordered by name.}
\label{tab_field}
\resizebox{0.65\linewidth}{!}{%
\begin{tabular}{cccccccccc}
\hline \hline
No. & Star & \begin{tabular}[c]{@{}c@{}}Spectral\\ Type\end{tabular} & \begin{tabular}[c]{@{}c@{}}$V$\\ {[}mag{]}\end{tabular} & \begin{tabular}[c]{@{}c@{}}$B-V$\\ {[}mag{]}\end{tabular} & \begin{tabular}[c]{@{}c@{}}$E(B-V)$\\ {[}mag{]}\end{tabular} & \begin{tabular}[c]{@{}c@{}}$Mv$\\ {[}mag{]}\end{tabular} & \begin{tabular}[c]{@{}c@{}}BC\\ {[}mag{]}\end{tabular} & \begin{tabular}[c]{@{}c@{}}dist.\\ {[}pc{]}\end{tabular} & \begin{tabular}[c]{@{}c@{}}$\logL$\\ {[}$\Lsun${]}\end{tabular} \\ \hline
40 & HD 886 & B2IV & 2.83 & -0.19 & 0.03 & -3.06 & -2.18 & 144 ± 9 & 3.99 ± 0.08 \\
65 & HD 3360 & B2IV & 3.69 & -0.20 & 0.02 & -2.26 & -2.03 & 150 ± 7 & 3.61 ± 0.07 \\
66 & HD 16582 & B2IV & 4.08 & -0.21 & 0.01 & -2.42 & -2.09 & 195 ± 11 & 3.70 ± 0.08 \\
49 & HD 22951 & B0.5V & 4.97 & -0.05 & 0.21 & -3.52 & -2.75 & 370 ± 18 & 4.40 ± 0.07 \\
38 & HD 24131 & B1V & 5.78 & -0.04 & 0.21 & -2.68 & -2.62 & 365 ± 11 & 4.02 ± 0.06 \\
77 & HD 24760 & B1.5III & 2.90 & -0.20 & 0.06 & -3.73 & -2.74 & 196 ± 9 & 4.48 ± 0.07 \\
31 & HD 29248 & B2III & 3.93 & -0.21 & 0.02 & -2.72 & -2.19 & 208 ± 11 & 3.86 ± 0.06 \\
83 & HD 30836 & B2III & 3.68 & -0.16 & 0.06 & -3.59 & -2.10 & 258 ± 19 & 4.17 ± 0.09 \\
74 & HD 34078 & O9.5V & 5.99 & 0.20 & 0.49 & -3.65 & -3.08 & 391 ± 5 & 4.59 ± 0.06 \\
3 & HD 34816 & B0.5V & 4.29 & -0.24 & 0.04 & -3.21 & -2.93 & 299 ± 16 & 4.35 ± 0.07 \\
11 & HD 35039 & B2IV & 4.72 & -0.17 & 0.04 & -3.12 & -1.89 & 350 ± 20 & 3.90 ± 0.06 \\
12 & HD 35299 & B2III & 5.69 & -0.21 & 0.02 & -2.17 & -2.35 & 361 ± 12 & 3.70 ± 0.06 \\
44 & HD 35337 & B2IV & 5.25 & -0.22 & 0.02 & -2.65 & -2.36 & 370 ± 12 & 3.90 ± 0.06 \\
89 & HD 35468 & B2V & 1.64 & -0.22 & 0.01 & -2.84 & -2.16 & 77 ± 3 & 3.90 ± 0.06 \\
55 & HD 36285 & B2III & 6.33 & -0.19 & 0.02 & -1.49 & -2.16 & 353 ± 7 & 3.35 ± 0.06 \\
8 & HD 36351 & B3II/III & 5.46 & -0.18 & 0.04 & -2.24 & -2.18 & 326 ± 12 & 3.66 ± 0.07 \\
16 & HD 36430 & B3V & 6.23 & -0.18 & 0.02 & -1.83 & -1.91 & 397 ± 11 & 3.39 ± 0.07 \\
75 & HD 36512 & O9.7V & 4.62 & -0.26 & 0.04 & -3.55 & -3.12 & 407 ± 23 & 4.56 ± 0.07 \\
84 & HD 36591 & B1IV & 5.34 & -0.19 & 0.06 & -3.01 & -2.66 & 426 ± 33 & 4.16 ± 0.09 \\
25 & HD 36629 & B2V & 7.65 & -0.01 & 0.21 & -1.22 & -2.23 & 429 ± 6 & 3.27 ± 0.07 \\
18 & HD 36959 & B1V & 5.67 & -0.23 & 0.02 & -2.17 & -2.59 & 358 ± 14 & 3.80 ± 0.05 \\
54 & HD 36960 & B0.5V & 4.78 & -0.25 & 0.02 & -3.20 & -2.82 & 382 ± 17 & 4.30 ± 0.06 \\
21 & HD 37017 & B2/3V & 6.57 & -0.14 & 0.05 & -1.36 & -1.82 & 360 ± 9 & 3.17 ± 0.08 \\
26 & HD 37020 & B0V & 6.73 & 0.02 & 0.29 & -3.03 & -2.96 & 379 ± 10 & 4.29 ± 0.08 \\
2 & HD 37023 & B1.5Vp & 6.70 & 0.09 & 0.39 & -3.37 & -3.07 & 439 ± 6 & 4.47 ± 0.07 \\
71 & HD 37042 & B0.7V & 6.38 & -0.09 & 0.17 & -2.68 & -2.86 & 418 ± 9 & 4.11 ± 0.07 \\
23 & HD 37209 & B2IV & 5.71 & -0.21 & 0.03 & -2.31 & -2.40 & 388 ± 14 & 3.78 ± 0.06 \\
28 & HD 37356 & B3V & 6.16 & -0.03 & 0.19 & -2.86 & -2.23 & 455 ± 12 & 3.93 ± 0.08 \\
72 & HD 37481 & B1.5IV & 5.96 & -0.22 & 0.02 & -1.91 & -2.32 & 365 ± 9 & 3.59 ± 0.06 \\
19 & HD 37744 & B2IV & 6.22 & -0.19 & 0.04 & -1.90 & -2.39 & 397 ± 11 & 3.61 ± 0.05 \\
32 & HD 38622 & B2IV-V & 5.28 & -0.16 & 0.03 & -2.12 & -1.61 & 289 ± 9 & 3.39 ± 0.08 \\
29 & HD 41753 & B3IV & 4.42 & -0.16 & 0.02 & -1.97 & -1.55 & 185 ± 7 & 3.31 ± 0.08 \\
1 & HD 44743 & B1II-III & 1.98 & -0.24 & 0.01 & -3.95 & -2.36 & 151 ± 5 & 4.42 ± 0.06 \\
37 & HD 45546 & B2V & 5.06 & -0.18 & 0.03 & -1.98 & -1.88 & 245 ± 17 & 3.44 ± 0.09 \\
86 & HD 46328 & B0.7IV & 4.34 & -0.25 & 0.02 & -4.04 & -2.68 & 461 ± 28 & 4.58 ± 0.07 \\
80 & HD 50707 & B1IV & 4.82 & -0.21 & 0.04 & -3.16 & -2.54 & 373 ± 15 & 4.18 ± 0.07 \\
90 & HD 51309 & B3Ib & 4.36 & -0.06 & 0.12 & -6.18 & -1.60 & 1053 ± 211 & 5.01 ± 0.20 \\
81 & HD 52089 & B1.5II & 1.50 & -0.21 & 0.02 & -4.04 & -2.25 & 124 ± 2 & 4.41 ± 0.06 \\
56 & HD 56139 & B2.5Ve & 4.01 & -0.15 & 0.05 & -3.40 & -1.88 & 281 ± 12 & 4.01 ± 0.08 \\
5 & HD 63578 & B2III & 5.22 & -0.15 & 0.08 & -3.34 & -2.44 & 459 ± 25 & 4.21 ± 0.08 \\
17 & HD 64722 & B2IV & 5.70 & -0.15 & 0.09 & -2.66 & -2.46 & 415 ± 10 & 3.95 ± 0.06 \\
35 & HD 64740 & B2V & 4.63 & -0.23 & 0.01 & -2.35 & -2.27 & 245 ± 7 & 3.74 ± 0.06 \\
36 & HD 68243 & B2III & 4.17 & -0.18 & 0.06 & -3.92 & -2.41 & 383 ± 18 & 4.43 ± 0.07 \\
78 & HD 74575 & B1.5III & 3.68 & -0.18 & 0.05 & -3.41 & -2.26 & 243 ± 12 & 4.16 ± 0.06 \\
70 & HD 93030 & B0Vp & 2.74 & -0.22 & 0.06 & -3.04 & -2.96 & 131 ± 13 & 4.30 ± 0.10 \\
58 & HD 106490 & B2IV & 2.79 & -0.19 & 0.04 & -3.06 & -2.27 & 139 ± 14 & 4.03 ± 0.10 \\
73 & HD 108248 & B0.5IV & 1.28 & -0.18 & 0.08 & -3.18 & -2.80 & 69 ± 3 & 4.29 ± 0.07 \\
67 & HD 108249 & B1V & 1.58 & -0.17 & 0.08 & -2.86 & -2.57 & 69 ± 3 & 4.07 ± 0.07 \\
30 & HD 109668 & B2IV & 2.69 & -0.18 & 0.04 & -2.76 & -2.12 & 116 ± 11 & 3.85 ± 0.11 \\
50 & HD 110879 & B2V+B3V & 3.31 & -0.18 & 0.03 & -1.70 & -2.10 & 96 ± 4 & 3.42 ± 0.08 \\
24 & HD 111123 & B1IV & 1.25 & -0.24 & 0.02 & -3.48 & -2.67 & 85 ± 7 & 4.36 ± 0.09 \\
6 & HD 112092 & B2IV-V & 4.03 & -0.18 & 0.03 & -1.53 & -2.02 & 125 ± 3 & 3.32 ± 0.07 \\
33 & HD 120324 & B2Vnpe & 3.47 & -0.17 & 0.06 & -2.38 & -2.23 & 136 ± 10 & 3.74 ± 0.09 \\
43 & HD 121743 & B2IV & 3.83 & -0.22 & 0.01 & -1.94 & -2.15 & 141 ± 5 & 3.53 ± 0.06 \\
15 & HD 121790 & B2IV-V & 3.87 & -0.21 & 0.02 & -1.76 & -2.13 & 131 ± 0 & 3.45 ± 0.06 \\
22 & HD 122980 & B2V & 4.36 & -0.20 & 0.02 & -1.63 & -2.06 & 154 ± 5 & 3.37 ± 0.04 \\
4 & HD 142669 & B2IV-V & 3.87 & -0.20 & 0.02 & -1.99 & -2.07 & 145 ± 4 & 3.52 ± 0.06 \\
41 & HD 143018 & B1V+B2:V: & 2.89 & -0.18 & 0.07 & -2.41 & -2.71 & 103 ± 7 & 3.94 ± 0.09 \\
68 & HD 143275 & B0.3IV & 2.29 & -0.12 & 0.15 & -4.09 & -2.91 & 151 ± 20 & 4.70 ± 0.13 \\
62 & HD 160578 & B1.5III & 2.39 & -0.17 & 0.05 & -3.63 & -2.22 & 148 ± 4 & 4.24 ± 0.07 \\
9 & HD 160762 & B3IV & 3.82 & -0.18 & 0.01 & -2.15 & -1.61 & 153 ± 5 & 3.40 ± 0.07 \\
48 & HD 180163 & B2.5IV & 4.43 & -0.15 & 0.03 & -3.35 & -1.47 & 344 ± 13 & 3.82 ± 0.08 \\
59 & HD 184171 & B3IV & 4.74 & -0.15 & 0.03 & -2.21 & -1.44 & 236 ± 5 & 3.36 ± 0.08 \\
52 & HD 188252 & B2III & 5.91 & -0.17 & 0.04 & -3.98 & -2.07 & 893 ± 40 & 4.32 ± 0.08 \\
34 & HD 202347 & B1V & 7.52 & -0.11 & 0.12 & -2.29 & -2.32 & 775 ± 24 & 3.74 ± 0.07 \\
85 & HD 205021 & B0.5IIIs & 3.23 & -0.20 & 0.05 & -3.54 & -2.66 & 210 ± 13 & 4.38 ± 0.08 \\
7 & HD 212978 & B2V & 6.16 & -0.14 & 0.07 & -2.53 & -1.94 & 498 ± 12 & 3.69 ± 0.08 \\
53 & HD 214263 & B2V & 6.84 & -0.13 & 0.07 & -1.93 & -1.98 & 510 ± 10 & 3.46 ± 0.07 \\
69 & HD 214680 & O9V & 4.89 & -0.21 & 0.09 & -3.70 & -3.13 & 457 ± 27 & 4.63 ± 0.07 \\
57 & HD 214993 & B2III & 5.25 & -0.14 & 0.10 & -3.00 & -2.41 & 386 ± 19 & 4.06 ± 0.07 \\
42 & HD 216916 & B2IV & 5.60 & -0.15 & 0.08 & -2.89 & -2.29 & 446 ± 14 & 3.97 ± 0.07 \\
13 & BD+56 576 & B1.5I & 9.40 & 0.28 & 0.51 & -4.01 & -2.21 & 2500 ± 125 & 4.39 ± 0.07 \\
88 & 3293-002 & B0.7 Ib & 6.67 & 0.03 & 0.26 & -5.97 & -2.35 & 2335 ± 57 & 5.22 ± 0.05 \\
87 & 3293-003 & B1 III & 7.58 & 0.05 & 0.27 & -5.10 & -2.18 & 2335 ± 57 & 4.81 ± 0.06 \\
61 & 3293-004 & B1 III & 7.99 & -0.01 & 0.22 & -4.53 & -2.33 & 2335 ± 57 & 4.64 ± 0.06 \\
64 & 3293-005 & B1 III & 8.08 & 0.04 & 0.26 & -4.57 & -2.19 & 2335 ± 57 & 4.60 ± 0.08 \\
82 & 3293-006 & B1 III & 8.18 & 0.02 & 0.25 & -4.44 & -2.33 & 2335 ± 57 & 4.60 ± 0.07 \\
79 & 3293-007 & B1 III & 8.21 & 0.13 & 0.36 & -4.74 & -2.25 & 2335 ± 57 & 4.69 ± 0.06 \\
60 & 3293-008 & B1 III & 8.54 & 0.00 & 0.23 & -4.02 & -2.31 & 2335 ± 57 & 4.43 ± 0.06 \\
63 & 3293-010 & B1 III & 8.70 & -0.05 & 0.18 & -3.71 & -2.32 & 2335 ± 57 & 4.31 ± 0.06 \\
27 & 3293-012 & B1 III & 8.91 & 0.02 & 0.26 & -3.73 & -2.42 & 2335 ± 57 & 4.35 ± 0.06 \\
10 & 3293-015 & B1 V & 9.09 & -0.06 & 0.19 & -3.33 & -2.56 & 2335 ± 57 & 4.25 ± 0.07 \\
76 & 3293-016 & BI III & 9.31 & -0.02 & 0.22 & -3.22 & -2.50 & 2335 ± 57 & 4.18 ± 0.06 \\
46 & 3293-019 & B1 V & 9.22 & -0.08 & 0.16 & -3.13 & -2.51 & 2335 ± 57 & 4.15 ± 0.05 \\
14 & 3293-023 & B1.5 III & 9.95 & -0.07 & 0.16 & -2.37 & -2.26 & 2335 ± 57 & 3.75 ± 0.06 \\
45 & 3293-024 & B1.5 III & 9.95 & -0.05 & 0.17 & -2.42 & -2.20 & 2335 ± 57 & 3.75 ± 0.06 \\
47 & 3293-028 & B2 V & 10.24 & 0.01 & 0.22 & -2.29 & -2.09 & 2335 ± 57 & 3.65 ± 0.10 \\
39 & 3293-030 & B2 V & 10.51 & -0.03 & 0.17 & -1.87 & -1.92 & 2335 ± 57 & 3.41 ± 0.10 \\
51 & 3293-031 & B2 V & 10.59 & 0.02 & 0.23 & -1.97 & -2.08 & 2335 ± 57 & 3.51 ± 0.09 \\
20 & 3293-038 & B2.5 V & 10.95 & -0.01 & 0.20 & -1.50 & -1.98 & 2335 ± 57 & 3.29 ± 0.10 \\ \hline
\end{tabular}}
\end{center}
\end{table*}

\section{Nitrogen enhancement factors}\label{app_cno}
\begin{figure*} 
	\centering
	\includegraphics[width=\linewidth]{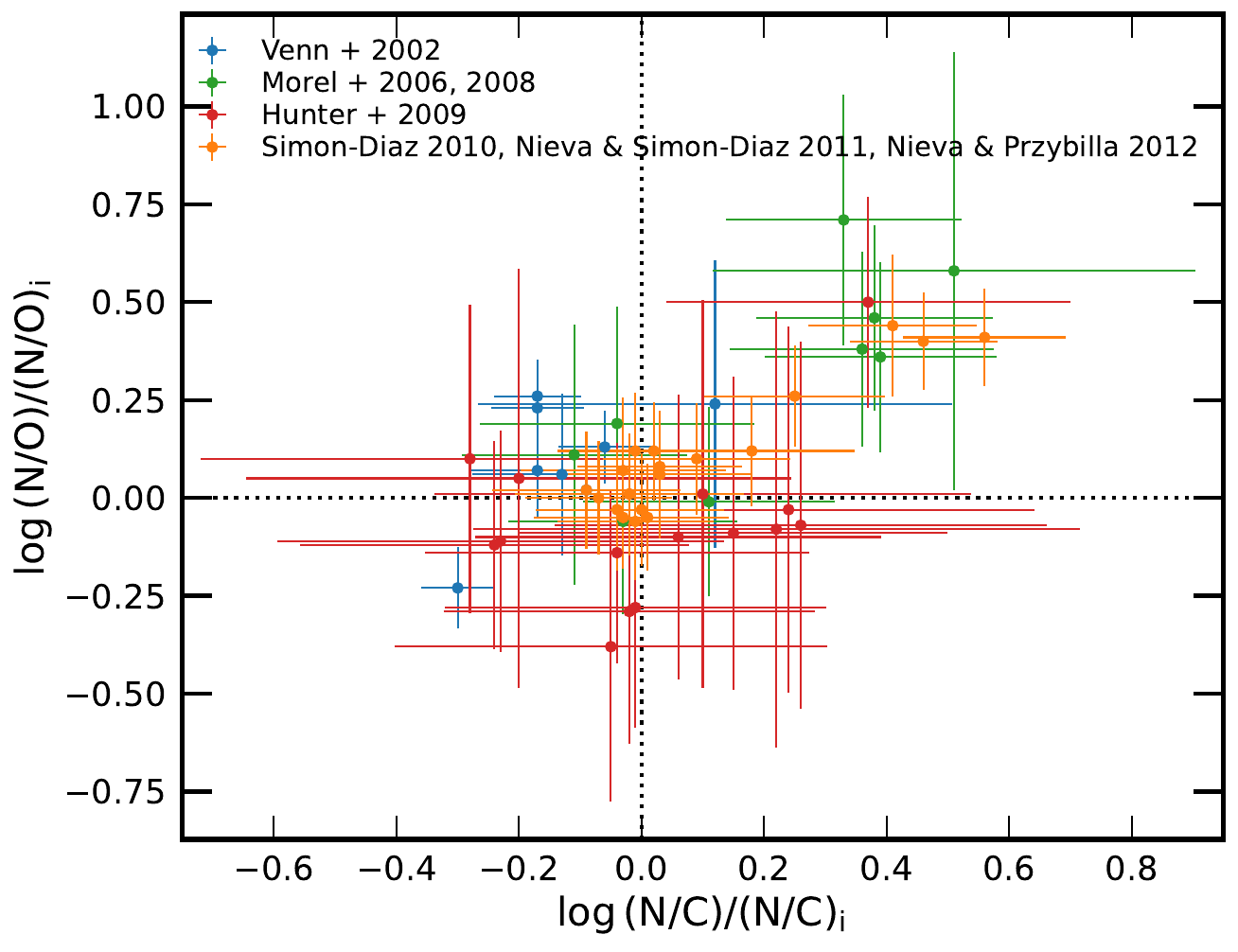}
	\caption{Nitrogen enhancement factors of the stars, defined as nitrogen abundance over carbon abundance (x-axis) and oxygen abundance (y-axis) with respect to their initial values (see Table~\ref{tab_cno}). The error bars show the original $\Delta \logNCNCi$ and $\Delta \logNCNCi$ before adopting the maximum of 0.4\,dex, or the original values as the uncertainties. Each data point is color-coded based on the reference from which the surface carbon, nitrogen, and oxygen abundance are obtained. The black dotted lines show the baselines of no nitrogen enhancement.}
	\label{fig_cno}
\end{figure*}

CNO abundances are available for $\sim \, 60$ stars in our sample. The abundances are compiled from several works, meaning that they are not homogeneously derived. To minimize any potential systematic difference between different abundance analysis methods, we use nitrogen abundance with respect to their carbon and oxygen abundance, $\logNC$ and $\logNO$, for our analysis. To identify nitrogen-enhanced stars, we plot the stars' $\logNC$ and $\logNO$ with respect to their initial values in \Fig{fig_cno}. The adopted initial values for NGC\,3293 stars and solar-neighborhood stars are presented in Table~\ref{tab_cno}. Some stars show nitrogen depletion, $\logNCNCi < 0 $ or $\logNONOi < 0 $, which is not predicted for the main sequence stars, possibly due to the old abundance analyses. 
To force no stars to show this, we force the minimum uncertainty of 0.4\,dex for $\Delta \logNCNCi$ and $\Delta \logNONOi$. Then, all the stars in the bottom ($\logNONOi < 0 $) or left ($\logNCNCi < 0 $) of the plane have error ellipses that contain the origin of $\logNCNCi=0$, $\logNONOi=0$.

Even with this generous uncertainty assumption, nine stars in the upper-rightmost corner show a clear sign of nitrogen enhancement; their error ellipses do not contain the origin. We identify these stars as nitrogen-enhanced stars and refer to them in the text.

One interesting feature in \Fig{fig_cno} is an apparent gap between the nitrogen-normal stars and the nitrogen-rich stars. This is in line with the findings of \citet{Herrero2004,Morel2008} as they also identified an indication of a gap at $\logNC \sim -0.3$ ($\logNCNCi \sim 0.3$ in our figure) in their samples, which mainly consist of slowly rotating stars. Such a gap could be consistent with the outcome of binary mass transfer, which can  change the surface nitrogen abundance rather abruptly \citep{Langer2010}. Indeed, six out of nine nitrogen-rich stars in our sample belong to the low $\Pi$ group.

\section{The correlation between $\pev$ and $\pmax$}\label{app_pev}
\begin{figure*} 
	\centering
	\includegraphics[width=\linewidth]{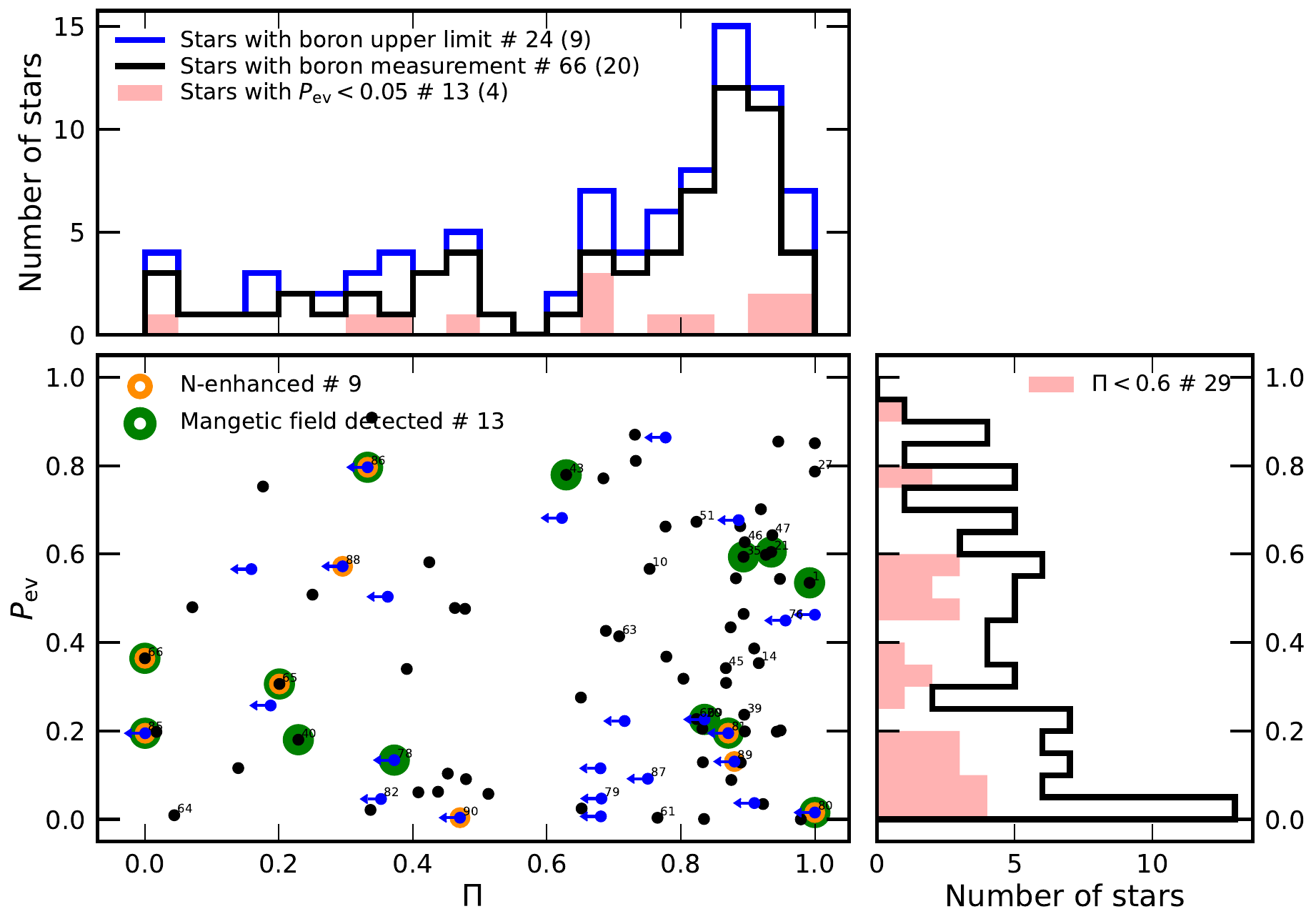}
	\caption{Same as \Fig{fig_pvsini}, but with $\pev$ in the ordinate instead of $\vsini$. The lower right panel shows the distribution of $\pev$, and the red histogram in the upper left panel shows the distribution corresponding to stars with $\pev < 0.05$.}
	\label{fig_pev}
\end{figure*}

The low value of $\pev$ for the two cases, 1. stars near the edge of the main sequence band and 2. stars showing mass discrepancy, is expected to be irrelevant to rotational mixing. We involve $\pev$ in Eq.~\ref{e3} to single out the sole effect of rotational mixing in $\pmax$. To check if $\pev$ is truly not correlated with $\pmax$, the quantity that signifies rotational mixing, we plot the two quantities in \Fig{fig_pev}. Data points are all over the plane: stars with small values of $\pev$ are spread over the whole $\pmax$ range, and stars in the low $\pmax$ group are also spread over the whole $\pev$ range. As expected, no correlation is found between $\pev$ and $\pmax$.

\section{Outliers in the low $\pmax$ group}\label{app_out}
\begin{figure*} 
	\centering
	\includegraphics[width=0.45\linewidth]{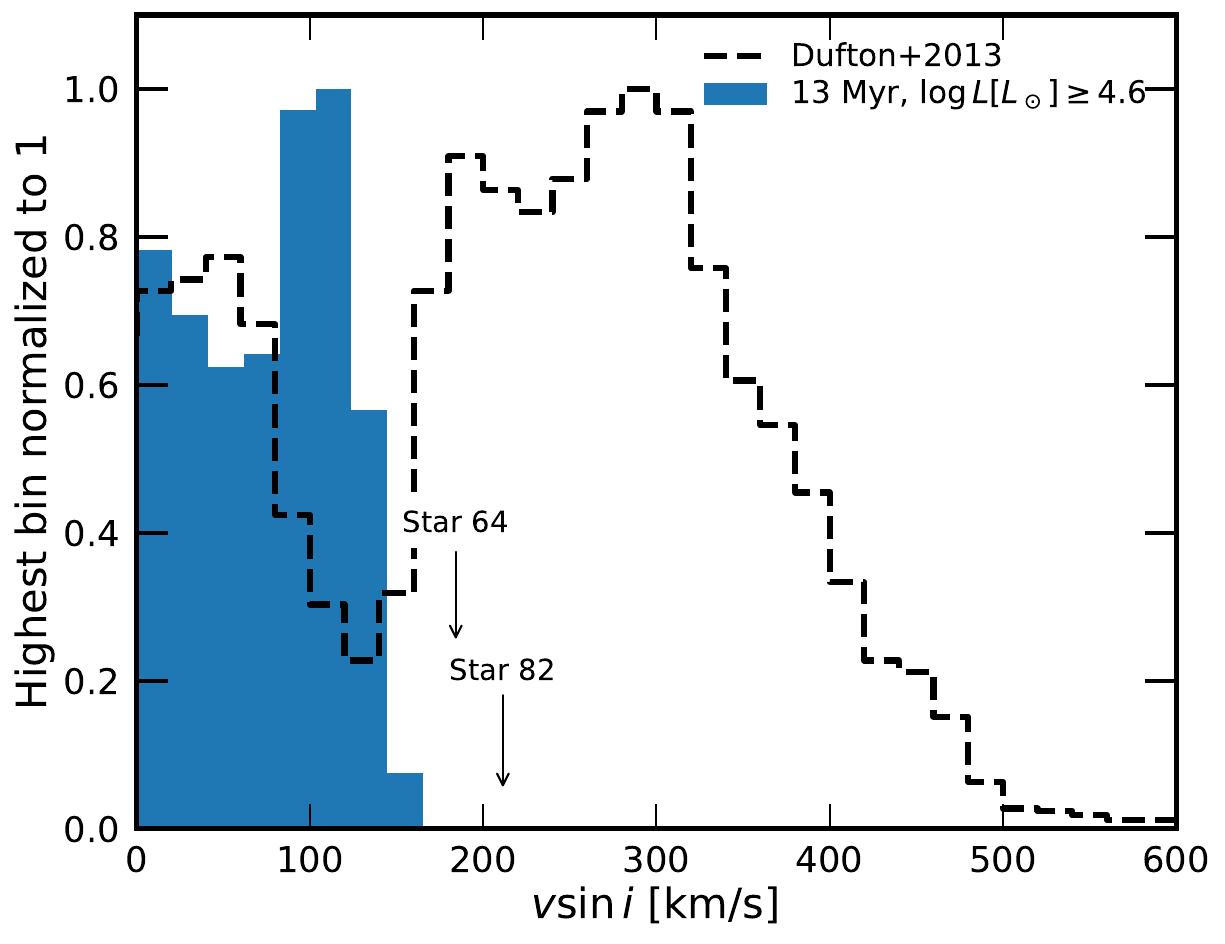}
	\includegraphics[width=0.45\linewidth]{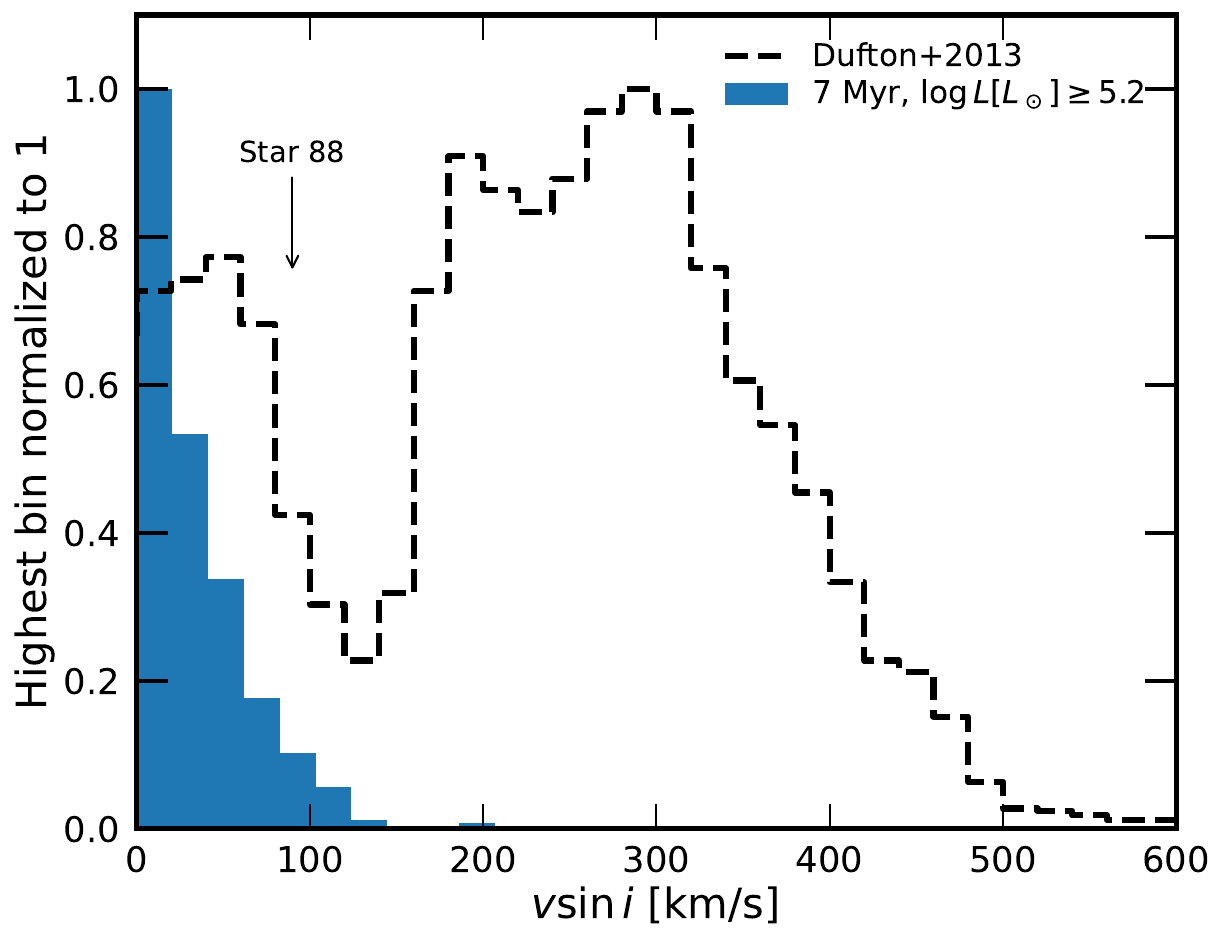}
	\caption{Distribution of $\vsini$ from population synthesis calculations for evolved massive main sequence stars based on our single star model  (blue histogram), for the ages and luminosities shown in the legend. The dashed line shows our adopted initial rotational velocity distribution from \citet{Dufton2013}. The $\vsini$ values of Stars\,64, 82, 88 are indicated by arrows in the corresponding panels of their approximate age and luminosity regime.}
	\label{fig_out}
\end{figure*}

The low $\pmax$ group is dominated by slow rotators, and only three stars (Stars\, 64, 82, 88) stand out in \Fig{fig_pvsini} with their $\vsini$ of $\gtrsim 90 \kms$. We argue in the text that these stars show low $\pmax$ not because of too much boron depletion for their rotation, like other slow rotators in the low $\pmax$ group, but because their rotation is not compatible with our models. In this section, we substantiate this argument by comparing the $\vsini$ of the stars and our models. In \Fig{fig_out}, the majority of our stellar models show lower $\vsini$ than the stars. This is because of the spin-down, in particular, in luminous, fast rotating models near the end of the main sequence. Most of the stellar models corresponding to our sample stars ($M\lesssim 20\mso$ and $\vini \lesssim 200 \kms$) do not experience significant angular momentum loss during the main sequence phase (\Fig{fig_grid}). However, the three outliers are among the most luminous, fastly rotating, and evolved, so the considered model parameter space is such that the angular momentum loss can be non-negligible. The incompatibility in the rotation of the three outliers and the models might come from a possible overestimation in angular momentum loss in our models. We do not investigate this further here since the majority of our sample stars are not affected by this.

\end{appendix}

\end{document}